\documentclass[twocolumn,prd,nofootinbib,showpacs,preprintnumbers]{revtex4}
\usepackage{graphicx}
\usepackage{color}
\usepackage{amsmath,amssymb,amsfonts,dsfont}
\usepackage{multirow}
\usepackage{sistyle}
\usepackage{ulem}

\pdfoutput=1 % if your are submitting a pdflatex (i.e. if you have
\newcommand{\lsim}{\mathrel{\mathop{\kern 0pt \rlap
  {\raise.2ex\hbox{$<$}}}
  \lower.9ex\hbox{\kern-.190em $\sim$}}}
\newcommand{\gsim}{\mathrel{\mathop{\kern 0pt \rlap
  {\raise.2ex\hbox{$>$}}}
  \lower.9ex\hbox{\kern-.190em $\sim$}}}

\def \aap  {A\&A}

\def \apjs  {ApJS}
\def \apjl  {ApJL}

\begin{document}

%\preprint{TTK-20-XX}

%\title{Detectability of a dark matter signal from the Galactic center by the {\it Fermi} Large Area Telescope}
\title{Investigating the {\it Fermi} Large Area Telescope sensitivity of detecting the characteristics of the Galactic center excess}

\author{Mattia Di Mauro,}\email{dimauro.mattia@gmail.com}
\affiliation{NASA Goddard Space Flight Center, Greenbelt, MD 20771, USA}
\affiliation{Catholic University of America, Department of Physics, Washington DC 20064, USA}

%\date{\today} 
\begin{abstract}
The center of the Milky Way is offering one of the most striking mystery in Astroparticle Physics.
An excess of $\gamma$ rays (GCE) has been measured by several groups in the data collected by the {\it Fermi} Large Area Telescope (LAT) towards the Galactic center region.
The spectrum and spatial morphology of the GCE have been claimed by some groups to be compatible with a signal from the Galactic halo of dark matter (DM). Instead, other analyses have demonstrated that the GCE properties, e.g., its energy spectrum, highly depend on the choice of the Galactic interstellar emission (IEM) model source catalogs and analysis techniques.

In this paper we investigate the sensitivity of {\it Fermi}-LAT to detect the characteristics of the GCE.
%a signal of $\gamma$ rays from the Galactic halo of DM, which has its brightest emission from the Galactic center.
In particular we simulate the GCE as given by DM and we verify that, with a perfect knowledge of the background components, its energy spectrum, position, spatial morphology and symmetry is properly measured. 
We also inspect two more realist cases for which there are imperfections in the IEM model. In the first we have an un-modeled $\gamma$-ray source, constituted by the low-latitude component of the {\it Fermi} bubbles. In the second we simulate the data with one IEM template and analyze the data with an other.
We verify that a mismodeling of the IEM introduces a systematics of about $10-15\%$ in the GCE energy spectrum between 1-10 GeV and about 5\% in the value of the slope for a NFW DM density profile, which is used to fit the GCE spatial morphology.
%Moreover, we present a model independent technique to inspect the spatial distribution of the GCE and test its spherical symmetry.
Finally, we show how the GCE would be detected in case of alternative processes such as $\gamma$-ray emission from a bulge population of pulsars or from electrons and positrons or protons injected from the Galactic center. We demonstrate that for each of these cases there is a distinctive smoking gun signature that would help to identify the real mechanism behind the origin of the GCE.

\end{abstract}

\maketitle
%\flushbottom

%%%%%%%%%%%%%%%%%%%%%%%%%%%%%%%%%%%%%%%%%%%%%%%%%%%%%%%%%%%%
\section{Introduction}
\label{sec:intro}
%%%%%%%%%%%%%%%%%%%%%%%%%%%%%%%%%%%%%%%%%%%%%%%%%%%%%%%%%%%%
The center of the Milky Way is one of the most intriguing regions for $\gamma$-ray Astrophysics.
The $\gamma$-ray emission from the Galactic center is mainly produced by the Galactic interstellar emission (IEM), which is due to the interactions of cosmic rays with the interstellar medium and radiation fields, and by individual sources, mainly supernova remnants and pulsar wind nebulae.
This region is also theorized to be the brightest for a putative $\gamma$-ray signal from dark matter (DM) particle interactions (see, e.g., \cite{Pieri:2009je}). 
DM is modeled in N-body simulations to have a main Galactic halo, of the size of about hundreds of kpc, and smaller subhalos distributed allover in the Galaxy. The same structure should also be present in other galaxies. Therefore, the Galactic center is the direction with the predicted highest $\gamma$-ray emission from DM.
 
Several groups have detected an excess of $\gamma$ rays in {\it Fermi}-LAT data measured from the Galactic center region (see, e.g., \cite{Goodenough:2009gk,Hooper:2010mq,Boyarsky:2010dr,Hooper:2011ti,Abazajian:2012pn,Gordon:2013vta,Abazajian:2014fta,Daylan:2014rsa,Calore:2014nla,Calore:2014xka,TheFermi-LAT:2015kwa,TheFermi-LAT:2017vmf}).
We will label in the rest of the paper this excess as the Galactic center excess (GCE).
The GCE has a peak, in the flux calculated as $E^2 dN/dE$, at around a few GeV, where $E$ is the $\gamma$-ray energy and $dN/dE$ is the flux in units of MeV$^{-1}$ cm$^{-1}$ s$^{-1}$.
Moreover, it has been detected as significant with respect to several choices for the IEM models, source catalogs, data selections and analysis techniques.

The GCE has been described by some groups (see, e.g., \cite{Daylan:2014rsa}) to be spherically symmetric and centered around the Galactic center.
Moreover, the flux would be compatible with DM particles annihilating through hadronic channels (for example into $b\bar{b}$) with a thermal annihilation cross section which provides a density of DM compatible with the observations  \cite{Aghanim:2018eyx}. This makes the existence of the GCE very appealing for DM and in general for new Physics searches.

Instead, Refs.~\cite{Calore:2014xka,TheFermi-LAT:2015kwa,TheFermi-LAT:2017vmf} have tested several IEM models and source catalogs and shown that the properties of the GCE, in particular its spectrum, are too uncertain to conclude that it is of DM origin. 
Moreover, other regions along the Galactic plane, where a DM signal is not expected to contribute, show excesses of similar amplitude relative to the local background \cite{Calore:2014xka,TheFermi-LAT:2017vmf}. 
These results cast serious doubts on the DM interpretation of the GCE.

Refs.~\cite{Bartels:2015aea,Lee:2015fea} have published compelling evidences for the existence of a faint population of sources located in the Galactic center region and with properties that can explain the GCE.
These results have been derived using two techniques, called wavelet analysis and non-Poissonian template fitting, which inspect the possible presence of sources shining under the {\it Fermi}-LAT detection threshold.
These faint sources could be interpreted as a population of millisecond pulsars located in the bulge of the Milky Way.
Very recently references \cite{Leane:2019uhc,Chang:2019ars} have casted doubts on the robustness of the results presented in \cite{Lee:2015fea}, and as a consequence on the fact that the GCE is due to a population of pulsars. 
They have shown that the non-Poissonian template fitting method can misattribute, to point sources or DM unmodeled point sources, imperfections in the modeling of {\it Fermi}-LAT data.
On the other hand, the authors of \cite{Zhong:2019ycb}  have used the latest 4FGL catalog released by the {\it Fermi}-LAT Collaboration and found that the excess is still present. However, when they apply a wavelet  method to the Galactic center region, similarly to what done in \cite{Bartels:2015aea}, they do not find any evidence for the existence of a faint population of un-modeled sources which can be attributed to the GCE.

An alternative interpretation for the GCE is associated to cosmic rays produced from the Galactic center during recent outbursts.
$\gamma$ rays are produced via neutral
pion ($\pi^0$) production (mainly from protons interacting with interstellar medium atoms) \cite{Carlson:2014cwa} or through electrons and positrons inverse Compton scattering on the interstellar radiation fields \cite{Petrovic:2014uda,Gaggero:2015nsa}.
However, the hadronic scenario (i.e., protons) predicts a $\gamma$-ray signal that is
significantly extended along the Galactic plane since the $\pi^0$ production is correlated with
the distribution of gas present in the Milky Way. 
This spatial shape is incompatible with the
observed characteristics of the GCE \cite{Petrovic:2014uda}.  
On the other hand, the case of a leptonic outburst leads to a signal that is more smoothly distributed
and spherically symmetric.
However, it requires a complicated scenario with at least two outbursts
to explain the morphology and the intensity of the excess with the older
outbursts injecting more-energetic electrons. 
%Finally, an additional population of supernova remnants near the GC that steadily injects
%CRs is also a viable interpretation for the GC excess~\cite{Gaggero:2015nsa,Carlson:2015ona}.

The puzzle of the GCE origin is thus far from being solved and recent publications (see, e.g, \cite{Leane:2019uhc,Chang:2019ars,Zhong:2019ycb}) have questioned recent claims that it is due to a population of pulsars in the Galactic bulge.
The main difficulty in finding a robust interpretation is due to model precisely a complicated region as the Galactic center. 
 
In this paper we investigate, for the first time using simulations, the sensitivity of {\it Fermi}-LAT to detect the GCE.
We first assume that the GCE is due to DM and we consider the ideal case for which the astrophysical backgrounds, i.e., point and extended sources and the IEM, are perfectly modeled. We demonstrate that, in this case, we are able to reconstruct properly all the GCE properties, i.e., flux, spatial morphology and position.
In particular we show that, using a technique that is independent by the specific choice of the GCE spatial template, we are able to derive with high precision its spatial morphology.
Then, we consider, in a more realistic scenario, that there are imperfections in the IEM model we use to analyze the data with respect to the one used to generate the simulations.
This is inspected in two ways. 
First, we simulate the data including also the low-latitude component of the {\it Fermi} bubbles and then we do not include it when we analyze the data. 
This circumstances is comparable to the case of a missing component in the data that could be degenerate with the GCE.
Second we simulate the data using one IEM model and then we analyze them with a different model.
%We also test the scenario for which we have a $\gamma$-ray component from the Galactic center that is not included in the model.
In both cases, we demonstrate how imperfections of the IEM model or the presence of an un-modeled component produce systematic uncertainties in the measured flux and spatial morphology of the GCE which are much larger than the statistical errors.
Finally, we generate simulations where the GCE is given by the $\gamma$-ray emission from a bulge population of pulsars or from cosmic-ray electrons and positrons or protons injected from the Galactic center.
We will calculate the properties of the GCE for each of this case and present a distinctive smoking gun signature for each of them that could help to identify the real origin of the GCE.  

The paper is organized as follows: in Sec.~\ref{sec:analysis} we explain the data selection and the IEM models and source catalog we choose for the analysis. We also present our analysis technique and simulation setup. Then, in Sec.~\ref{sec:idealcase} we apply our analysis to the idea case where all the $\gamma$-ray components are perfectly modeled. In Sec.~\ref{sec:realisticcase} we present a more realistic case for which we do not have a perfect modellization of the background components and show how the GCE properties are reconstructed.
Finally, in Sec.~\ref{sec:psdiffusive} we will consider other origins for the GCE and present the flux and spatial morphology would be measured in these cases.

%%%%%%%%%%%%%%%%%%%%%%%%%%%%%%%%%%%%%%%%%%%%%%%%%%%%%%%%
\section{Analysis Framework}
\label{sec:analysis} 
%%%%%%%%%%%%%%%%%%%%%%%%%%%%%%%%%%%%%%%%%%%%%%%%%%%%%%%%%%%%

\subsection{Galactic interstellar emission models and data selection}
\label{sec:iemdata}

The main components of the $\gamma$-ray IEM are due to inelastic hadron collisions and subsequent decay of $\pi^0$ particles, inverse Compton scattering of electrons and positrons on the Galactic interstellar radiation fields and, at energies lower than 10 GeV, bremsstrahlung emission from electrons and positrons interacting with interstellar gas.

The templates we use for the IEM have been produced with the GALPROP code\footnote{\url{http://galprop.stanford.edu}} \cite{1998ApJ...493..694M,2000ApJ...537..763S,2004ApJ...613..962S} which calculates the propagation and interactions of cosmic rays in the Galaxy by numerically solving the transport equations given a model for the cosmic-ray source distribution, injection spectrum, and interaction targets.
We follow Ref.~\cite{TheFermi-LAT:2017vmf} for the choice of the IEM models.
We recall here the main properties and refer to \cite{TheFermi-LAT:2017vmf} for a complete description.

The reference model we use is labelled as {\tt Baseline} and it is taken from one of the models in \cite{2012ApJ...750....3A}.
It assumes a cosmic-ray source distribution traced by the measured distribution of pulsars from \cite{Lorimer:2006qs}, the cosmic-ray confinement volume with a height of 10 kpc and a radius of 20 kpc.
This model assumes HI column densities derived from the 21-cm line intensities for a spin temperature of 150 K. The dust reddening map of \cite{1998ApJ...500..525S} is used to correct the HI maps to account for the presence of dark neutral gas not traced by the combination of HI and CO surveys \cite{2012ApJ...750....3A}. 
%We neglect the contribution from ionized gas. 
Furthermore, this model includes a unique inverse Compton component derived from the interstellar radiation field model reported in \cite{Porter_2008} and takes into account the emission due to the CMB, dust infrared emission, and starlight. 
The model also contains the Loop I, Sun, Moon and {\it Fermi} bubbles emissions.
For the latter we include two components: the low-latitude, closer to the Galactic center, and the high-latitude part.

We also employ other 6 IEM models which make different assumptions on the source distribution, gas component or inverse Compton template.
Each of these models modify only one of the ingredients written above with respect to the {\tt Baseline} model.
%Below we explain the change in the assumptions that each of this alternative IEM models have.
In order to account for different tracers of cosmic rays we use different source distributions: an alternative pulsar distribution (\cite{2004A&A...422..545Y}, hereafter referred to as {\tt Yusifov}), the distribution of supernova remnants \cite{1998ApJ...504..761C} (labelled as {\tt SNR}), and the distribution of OB stars \cite{2000A&A...358..521B} ({\tt Obstars}).
We also test more freedom for the inverse Compton emission separating the template into the three components of the interstellar radiation field ({\tt ICSsplit}).
In order to account for different gas models we use templates generated by information from starlight extinction due to interstellar dust (labelled as {\tt SLext}) and using the high-resolution maps from the the GASS survey \cite{2010A&A...521A..17K} and the dust extinction map from extinction map from \cite{2014A&A...571A..11P} that is built using IRAS and Planck data (labelled as {\tt PlanckGASS}).
As presented in \cite{TheFermi-LAT:2017vmf}, these models provide different results for the GCE flux and they provide an appropriate framework, even if not exhaustive, to bracket the effect of the choice of different IEM models in the measured GCE properties.

In our simulations we select a time range of 11 years, from 2008 August 4 to 2019 August 4, and we simulate $\gamma$-ray events in the energy range $E=[0.1,1000]$ GeV, passing standard data quality selection criteria\footnote{\url{https://fermi.gsfc.nasa.gov/ssc/data/analysis/documentation/Cicerone/Cicerone_Data_Exploration/Data_preparation.html}}.
We also select in some parts of the simulations smaller energy bins between $E=[0.1,1000]$ GeV to test a possible energy dependence in the results. 
In other parts we consider energies between $1-10$ GeV where the GCE is more significant and the LAT energy and spatial resolution is better than at lower energies\footnote{\url{https://www.slac.stanford.edu/exp/glast/groups/canda/lat_Performance.htm}}.
We use a region of interest (ROI) which is $30^{\circ}\times30^{\circ}$ centered around the Galactic center and we include in the model all the sources detected in the 4FGL {\it Fermi}-LAT catalog \cite{Fermi-LAT:2019yla} with an angular distance less than $18^{\circ}$ from the Galactic center.
The ROI size we select is motivated by the fact that within $10^{\circ}$ from the center of the Milky Way it is contained more than $90\%$ the GCE (see, e.g., \cite{Calore:2014xka,TheFermi-LAT:2017vmf}).
We consider events belonging to the Pass~8 {\tt SOURCEVETO} class, and use the corresponding instrument response functions {\tt P8R3\_SOURCEVETO\_V2} which has the same background rate than the {\tt SOURCE} class background rate up to 10 GeV but, above 50 GeV, its background rate is the same as the {\tt ULTRACLEANVETO} one while having 15\% more acceptance\footnote{\url{https://fermi.gsfc.nasa.gov/ssc/data/analysis/documentation/Cicerone/Cicerone_Data/LAT_DP.html#PhotonClassification}.}.
We bin the data assume 8 energy bins per decade and pixels with size $0.08^{\circ}$.

%%%We investigate how our results change making different selection of the data by selecting {\tt SOURCE} class (this case is labelled as {\tt SOURCE}), {\tt ULTRACLEANVETO} class or still using selecting only the {\tt SOURCEVETO} but with the cleanest PSF2 an 3 ({\tt SOURCEVETO PSF23}).

\subsection{Analysis Method}
\label{sec:analysismethod}

\begin{figure*}[t]
\includegraphics[width=0.49\textwidth]{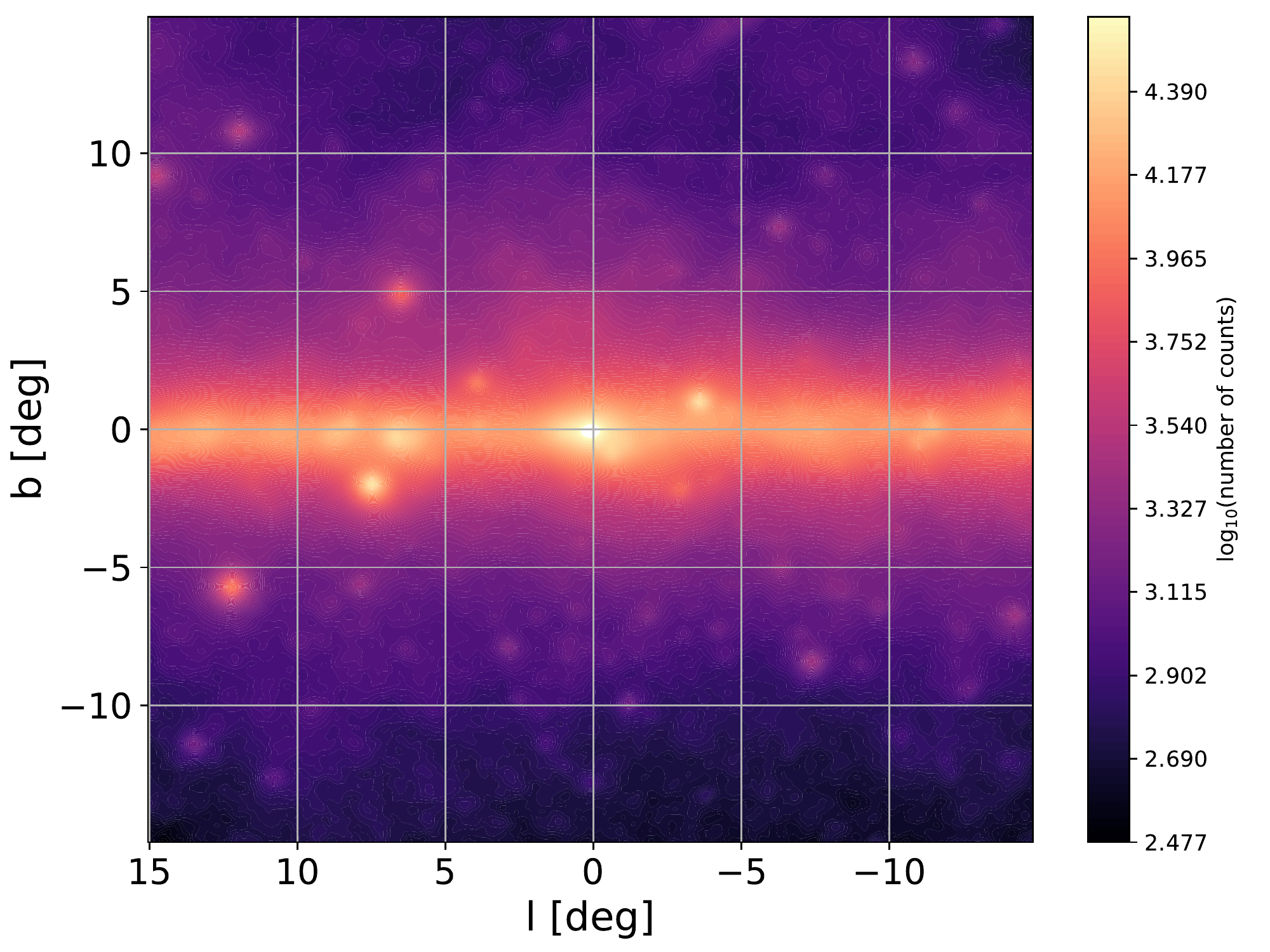}
\includegraphics[width=0.49\textwidth]{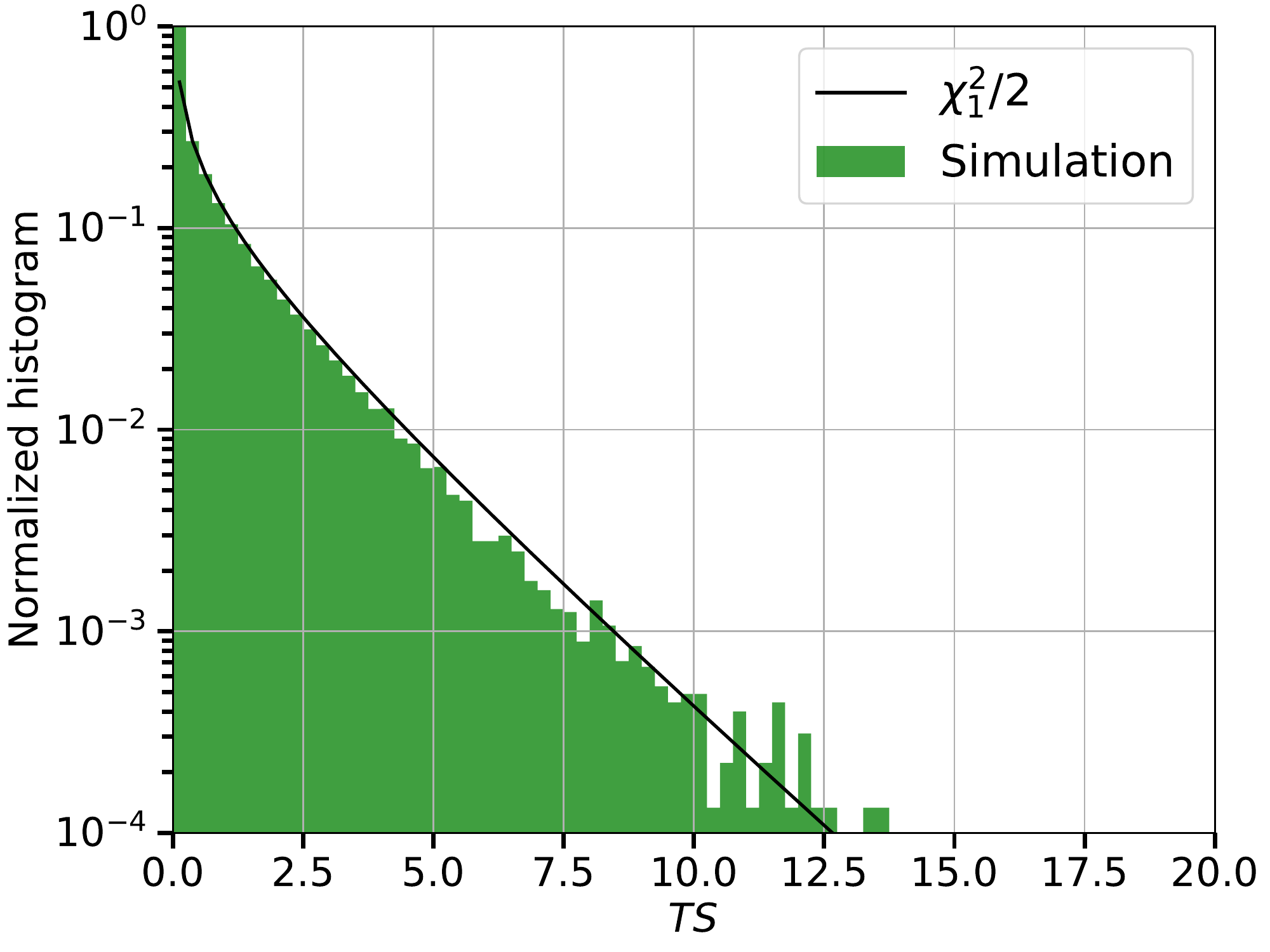}
\caption{Left Panel: Count map for one specific simulation generated with the {\tt Baseline} IEM model and considering the energy range between 1-10 GeV. The color bar represents the logarithm in base 10 of the number of counts.
Right Panel: Normalized histogram of the $TS$ values in the ROI derived after running the fitting procedure explained in Sec.~\ref{sec:analysis}.
We use in this simulation the {\tt Baseline} IEM model and we consider the energy range between 1-10 GeV. We also show the normalized $\chi_1^2/2$ distribution for comparison.}%, with consistent color code as the right panel.}  
\label{fig:TSdistr}
\end{figure*}

%The goal of this paper is to show the potential of using {\it Fermi}-LAT data to detect a DM signal towards the center of our Galaxy.
%In order to accomplish properly this task we need to generate realistic simulations. 
%This means that the flux of simulated point sources, interstellar emission and isotropic templates should be close to the real data.
%In addition the DM signal from the Galactic center should mimic the one detected for the GCE.

We have implemented an analysis pipeline based on {\tt FermiPy}\footnote{See \url{http://fermipy.readthedocs.io/en/latest/}}, a Python package that automates analyses with the {\tt Fermitools}\footnote{The {\tt Fermitools} are the official {\it Fermi} Science Tools \url{https://fermi.gsfc.nasa.gov/ssc/data/analysis/documentation/}} \citep{2017ICRC...35..824W}.
{\tt FermiPy} includes tools that generate simulations of the $\gamma$-ray sky, detect sources, calculate SED, find the extension of sources and much more.
We employ the version {\tt 18.0.0} of {\tt Fermipy} and  {\tt 1.1.7} of the {\tt Fermitools}.

The model we use to fit the ROI includes: the IEM, point and extended sources, isotropic template, the Sun, Moon and Loop I components, the {\it Fermi} bubbles divided into low and high-latitude parts and the DM template.
The template for all the components, except for the DM one, are modeled as explained in Sec.~\ref{sec:iemdata} and we remind are taken from Ref.~\cite{TheFermi-LAT:2017vmf}.

Instead, the DM flux is calculated as:
\begin{equation}
\frac{dN}{dE} = \mathcal{J} \times \frac{dN_\gamma}{dE},
\label{eq:flux}
\end{equation}
where $\mathcal{J}$ is the geometrical factor and represents the integral performed along the line of sight (l.o.s., $s$) of the squared DM density distribution $\rho$:
\begin{equation}
\mathcal{J} = \int_{l.o.s.} \rho^2 ds.
\label{eq:geom}
\end{equation}
We parametrize $\rho$ with a generalized NFW DM density function \cite{1997ApJ...490..493N}:
\begin{equation}
\rho_{\rm{NFW}} = \frac{\rho_0}{\left( \frac{r}{r_s} )\right)^{\gamma} \left( 1 + \frac{r}{r_s} \right)^{3-\gamma}} ,
\label{eq:NFW}
\end{equation}
with index $\gamma = 1.2$ and scaling radius $r_s = 20$ kpc. 
%The $\gamma$-ray flux produced by DM annihilation is calculated with (see, e.g., \cite{1990NuPhB.346..129B}):
%\begin{equation}
%\Phi_\gamma(E) = \frac{1}{4\pi} \frac{\langle\sigma v\rangle}{m_\chi^2} \frac{dN_\gamma}{d E} \frac{1}{2}\mathcal{J}, 
%label{eq:flux_gamma}
%\end{equation}
%where $m_\chi$ is the DM mass, \sigmav defines the annihilation cross section times the relative
%velocity, averaged over the Galactic velocity distribution function, $dN_\gamma/d E$ is the $\gamma$-ray production source spectrum per DM annihilation event and $\mathcal{J}$ is the integral performed along the line of sight of the squared DM density distribution.

%For the continuum $\gamma$-ray flux, we consider here $i)$ prompt emission, 
%where the photons are found in final-state showers or hadronic decays of the annihilation products,
%and $ii)$ ICS by energetic electrons and positrons --  produced in the same way -- off the interstellar radiation field (ISRF). 
%We have calculated  the  spectrum for the prompt emission of both photons and $e^\pm$ with the Pythia Montecarlo code (version 8.162) \cite{pythia}.  

$dN_\gamma/d E$ is the photon spectrum that is usually modeled with a specific particle physics theory ruling the annihilation process.
%In this paper we are not interested in testing a specific DM model for the particle interactions and we will assume, 
%for $dN_\gamma/d E$ a power-law function or a log-parabola.
The measured spectrum of the GCE (see, e.g., \cite{TheFermi-LAT:2017vmf}) has a peak at a few GeV and a low energy and high energy cutoff. 
Therefore, by choosing the SED measured for the GCE we would not be able to test the detectability of the GCE below 1 GeV and above 10 GeV. 
Moreover, we are not interested to test specific DM particle physics models. 
We rather prefer to use a simple DM spectrum which covers all the energy range considered in our analysis.
Therefore, we assume, otherwise differently stated, a power-law shape with a spectral index of $\Gamma=2.0$ normalized at 1 GeV to $7\times 10^{-7}$ MeV$^{-1}$ cm$^{-2}$ s$^{-1}$, which is compatible with the GCE spectrum at the same energy\footnote{The flux of the GCE published in \cite{TheFermi-LAT:2017vmf} is $2\times 10^{-7}$ MeV$^{-1}$ cm$^{-2}$ s$^{-1}$ for an ROI defined with an angular distance from the Galactic center of $10^{\circ}$. This flux rescaled to our ROI is roughly $7\times 10^{-7}$ MeV$^{-1}$ cm$^{-2}$ s$^{-1}$.}.

In order to produce simulations that are compatible with the real sky, we configure our model from the results of the analysis in \cite{TheFermi-LAT:2017vmf}.
Then we simulate the ROI using the tool {\tt gta.simulate\_roi} which takes the current best-fit model and replace the data counts cube with the simulated data. 
The simulation is created by generating an array of Poisson random numbers with expectation values drawn from the model cube of the binned analysis instance, i.e., from the source map.
In the left panel of Fig.~\ref{fig:TSdistr} we show the count map of one simulation generated with the {\tt Baseline} IEM model between 1-10 GeV.
%We check that these calibrate our model on the real data.
%This is achieved by performing a fit to the data using the {\tt Fermitools} tools implemented in {\tt FermiPy}.

We analyze the simulated data using the following pipeline.
We first run the {\tt gta.optimize()} tool that performs an automatic optimization of the ROI by fitting all sources with an iterative strategy. 
First, it simultaneously fits the normalizations of the brightest components and sources. Then, it individually fits the normalizations of all sources that are not included in the first step.  
Finally, it individually fits the shape and normalization parameters of all sources. 
{\tt gta.optimize()} is a fast method to reach a good agreement between model and data and it is close to a perfect fit for regions of the sky that do not contain large degeneracies between the different components. This is particularly try when analyzing ROIs at high latitudes and selecting data above 1 GeV.
However, the Galactic center is a very crowded and complicated region so we have to perform a fit where all the SED parameters of the sources in the model are free to vary at the same time.
This is done by running the {\tt gta.fit()} tool which is a wrapper of the pyLikelihood fit method implemented in the {\tt Fermitools}. This tool returns the best fit and error of the SED parameters and the full covariance matrix.

In the last step of the analysis, we delete from the ROI sources detected with a test statistics $TS$\footnote{The Test Statistic ($TS$) is defined as twice the difference in maximum log-likelihood between the null hypothesis (i.e., no source present) and the test hypothesis: $TS = 2 ( \log\mathcal{L}_{\rm test} -\log\mathcal{L}_{\rm null} )$~\cite{1996ApJ...461..396M}.} lower than 25. This is the usual cut in $TS$ that is used to include or not sources in {\it Fermi}-LAT catalogs and it corresponds roughly to $4.6\sigma$ significance for a power-law SED with two free parameters.
%Finally, we redo an other fit to have the final model.
Finally, we calculate the GCE SED using the {\tt gta.sed} tool which performs independent fits for the flux normalization in bins of energy using a power-law spectral parameterization with a fixed spectral index of $\Gamma = 2.0$ (for $dN/dE\sim E^{-\Gamma}$).
This assumption is appropriate considering small enough energy bins as we do in our analysis.
Therefore, the results obtained with {\tt gta.sed} is independent from the SED model assumed initially for the DM template.

We test that our fitting procedure recovers properly the injected signal by running the following simulations.
We take the ROI models calibrated on the real data from \cite{TheFermi-LAT:2017vmf} and we generate simulated data.
Then, we change the values of the normalizations and spectral indexes of each source and IEM component into a random number between 0.7 and 1.3 times its reference value. 
With this procedure, we thus create a new model with values for the SED parameters which are not exactly equal to the simulated ones.
%and we test if the fitting procedure is able to recover correctly the simulated ones.
Then, we run the pipeline and find that most of the final SED parameters and fluxes are consistent with the initial model within $1\sigma$ errors.
This is particularly true for the SED parameters of DM. The normalization and spectral index, and the DM SED points we find with our analysis are perfectly compatible with the spectrum of the injected signal.
We test this with several simulations and trying the IEM models listed in Sec.~\ref{sec:analysis}.

\begin{figure*}
\includegraphics[width=0.49\textwidth]{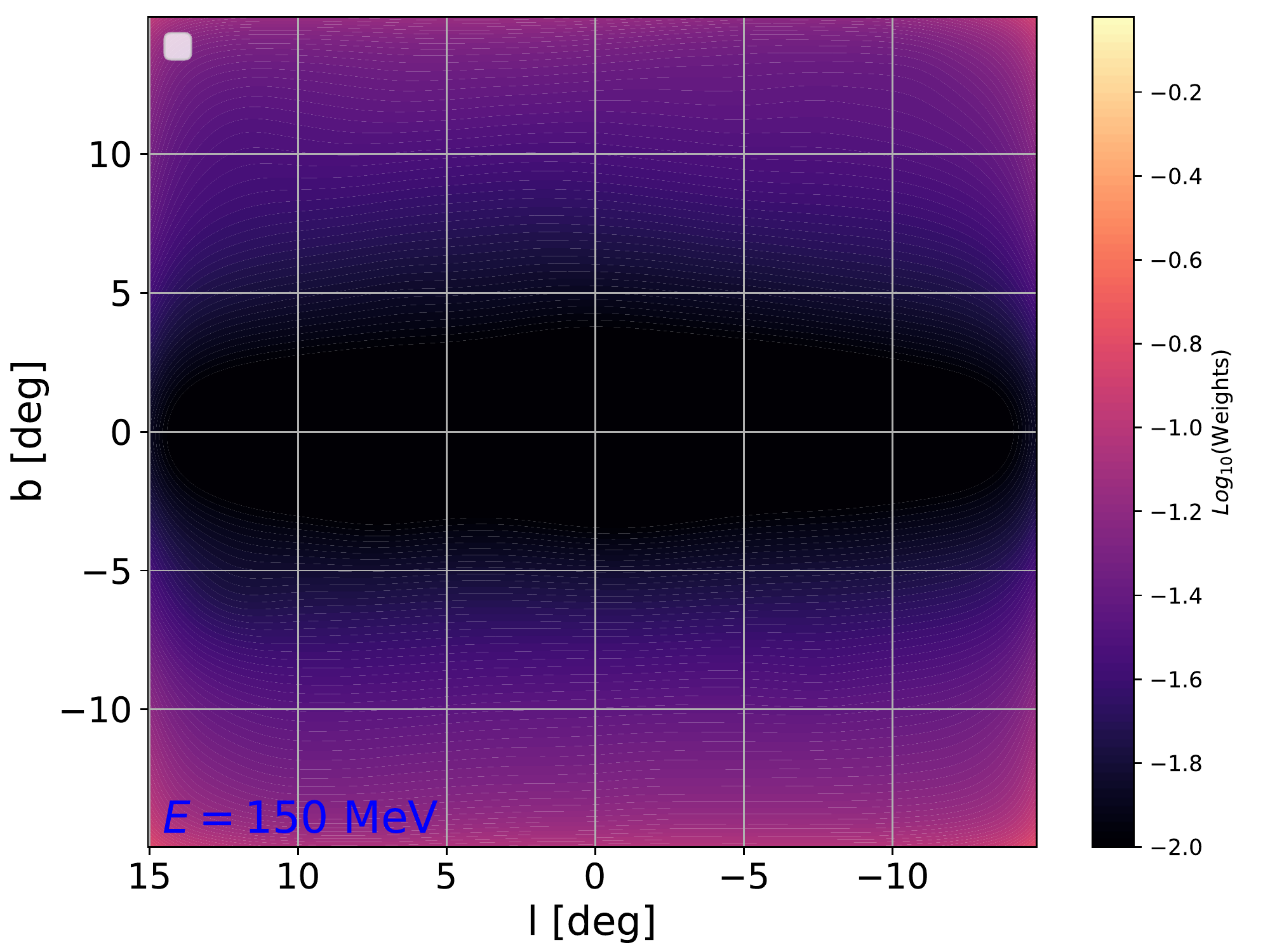}
\includegraphics[width=0.49\textwidth]{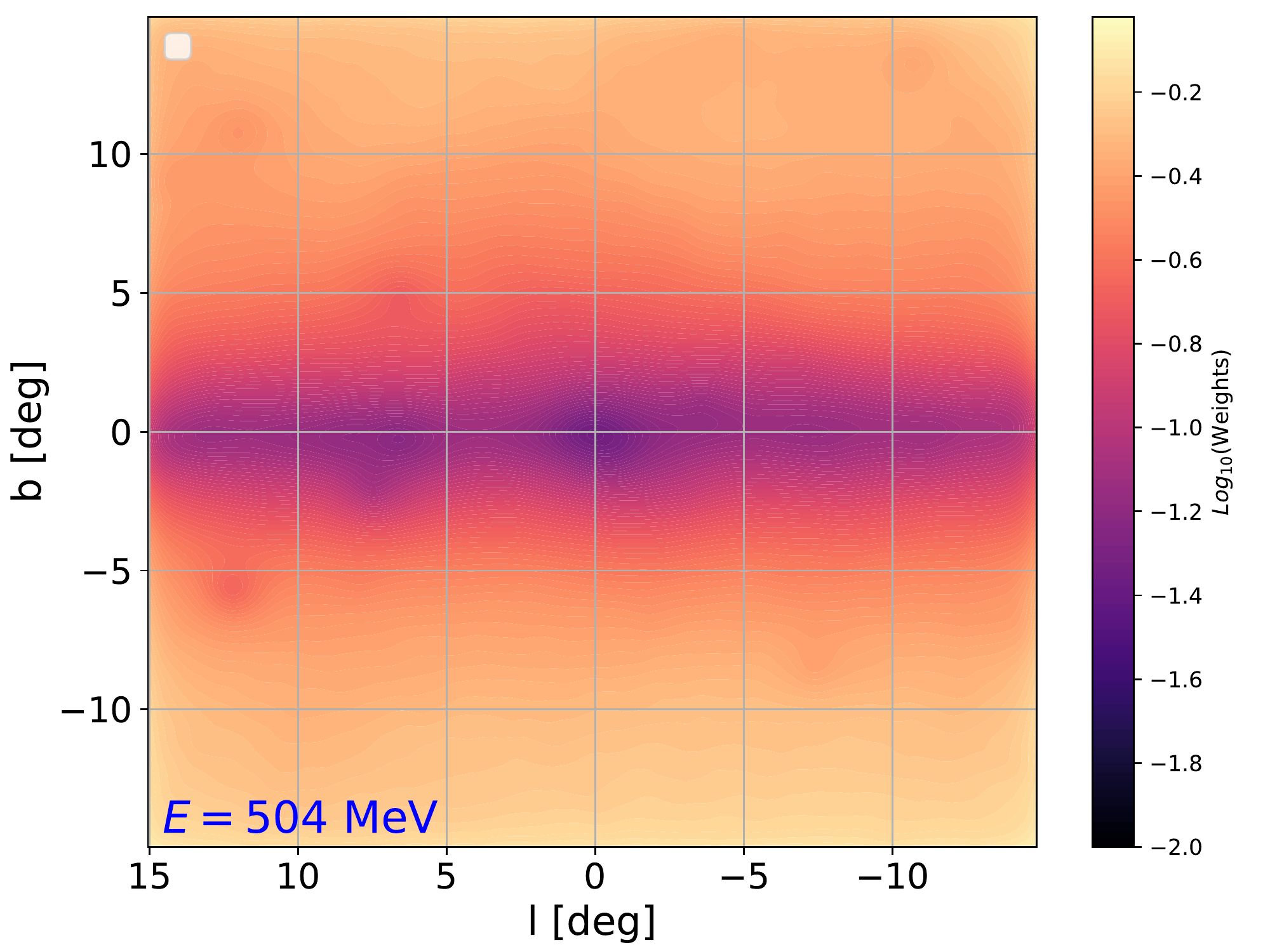}
\includegraphics[width=0.49\textwidth]{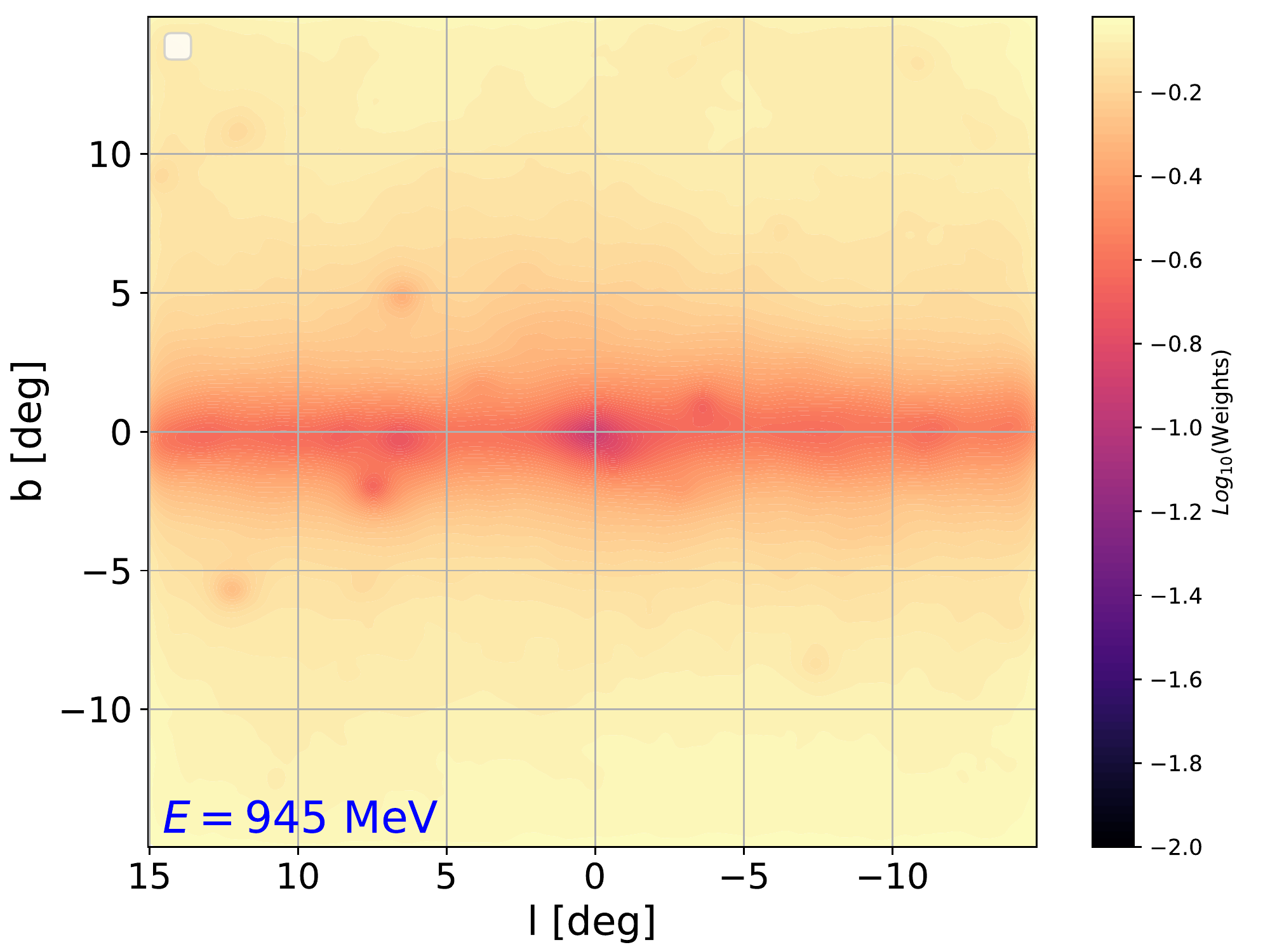}
\includegraphics[width=0.49\textwidth]{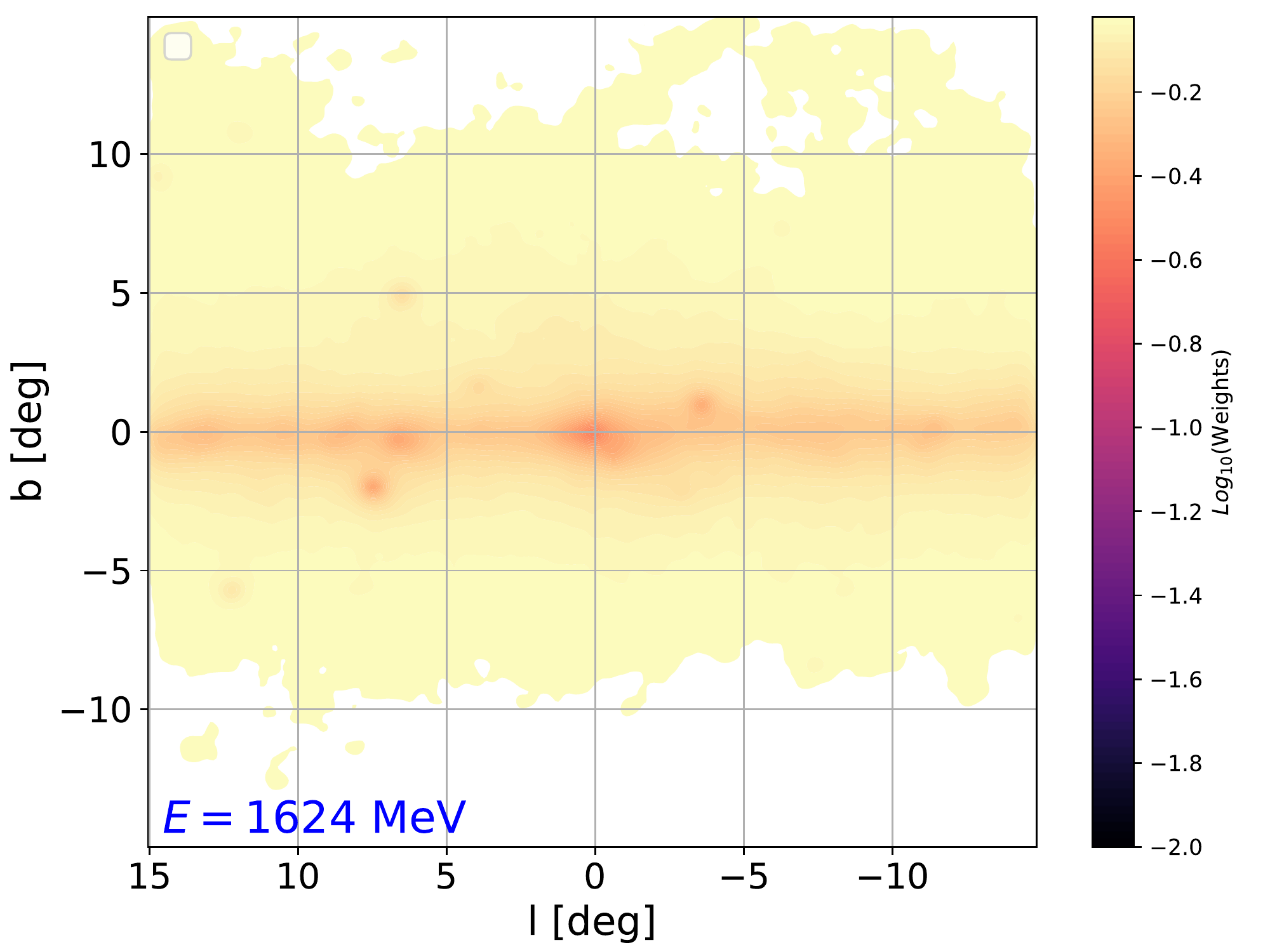}
\caption{Weight maps generated for $E=150$ MeV (top left panel), $E=504$ MeV (top right panel), $E=945$ MeV (bottom left panel) and $E=1624$ MeV (bottom right panel). The color bar represents the $\rm{Log}_{10}$ of the weight values. See the main text for further details on the weighted likelihood technique.}%, with consistent color code as the right panel.}  
\label{fig:mapweights}
\end{figure*}

\begin{figure}
\includegraphics[width=0.49\textwidth]{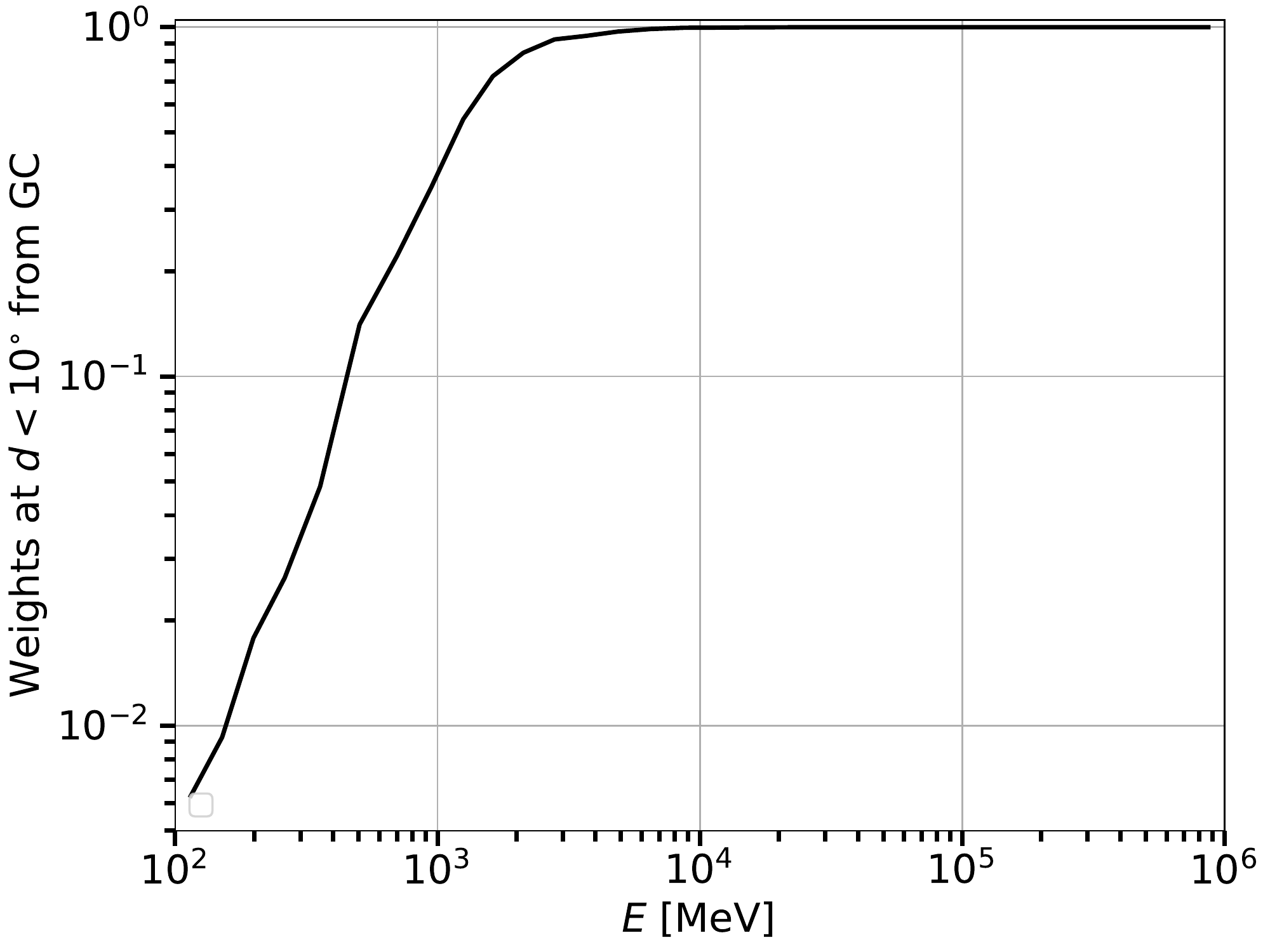}
\caption{Average of the weights at an angular distance $< 10^{\circ}$ from the Galactic center as a function of energy.}%, with consistent color code as the right panel.}  
\label{fig:weightsenergy}
\end{figure}

The most complicated region due to the uncertainty of the IEM is close to the Galactic plane ($b\sim0^{\circ}$).
In order to mitigate these uncertainties, we apply in our analysis a technique called weighted likelihood that has been recently included in the {\tt Fermitools}.
This method introduces weights for every pixel of the sky according to the number of counts.
The weights are then multiplied to the $\rm{Log} (\mathcal{L})$ found in each pixel in the template fitting of the maximum likelihood analysis. 
This procedure thus penalizes pixels with a very large number of photons and in which the systematics for the choice of the IEM could be larger\footnote{A technical document explaining the weighted likelihood is available at this link \url{https://fermi.gsfc.nasa.gov/ssc/data/analysis/scitools/weighted_like.pdf}.}.
We use as, in the 4FGL catalog paper, a systematic level of $\epsilon=3\%$. This value is motivated by the study performed in Ref.~\cite{Fermi-LAT:2019yla} with the relative spatial and spectral residuals in the Galactic plane where the diffuse emission is strongest. 
We show in Fig.~\ref{fig:mapweights} the weight maps derived at energies between 0.1 GeV to 1.6 GeV and in Fig.~\ref{fig:weightsenergy} the average value of the weights found as a function of energy and for an angular distance $< 10^{\circ}$ from the Galactic center as a function of energy.
At 150 MeV the weights are very small for almost all the ROI we consider, thus meaning that these energies are not important in the fitting procedure. For $E=500$ MeV the weights are particularly small, i.e.~less than 0.1, only for $|b|<3^{\circ}$. This implies that all these pixels are much less constraining in the fitting procedure than the higher latitude ones where the weights are much closer to 1.
Instead, at $E>1$ GeV most of the ROI has weights close to 1 thus most of pixels in the analysis have the same weights.
The effect of the weighted likelihood is thus important below 1 GeV and in the inner few degrees from the Galactic plane where the IEMs differ the most.
We will discuss the effect of the weighted likelihood analysis in deriving the spectrum of the GCE in Sec.~\ref{sec:wrongmodel} in the presence of imperfection in the IEM.

Finally, we apply the energy dispersion to all the components of our model using the method implemented in the {\tt Fermitools}\footnote{For a complete description see \url{https://fermi.gsfc.nasa.gov/ssc/data/analysis/documentation/Pass8_edisp_usage.html}.}.
Specifically, we select {\tt apply\_edisp=true} and {\tt edisp\_bins=-1} which applies the energy dispersion only on the spectrum accounting for one extra bin. 

We also check for the presence of significant residuals by running {\tt gta.tsmap} to the ROI model obtained after the analysis pipeline.
This tool generates a $TS$ map by running a fit in each ROI pixel for a test source with spectral index of 2.0. 
We show in the right panel of Fig.~\ref{fig:TSdistr} the histogram of the $TS$ values obtained in the $TS$ map for the analysis of one simulation where we use the {\tt Baseline} model in the energy range between 1-10 GeV.
The distribution is compatible with the chi-square distribution for one degree of freedom ($\chi_1^2/2$ ).
This is the result expected in case the model obtained in the fitting procedure properly represents the simulated data and no significant residuals are present.
We find this result because the same components included in the model for the fitting procedure are also used to generate the simulated data.
This scenario represents thus the ideal case where we have a perfect knowledge of the Galactic center emission.
We will show later how the $TS$ histogram changes in case we have imperfections in the IEM model or there is an un-modeled component in the data.

%\subsection{Simulation Setup}
%\label{sec:simulations}
%Therefore, we perform a fit to the real data, with the strategy presented in Sec.~\ref{sec:analysis}, and find the SED parameters of the different background components and the GCE.
%We apply this analysis specifically to the {\tt Baseline} model and then we will use as input the other interstellar emission model to see how imperfections of this component could affect the search for DM.
%Once we find the SED parameters for the different model components we perform simulations using the {\tt gta.similate\_roi} tool of {\tt Fermipy}. This tool generate simulated data for the an ROI using the best-fit model and replace the data counts cube with the simulated counts. The simulation is created by generating an array of Poisson random numbers with expectation values drawn from the model cube of the binned analysis instance. 
%We show in Fig.~\ref{fig:countmap} the count map of one simulations for photons in the energy range between 1-10 GeV.
%\textcolor{blue}{Maybe add the map of the residuals and spectrum plots from the fit to the real data.}

\subsection{Inspecting the spatial morphology of the excess with a model independent approach}
\label{sec:annulianalysis}

\begin{figure}[t]
\includegraphics[width=0.45\textwidth]{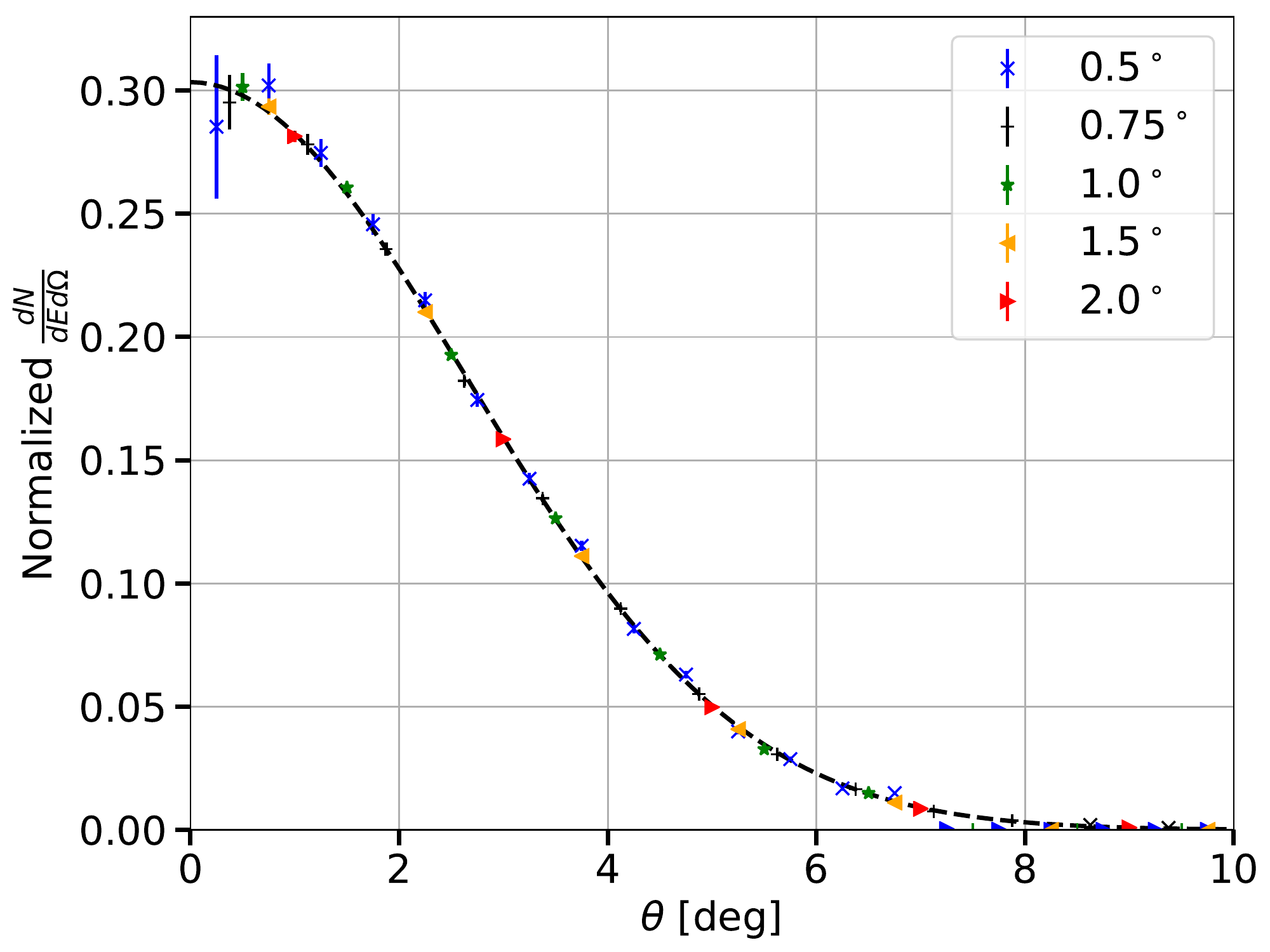}
\caption{Surface brightness data normalized to 1 found with the analysis explained in Sec.~\ref{sec:annulianalysis}. We show the data for different choices of the annulus size from 0.5$^{\circ}$ to 2$^{\circ}$. Together with the data we show the angular profile of a 1D Gaussian with standard deviation of $2.64^{\circ}$ which is equivalent to a standard deviation of $4^{\circ}$ for a 2D Gaussian template.}%, with consistent color code as the right panel.}  
\label{fig:SB110}
\end{figure}

\begin{figure*}[t]
\includegraphics[width=0.45\textwidth]{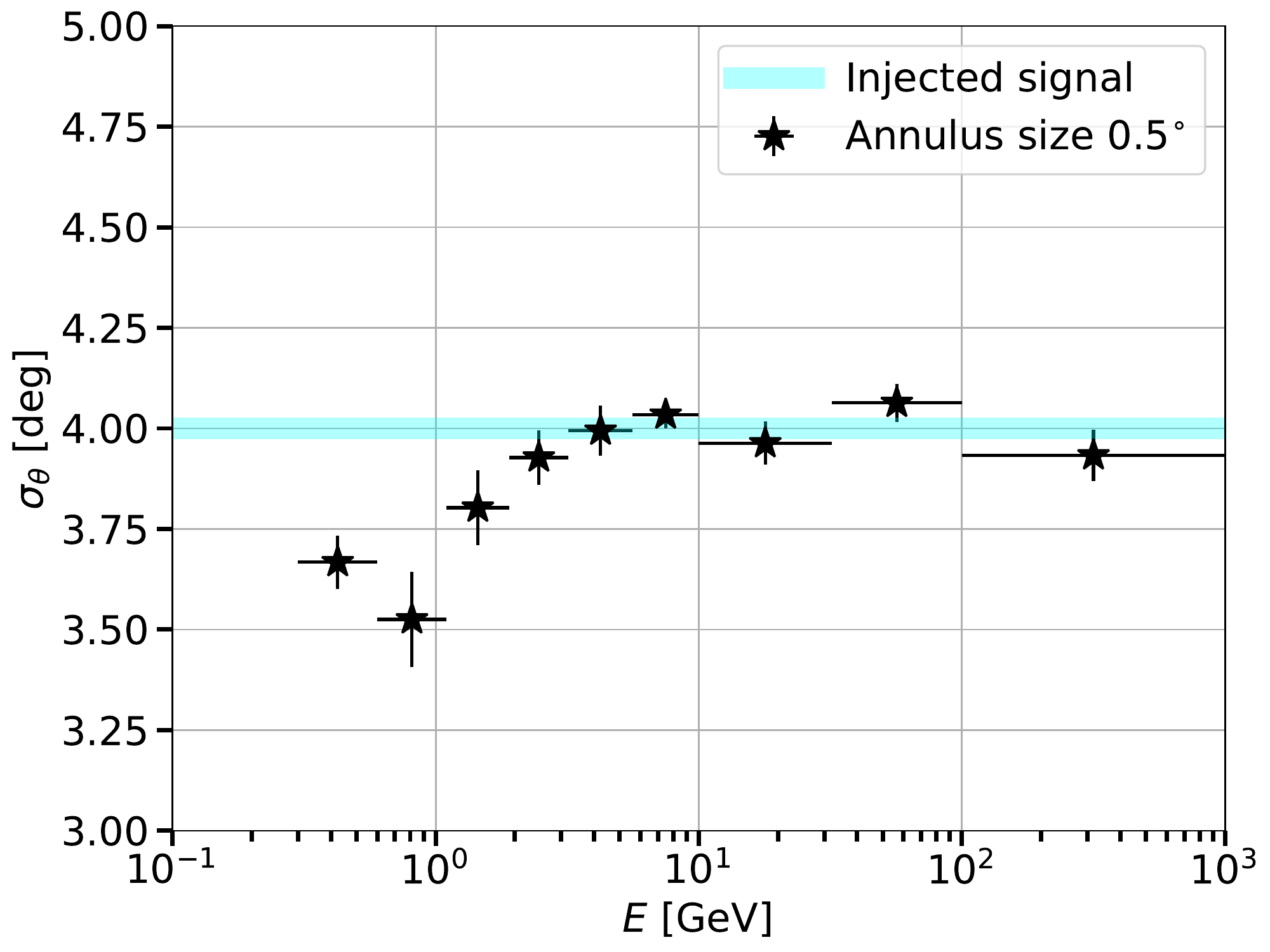}
\includegraphics[width=0.45\textwidth]{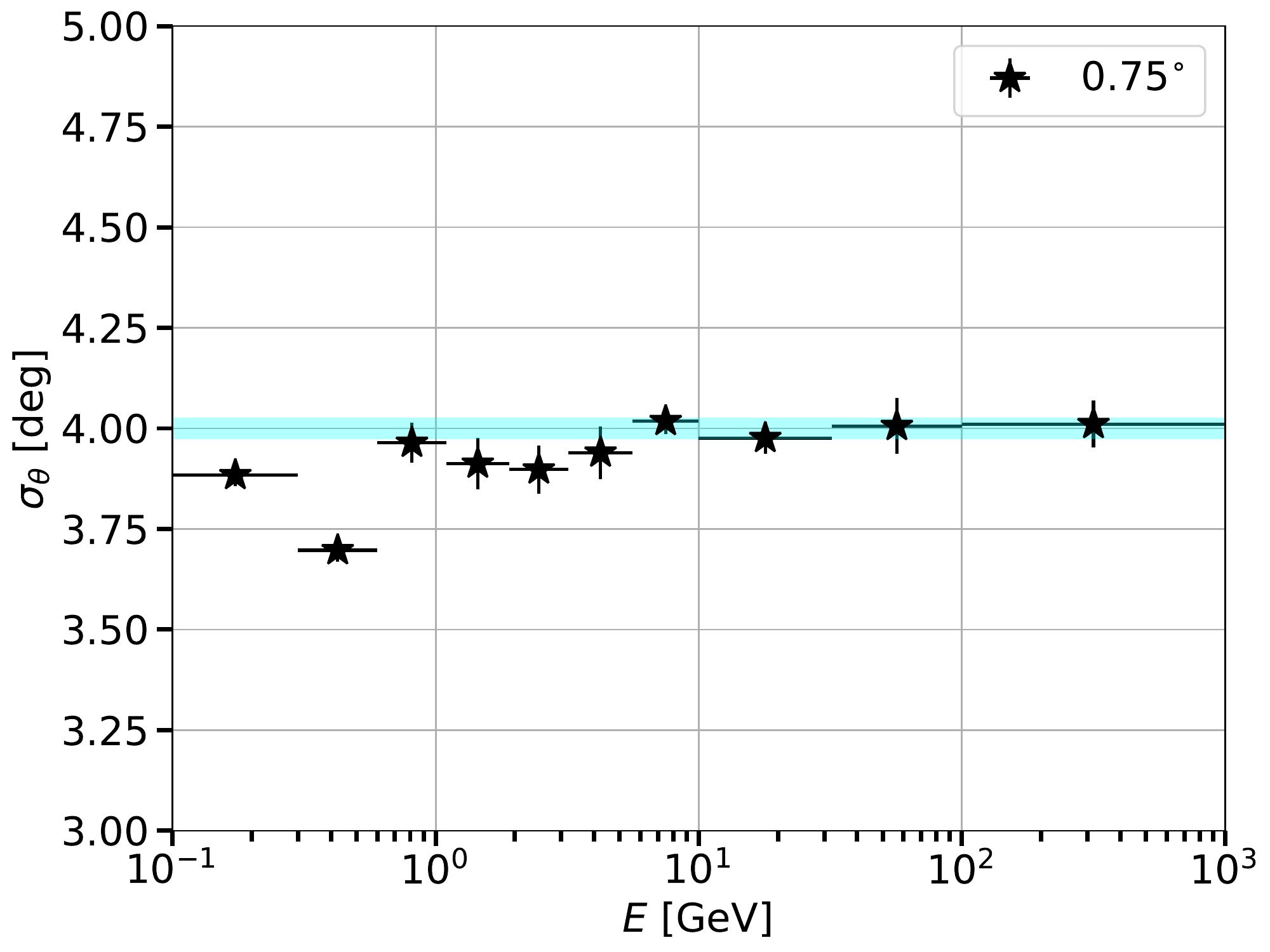}
\includegraphics[width=0.45\textwidth]{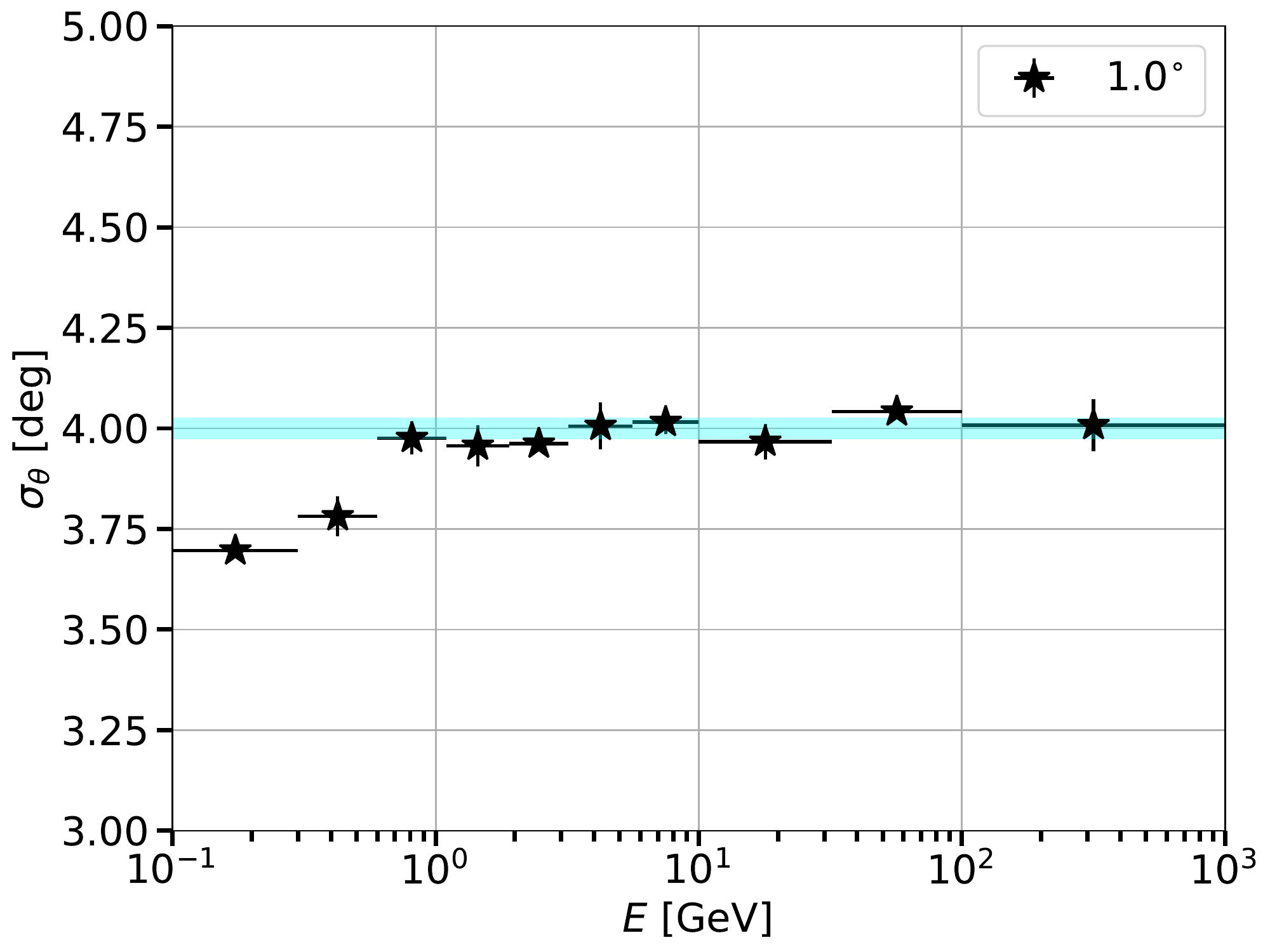}
\includegraphics[width=0.45\textwidth]{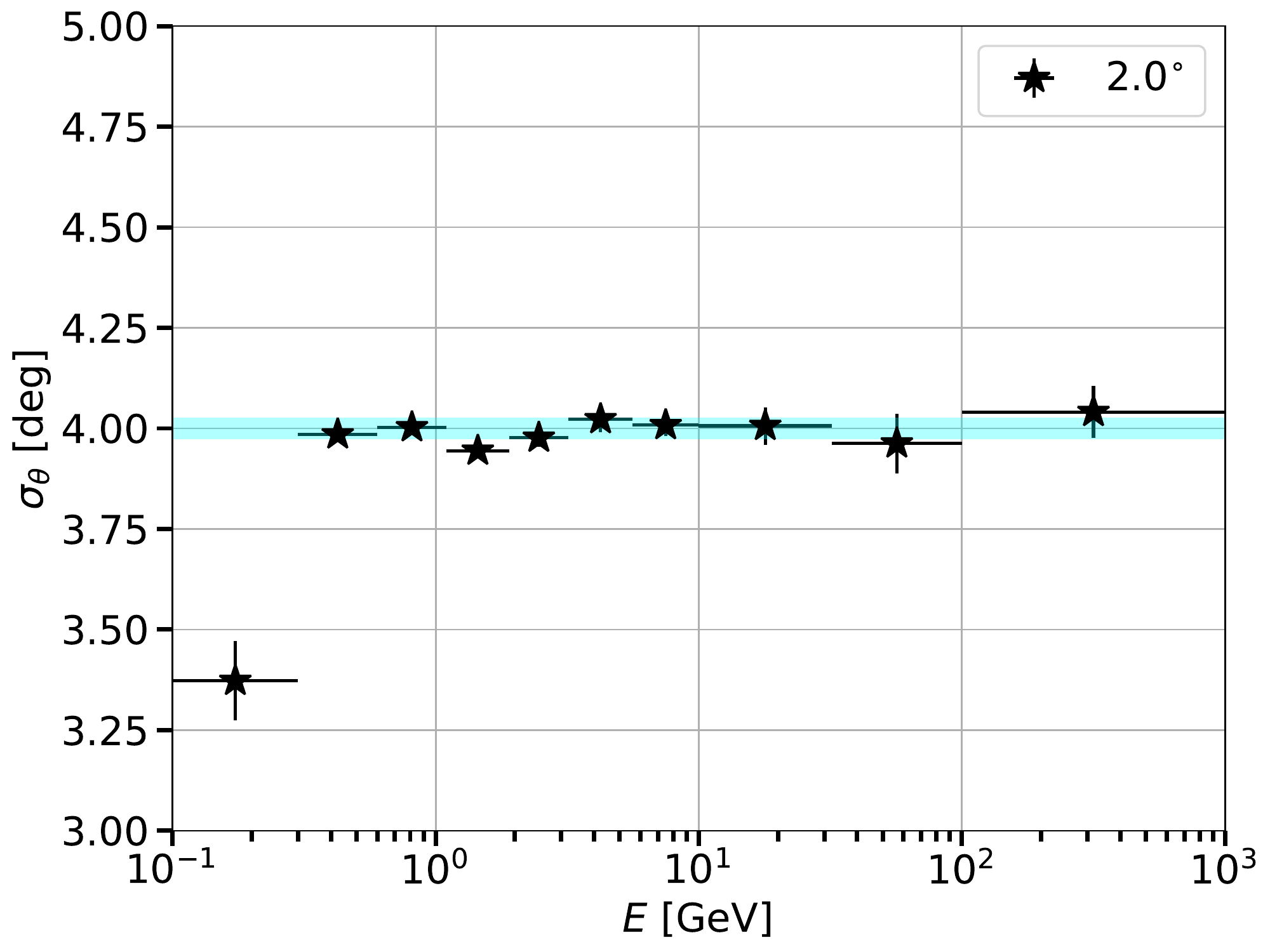}
\caption{Size of the standard deviation for a 2D Gaussian template ($\sigma_{\theta}$) that we find by fitting the surface brightness data (see Sec.~\ref{sec:annulianalysis} for further details on the analysis). We show $\sigma_{\theta}$ for the analysis applied in different energy bins from 0.1 to 1000 GeV and choosing different sizes for the annulus size from $0.5^{\circ}$ (top left) to $2.0^{\circ}$ (bottom right).}%, with consistent color code as the right panel.}  
\label{fig:SBenergies}
\end{figure*}

The spatial morphology of the GCE is one of the most important characteristic that could provide us an hint of its origin.
In this section we present an approach to derive the GCE spatial morphology that is independent from a specific choice of the excess template.
The results of this method can thus be used, as we will show, to derive the DM density profile needed to fit the GCE or to verify if other interpretations are more suitable to explain it.

We include in the model concentric and uniform annuli and we fit them to the data.
Each annulus has an SED given by a power law. The free parameters for each annulus are thus the normalization and the power-law index.
We demonstrate the feasibility of this method with a simulation of a 2D Gaussian template with a standard deviation of $\sigma_{\theta} = 4^{\circ}$.
We choose to use a power-law SED shape with spectral index $\Gamma=2.0$ and normalized at 1 GeV to $7\times 10^{-7}$ MeV$^{-1}$ cm$^{-2}$ s$^{-1}$, which is compatible to the GCE spectrum at the same energy (see, e.g., \cite{TheFermi-LAT:2017vmf}).

We employ the fitting procedure explained in Sec.~\ref{sec:analysis}.
We extract from the fit the energy flux of each annulus in units of MeV cm$^{-2}$ s$^{-1}$. 
We compute the GCE surface brightness $dN/(dEd\Omega)$ by dividing the annulus energy fluxes by their solid angles $d\Omega$.
Then, we perform a fit to the surface brightness data with a 1D Gaussian template and find its standard deviation $\sigma_{\theta,1D}$.
Finally, we convert the standard deviation of the 1D Gaussian to the standard deviation of the 2D Gaussian $\sigma_{\theta}$ by multiplying $\sigma_{\theta,1D}$ for a conversion factor which is roughly 1.515.

We show in Fig.~\ref{fig:SB110} the surface brightness data obtained for the energy range 1-10 GeV and using annulus widths between 0.5$^{\circ}$ to 2$^{\circ}$.
We test different annuli sizes to verify whether the analysis provides significantly different results according to the value of this parameter.
The best fit values we find for the standard deviation of the 2D Gaussian are $[3.93\pm0.05,3.97\pm0.04,3.96\pm0.05,3.98\pm0.03,3.98\pm0.03]$ for the following annulus sizes $[0.5^{\circ},0.75^{\circ},1.0^{\circ},1.5^{\circ},2.0^{\circ}]$. 
The values of $\sigma_{\theta}$ are thus compatible within $1\sigma$ errors with the one of the injected signal. 
$\sigma_{\theta}$ obtained with a size of the annuli of $0.5^{\circ}$ seems to be the one that most deviates from the initial value with a difference of about 2\%. However, this discrepancy does not appear to be significant. 

We also apply this analysis in the following energy bins: $0.1-0.3$ GeV, $0.3-0.6$ GeV, $0.6-1.1$ GeV, $1.1-3.2$ GeV, $3.2-5.6$ GeV, $5.6-10$ GeV, $10-32$ GeV, $32-100$ GeV, $100-1000$ GeV.
This is done to check how the proposed technique performs selecting different energies.
The results are reported in Fig.~\ref{fig:SBenergies} for different annulus sizes.
In case the annulus size is in the range $0.75^{\circ}-1.5^{\circ}$, the energy range for which the value of $\sigma_{\theta}$ is not compatible with the injected signal is $E<0.6$ GeV, i.e.~in the first two energy bins considered in the analysis. For an annulus size of $0.5^{\circ}$ the results are correct for energies larger than a few GeV. 
This is due to the fact that the {\it Fermi}-LAT PSF below 1 GeV is too large to be able to disentangle so narrow annuli.
Instead, for energies above 1 GeV the value of $\sigma_{\theta}$ calculated from the fit to the surface brightness is compatible within the statistical errors to the shape of the injected signal.
For an annulus size of $2^{\circ}$ we can reconstruct properly the GCE size also for energies between 0.3-0.6 GeV, i.e., at slightly lower energies with respect to the other cases.
To conclude, this exercise demonstrates that taking an annulus size larger than $0.75^{\circ}$ and an energy range above 600 MeV are safe choices to analyze the surface brightness of the GCE and to find its spatial distribution.

\section{Ideal case: perfect knowledge of background components}
\label{sec:idealcase}

We start by showing the capability of our analysis method on simulated {\it Fermi}-LAT data in case we have a perfect knowledge of background components.
This is done by generating simulations of the Galactic center region using a model that contains the components explained in Sec.~\ref{sec:iemdata} and then by employing the same model also in the analysis of the data.
We use the full energy range between 0.1-1000 GeV to test also the possibility to find reliable estimate of the low and high-energy tails of the GCE spectrum.
We remind that the GCE is simulated using a DM template which follows a generalized NFW density profile with index $\gamma = 1.2$ and scaling radius $r_s = 20$ kpc.
%This DM template has been found to be consistent with the GCE spatial distribution (see, e.g., \cite{TheFermi-LAT:2017vmf}). 
For the DM SED we use a power-law index with $\Gamma=2.0$ normalized at 1 GeV to $7\times 10^{-7}$ MeV$^{-1}$ cm$^{-2}$ s$^{-1}$.
%, which is compatible with the GCE spectrum at the same energy.
Then, the simulated data are analyzed following the technique explained in Sec.~\ref{sec:analysismethod} and \ref{sec:annulianalysis}.
%The fit does not leave any significant residuals since the model is the same used to generated the simulated data. Therefore the $TS$ distribution follows the $\chi^2/2$ (see, e.g., Fig.~\ref{fig:TSdistr} left panel).
In the next sections we will show how our analysis is able to reconstruct the DM energy spectrum, spatial morphology and position in this ideal case.
%\textcolor{blue}{Not sure if I have to include the correlation parameters here.}

\subsection{Spectrum of the Galactic center excess}
\label{sec:spectrumDM}

\begin{figure}
\includegraphics[width=0.45\textwidth]{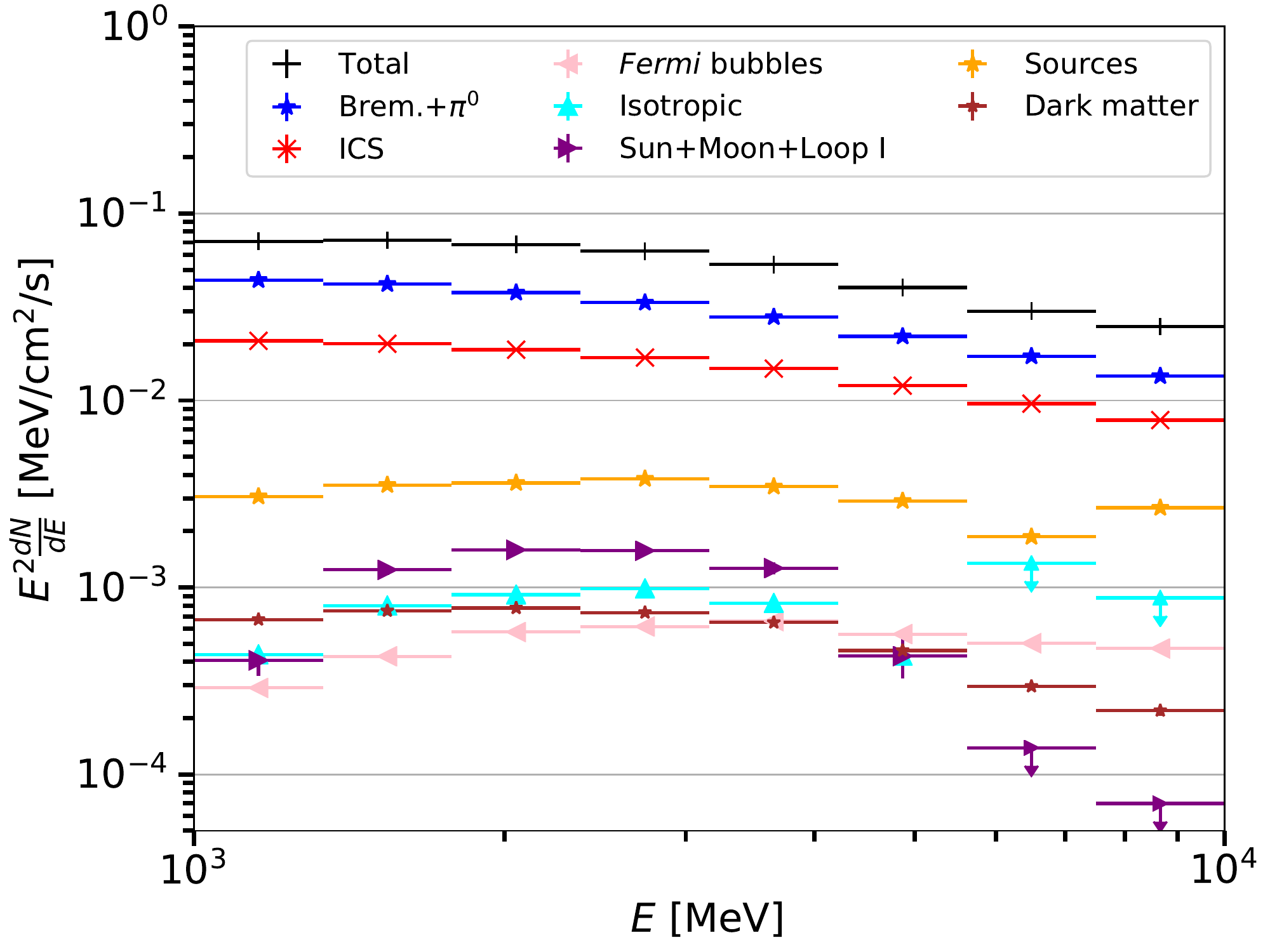}
\caption{Energy spectrum between 1 and 10 GeV of the different components included in our analysis using the Baseline IEM (see Sec.~\ref{sec:iemdata} for further details).}%, with consistent color code as the right panel.}  
\label{fig:spectrumcomp}
\end{figure}

\begin{figure}
\includegraphics[width=0.45\textwidth]{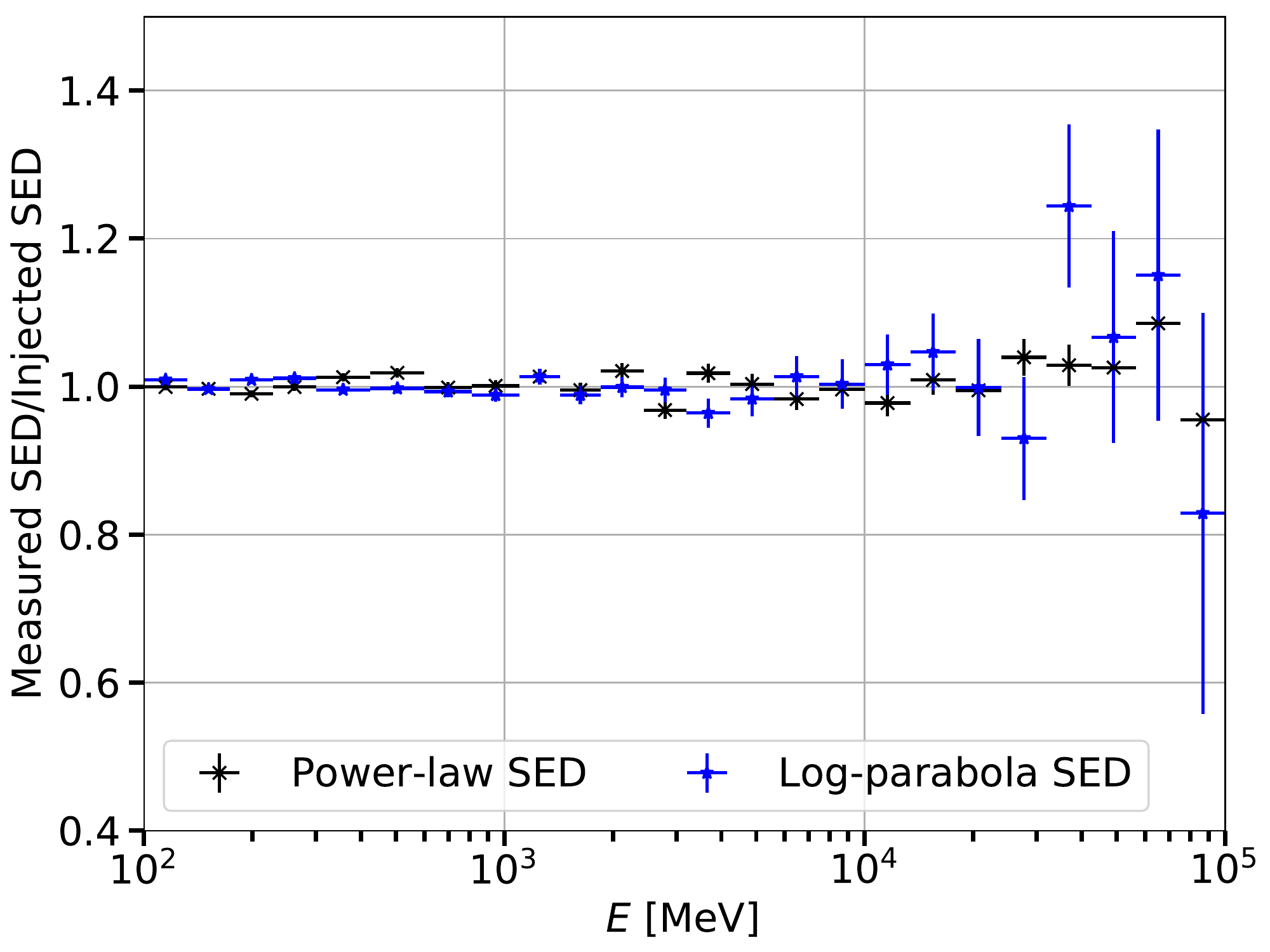}
\includegraphics[width=0.45\textwidth]{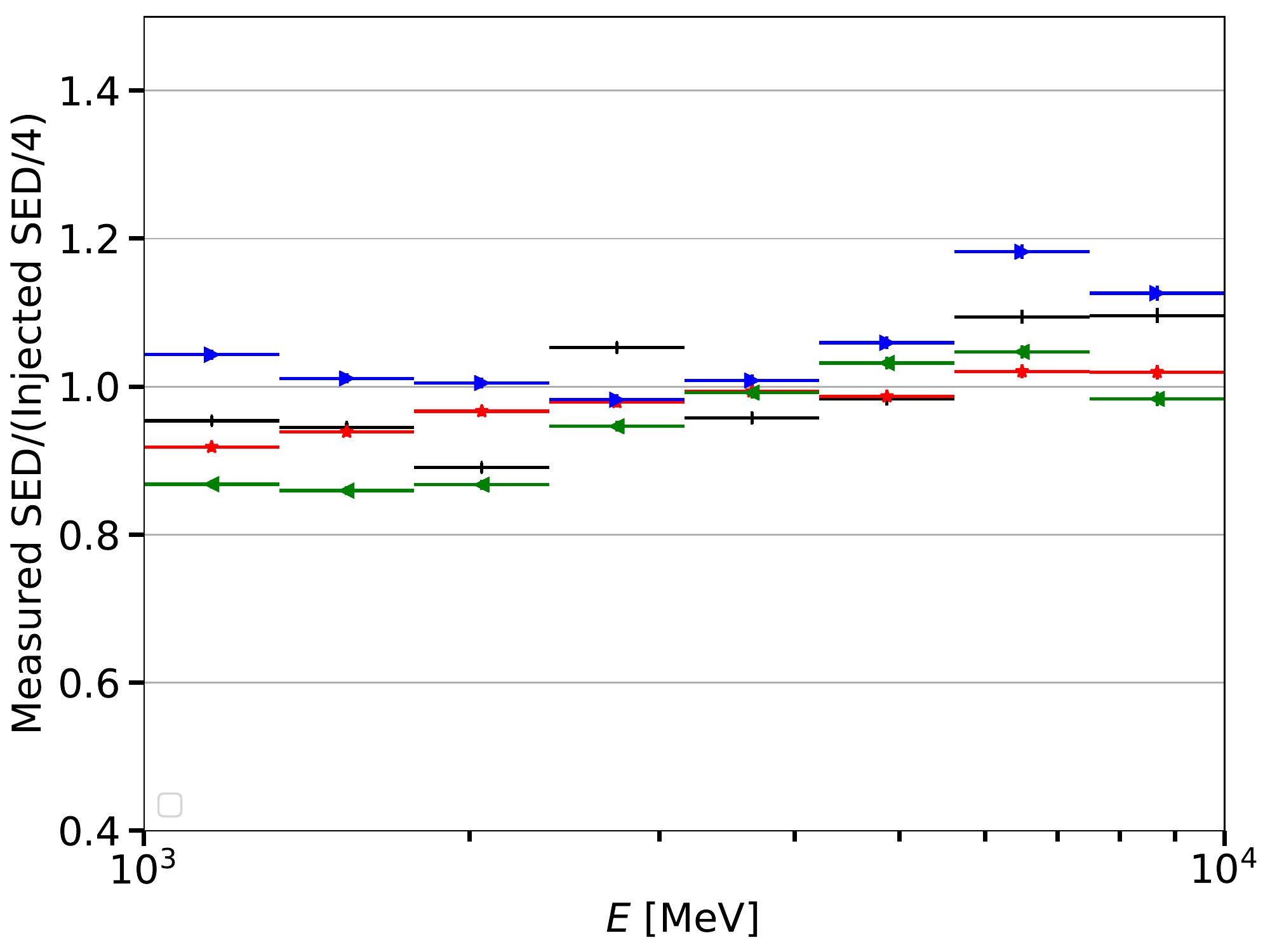}
\caption{Top Panel: Ratio between the SED of DM found by analyzing the simulated data and the SED of the injected GCE signal (see Sec.~\ref{sec:spectrumDM}). We show the results for the case of a power-law and log parabola SED. Bottom Panel: Ratio between the SED of the quadrants and the injected DM SED divided by 4.}%, with consistent color code as the right panel.}  
\label{fig:DMenergy}
\end{figure}

We first perform a fit to the ROI with the DM and all the other components free to vary and then we calculate the GCE SED using the {\tt gta.sed} tool implemented in {\tt Fermipy}.
In Fig.~\ref{fig:spectrumcomp} we show the energy spectrum of the different components of the IEM, point sources, isotropic template, Sun-Moon-LoopI template and dark matter. This is calculated using the {\tt Baseline} IEM (see Sec.~\ref{sec:iemdata}). The largest contribution is provided by the bremsstrahlung and $\pi^0$ components while the GCE contributes roughly at the $\%$ level.
%This tool performs independent fits for the flux normalization of DM in bins of energy. The normalization in each bin is fit using a power-law spectral parameterization with a fixed index. 
%Therefore, the results obtained with {\tt gta.sed} is independent by the SED model assumed initially for DM.
The comparison between the injected and observed DM SED is presented in the top panel of Fig.~\ref{fig:DMenergy} as ratio of the two.
The ratio is compatible to 1 for all the considered energies within the statistical errors even at low energy where the very high flux of the IEM model and the poor {\it Fermi}-LAT spatial and energy resolution makes the detection of the GCE more challenging.
The good agreement we find below 1 GeV is partially due to the fact that we include corrections due to energy dispersions with the tools implemented in the {\tt Fermitools}\footnote{See for more details \url{https://fermi.gsfc.nasa.gov/ssc/data/analysis/documentation/Pass8_edisp_usage.html}}.  
%This effect is particularly important at these low energies.
%$E>1$ while it is systematically lower below these energies.
%The maximum difference is of about 25\% at 100 MeV.
%This discrepancy is probably due to the non perfect implementation of the energy dispersion effect in the {\it Fermi}-LAT Science Tools. Indeed, the implementation that we use is the standard applied in {\it Fermi}-LAT analysis and it neglects the change of PSF in the energy bins included in the energy dispersion corrections. A recent attempt has been done by the {\it Fermi}-LAT team to include in the energy dispersion handling also the change of the PSF across the energies. 
%Checks that have been performed on simulations prove that this new method provides in a lot of cases results not compatible with expectations with respect to the old energy dispersion corrections. The reason for this is still under investigation in the LAT team so we have decided to not use it. 
%See this page for further information on the implementation of the energy dispersion in the {\it Fermi}-LAT Science Tools  \url{https://fermi.gsfc.nasa.gov/ssc/data/analysis/documentation/Pass8_edisp_usage.html}.\
%The results presented here tell us that looking to the SED of the GCE below 1 GeV might provide results that are systematically lower than the real one. Therfore the most reliable energies are $E>1$ GeV.
We also test an SED given by a Log Parabola $dN/dE \sim E^{-(\gamma+\beta \log(E))}$. We normalize the SED, as before, at 1 GeV to $7\times 10^{-7}$ MeV$^{-1}$ cm$^{-2}$ s$^{-1}$ and we choose $\Gamma=2.0$. Moreover, we fix $\beta=0.05$ which produces a small curvature at energies larger than 1 GeV.
The results we find with this SED are practically identical to the ones reported in Fig.~\ref{fig:DMenergy} for the power-law shape, i.e., the ratio between injected and observed spectrum is compatible with 1.

We perform an additional test by dividing the DM template into 4 quadrants separated by the Galactic plane ($b=0^{\circ}$) and the vertical direction from the Galactic center ($l=0^{\circ}$). 
This exercise tests if the GCE given by DM would be detected with the same spectrum in the different portions of the Galactic center region.
%We try two different choices for the orientation of the quadrants.
%The first one is with quadrants separated by the Galactic plane (i.e., $l=0^{\circ}$) and the vertical direction from the Galactic center. 
%We will call this as {\tt first configuration}. The second one is with quadrants separated by the angles $\pm 45^{\circ}$, $\pm135^{\circ}$ from the Galactic plane. This is labelled as {\tt second configuration} and it the first one rotated by $+45^{\circ}$.
The quadrants are added in the model and fitted to the ROI. Then, we compute their SED.
In the bottom panel of Fig.~\ref{fig:DMenergy} we show the ratio between the SED of each quadrant with the SED divided by 4 of the DM template considered as a whole.
%for each of them and we compare the spectrum of each quadrant with each other and with the one of the DM template when it is considered as a unique component.
The SED of the quadrants are different from each other by about $15\%$ and they differ from the total one of DM by about the same amount.
This is the value of the systematic error we should consider when analyzing the real data with a DM template divided into quadrants.
%\textcolor{blue}{What happens in case we run with no randomization?? Running a test here \url{/u/gl/mdimauro/dmcat/workdir/mattia/GC_DMPSR/try/GCexcess_extension_energy/sim_DM/morebins_gamma1p20/test_quadrants/Baseline_1-10GeV_true}}

\subsection{Spatial morphology}
\label{sec:spatialDM}

\begin{figure}[t]
\includegraphics[width=0.45\textwidth]{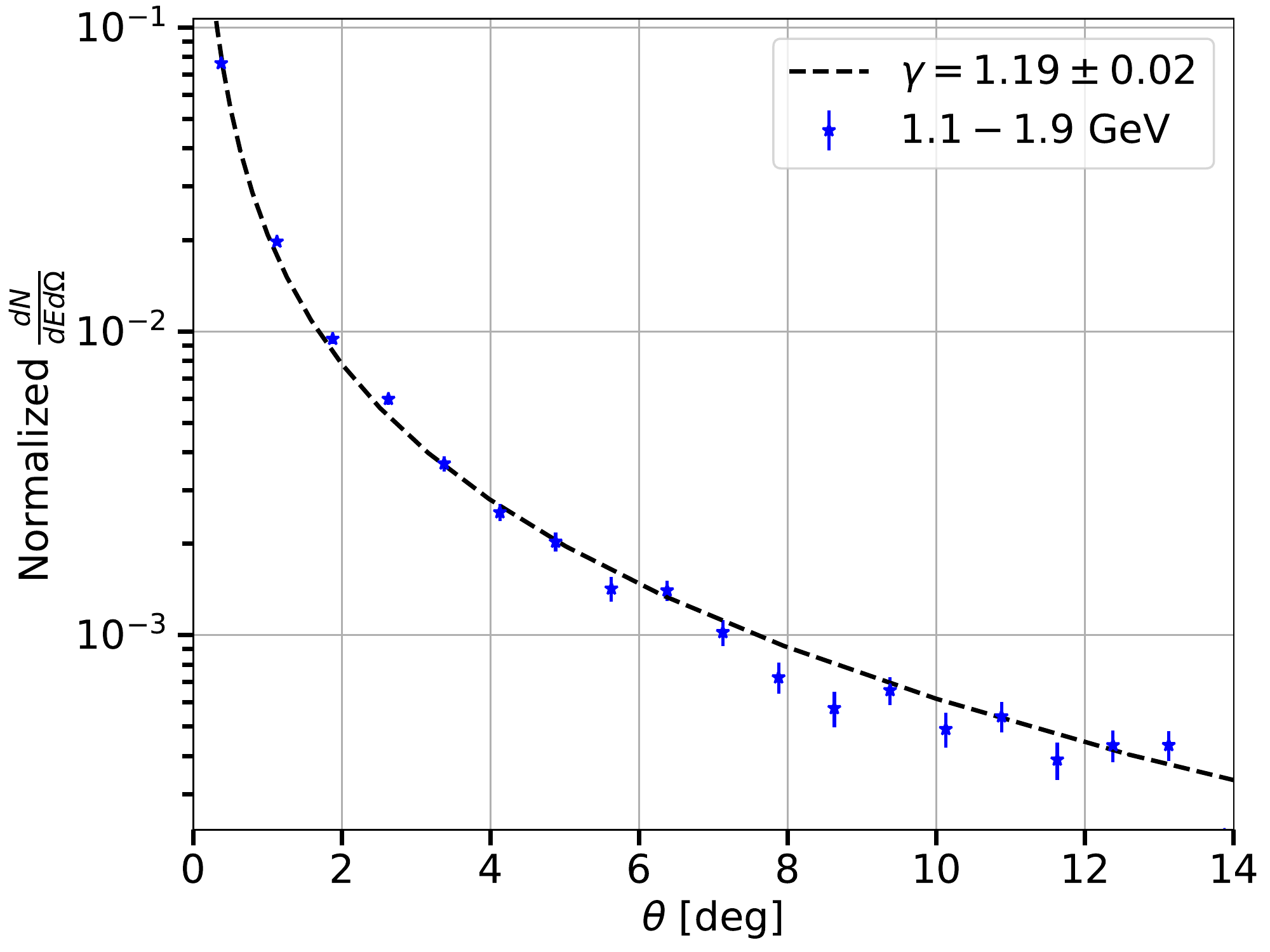}
\caption{Surface brightness data derived for the energy range $1.1-1.9$ GeV and with annulus size $0.75^{\circ}$. We also show the surface brightness for a DM template for the case of the best fit with $\gamma=1.19$.}%, with consistent color code as the right panel.}  
\label{fig:SBDM}
\end{figure}

\begin{figure*}[t]
\includegraphics[width=0.45\textwidth]{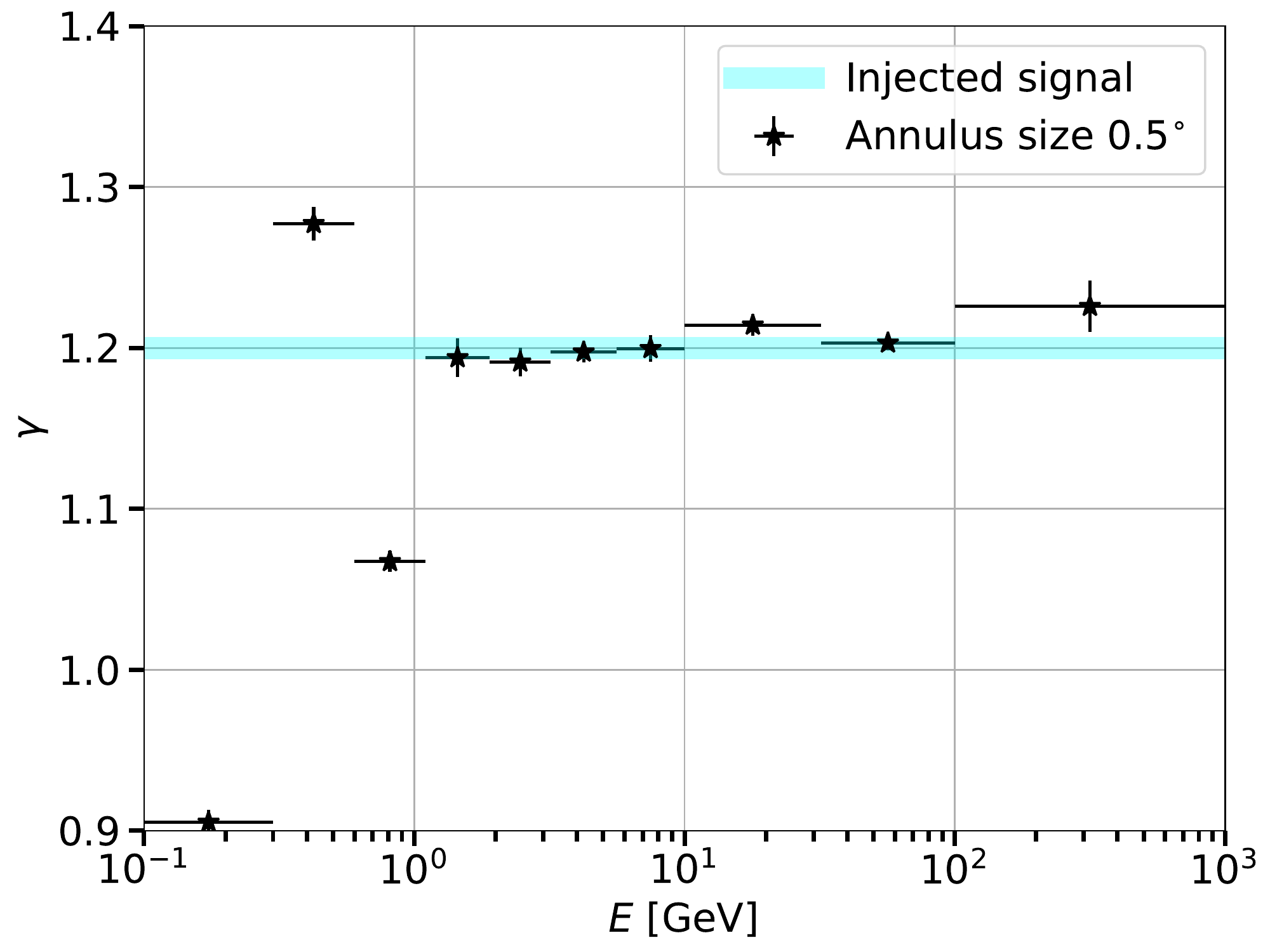}
\includegraphics[width=0.45\textwidth]{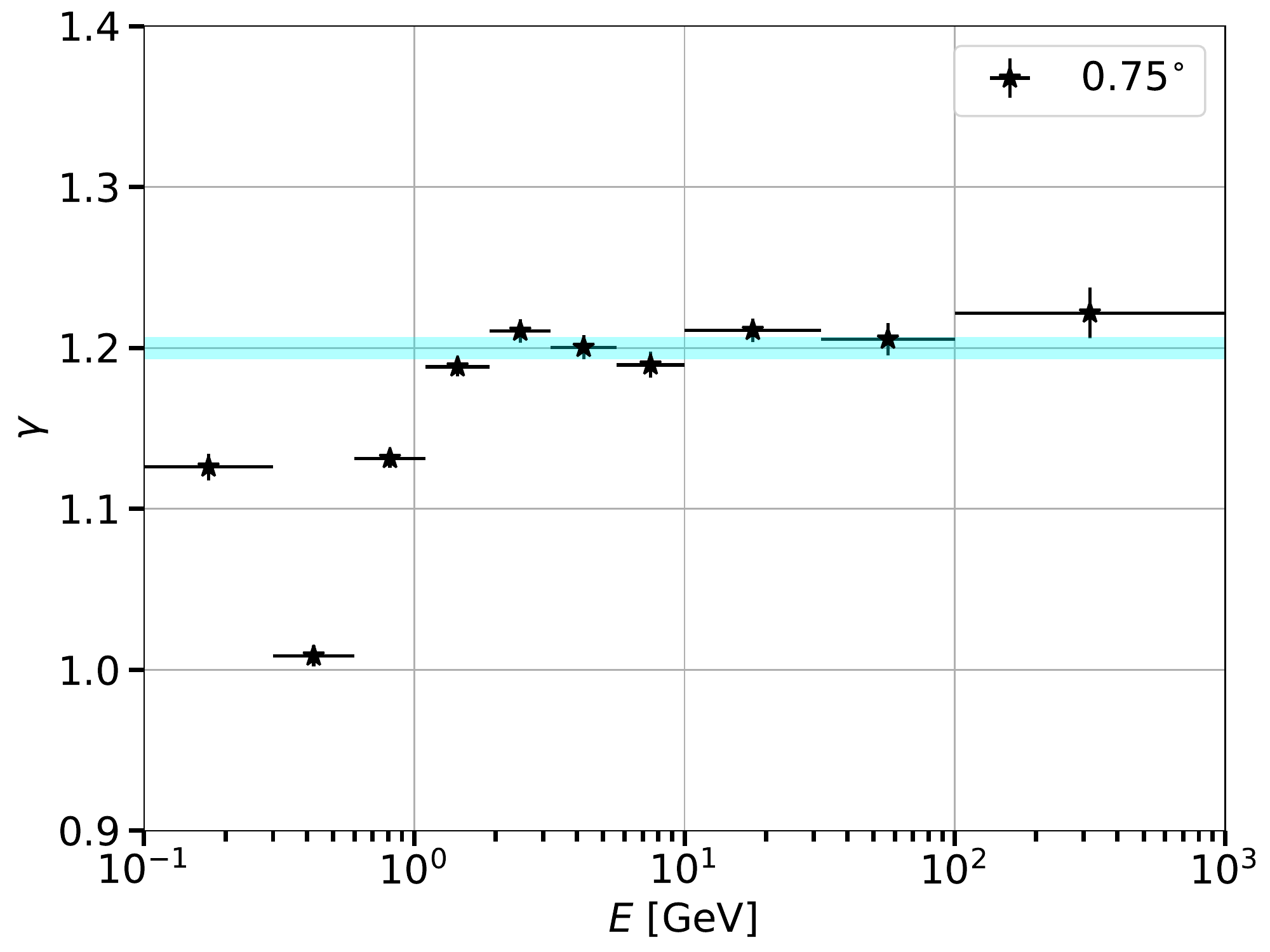}
\includegraphics[width=0.45\textwidth]{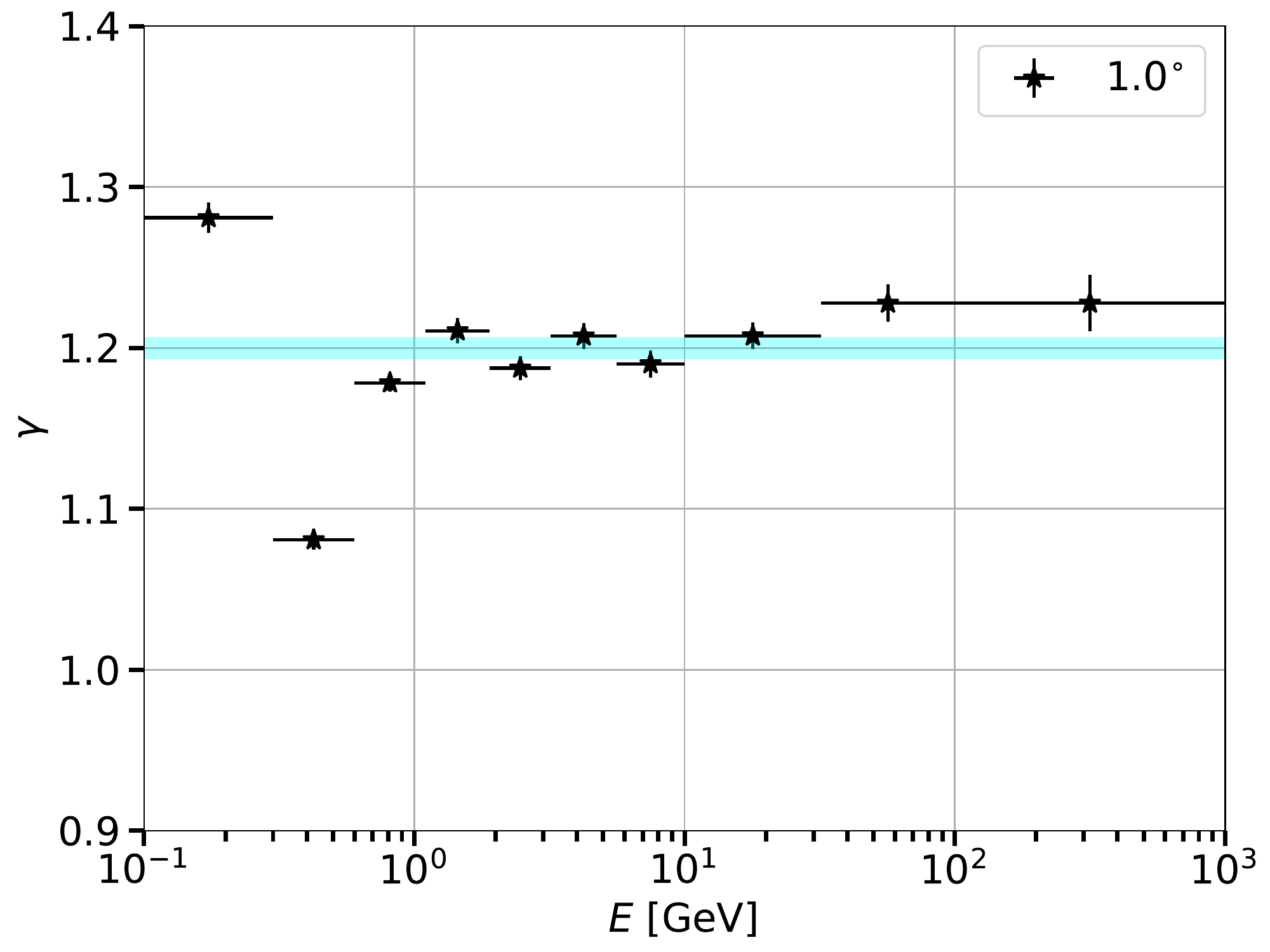}
\includegraphics[width=0.45\textwidth]{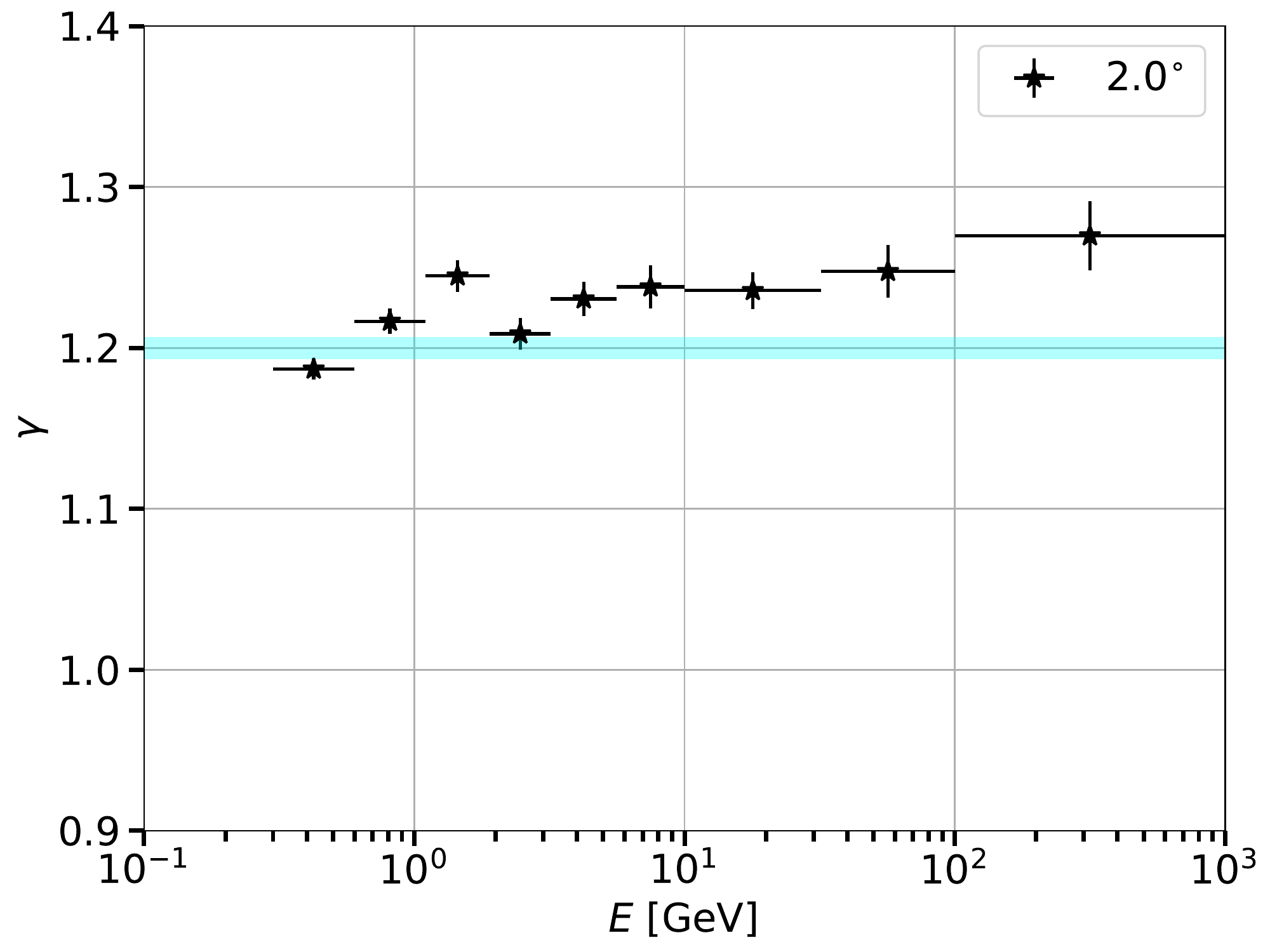}
\caption{Best-fit values of $\gamma$ (the slope of the DM density distribution for a NFW profile) found by fitting the surface brightness data for the simulation with an injected DM signal with $\gamma=1.2$. We show in each panel the best fit derived for the analysis applied in different energy bins and for different annulus sizes from $0.5^{\circ}$ (top left) to $2.0$ (bottom right). }%, with consistent color code as the right panel.}  
\label{fig:SBDMenergies}
\end{figure*}

%The spatial morphology of the GCE is one of the most important aspect that could provide us hints of its origin.
%In this section we apply two methods to inspect the morphology of the GCE in case it is given by DM.

We apply in this section the model independent approach presented in Sec.~\ref{sec:annulianalysis} to calculate the spatial distribution of the GCE in case we have a perfect knowledge of the background components.
We add in the model concentric and uniform annuli and we fit their SED parameters to the ROI.
Then, we extract the energy fluxes of the annuli and divide them by their solid angles.
As a result of this procedure we measure the surface brightness of the GCE.
We run this analysis in the following energy bins: $0.1-0.3$ GeV, $0.3-0.6$ GeV, $0.6-1.1$ GeV, $1.1-3.2$ GeV, $3.2-5.6$ GeV, $5.6-10$ GeV, $10-32$ GeV, $32-100$ GeV, $100-1000$ GeV.
In Fig.~\ref{fig:SBDM} we show, for example, the surface brightness data derived with our analysis in the energy range 1.1-1.9 GeV with annuli of width $0.75^{\circ}$. The surface brightness is vey peaked towards the Galactic center, as observed for the GCE. It decreases by roughly a factor of 100 considering an angular distance of $10^{\circ}$ away from the Galactic center. This justifies our choice of an ROI of size $30^{\circ}\times30^{\circ}$.

Once we have derived the surface brightness data in each energy bin and for the different choices of the annulus size, we perform a fit to the data by using the expected signal for $\gamma$ rays produced by DM. 
Specifically, we employ Eq.~\ref{eq:flux} where we evaluate the geometrical factor $\mathcal{J}$ as a function of the angular distance from the Galactic center. We leave free in the fit the normalization of the DM density profile $\rho_0$ and the slope $\gamma$ which changes the spatial profile of the $\gamma$-ray signal.
%In Fig.~\ref{fig:SBDM} we show together with the surface brightness data derived for $1.1-1.9$ GeV also the best fit with a DM template with $\gamma=1.19\pm0.02$ which is thus perfectly compatible with the injected DM signal.
In Fig.~\ref{fig:SBDMenergies} we display the results of the fit to the data for the different energy bins and using different sizes of the annuli from $0.5^{\circ}$ to $2.0^{\circ}$.
Inspecting those figures we can draw the conclusion that if the annulus size is smaller than $1^{\circ}$, we have results for the GCE profile that are not compatible with the injected signal if $E<1$ GeV.
%Indeed, for this energy range we find $\gamma$ best-fit values not compatible with $1.20$ regardless the annuli sizes.
Instead, for $E>1$ GeV the results for the best-fit values of $\gamma$ are compatible with 1.2 within the statistical errors.
For annuli of $1.0^{\circ}$ width the results are compatible with 1.2 for $E>0.6$ GeV, i.e.~at slightly smaller energies found for the previous cases.
Finally, if the annuli width is $2.0^{\circ}$ the best-fit values for $\gamma$ are systematically larger than 1.2.
This is due to the fact with such large annuli the surface brightness is not able to capture the real steepness of the DM emission.
To conclude an annulus size in the range $0.75^{\circ}-1.5^{\circ}$ and energies above 0.6 GeV are optimal choices to reconstruct properly the GCE spatial extension.
This conclusions is similar to the one we have drawn in Sec.~\ref{sec:annulianalysis} with the simulation of a Gaussian signal.
There are other two reasons for choosing these annulus sizes and energies.
The GCE spectrum for most of the analysis performed in the past has a peak at a few GeV and a low and high energy cutoff below 1 GeV and above 10 GeV.
So the most promising energy range to apply this technique is probably between 1-10 GeV.
Moreover, with annuli of width $0.5^{\circ}$ and an ROI of $30^{\circ}\times30^{\circ}$ there are 30 annuli in the model.
Since each annulus has two free SED parameters for a power-law shape, there are a total of 60 free parameters associated to annuli of $0.5^{\circ}$ width in addition to the ones of the other background sources.
Fits performed with so many parameters are very time consuming and challenging to perform.
It is thus recommended to choose broader annuli, between $0.75^{\circ}$ and $1.5^{\circ}$ of width, for which the number of free parameters is significantly lower and the results are still comparable with the injected signal (see Fig.~\ref{fig:SBDMenergies}).

\begin{figure}[t]
\includegraphics[width=0.45\textwidth]{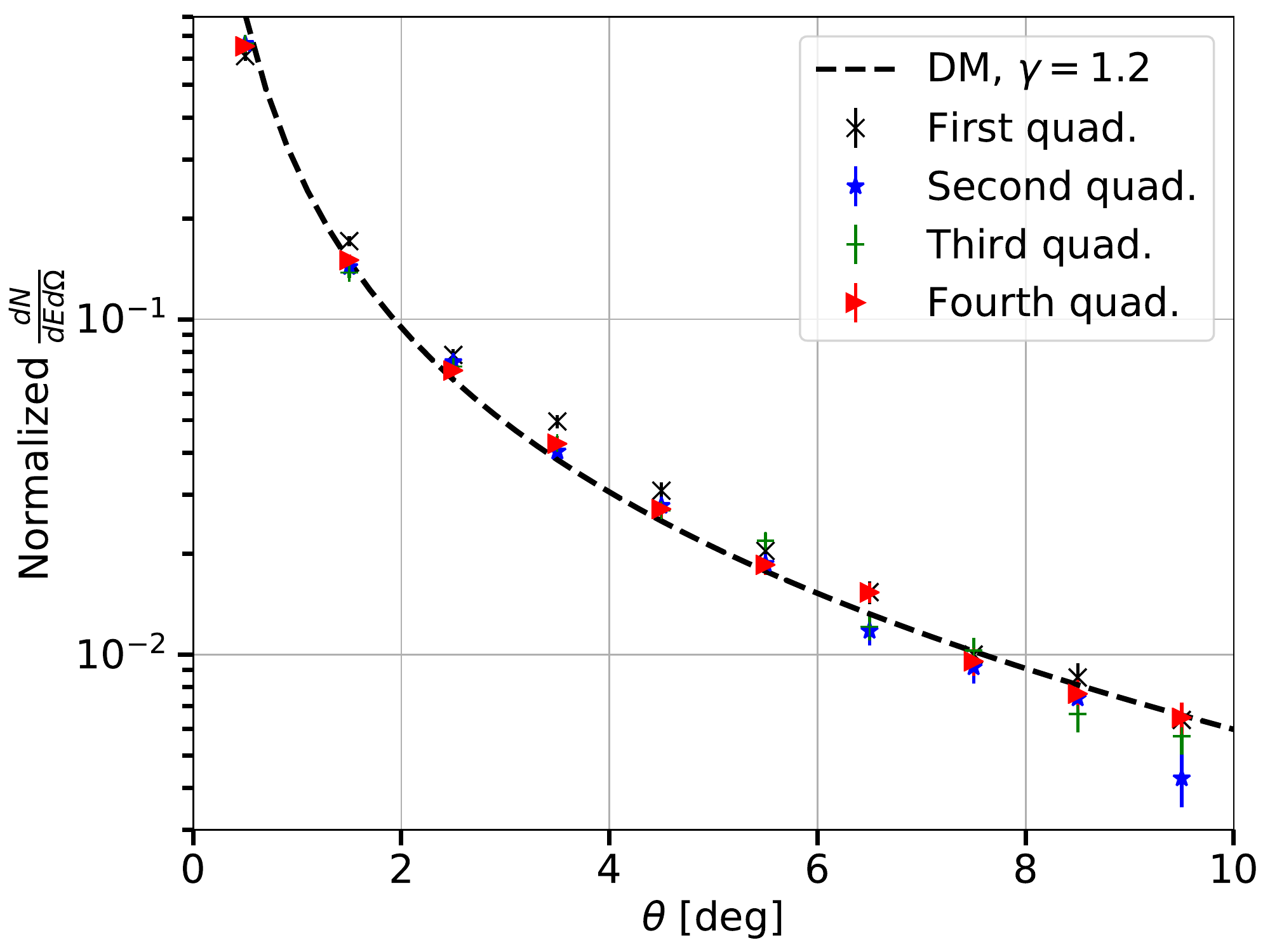}
\caption{Surface brightness of each quadrant of the GCE excess when it is due to a DM signal (see Sec.~\ref{sec:idealcase} for further details of the analysis). We show the data measured for each quadrant and the prediction for a DM signal with $\gamma=1.2$.}%, with consistent color code as the right panel.}  
\label{fig:SBDMquad}
\end{figure}

Our analysis would be able to find if the GCE spatial distribution is compatible with a DM signal.
An other important aspect to verify is whether the GCE is spherically symmetric.
In order to check this, we divide the annuli in four quadrants separated by the Galactic plane ($b=0^{\circ}$) and the vertical direction from the Galactic center ($l=0^{\circ}$). 
We consider for this test the energy range between 1-10 GeV since, as we have demonstrated before, this is the most promising energy range to study the GCE. 
We consider annuli of $1^{\circ}$ width that provide a good compromise between precision in the study of the GCE spatial distribution and a reasonable number of free parameter in the fit.
%We add each quadrant of all annuli as independent sources in the model.
Therefore, the model contains a total of 15 annuli divided into 4 quadrants. 
This brings to a total of 60 independent components added in the model.
%In addition to them there are the other background sources and IEM components free to vary in the fit.
We run the same analysis as before to find the surface brightness of each quadrant.
Then, we fit the surface brightness of each quadrant with a DM signal leaving free to vary $\rho_0$ and the slope $\gamma$ free for each quadrant.
We show in Fig.~\ref{fig:SBDMquad} the surface brightness data obtained for each quadrant together with the prediction for a DM signal with $\gamma=1.2$.
The data measured for each quadrant are all compatible with each other and with the DM signal.
Indeed, we find that $\gamma$ is equal to $1.20\pm 0.04$, $1.20\pm 0.03$, $1.18\pm 0.04$ and $1.21\pm 0.03$ in the four quadrants meaning that we find that the GCE is symmetric and each quadrant is compatible with the injected DM signal.
%We have tested in Sec.~\ref{sec:spectrumDM} the spectrum found in each quadrant and found that also this is compatible with the injected signal.

\begin{figure}[t]
\includegraphics[width=0.45\textwidth]{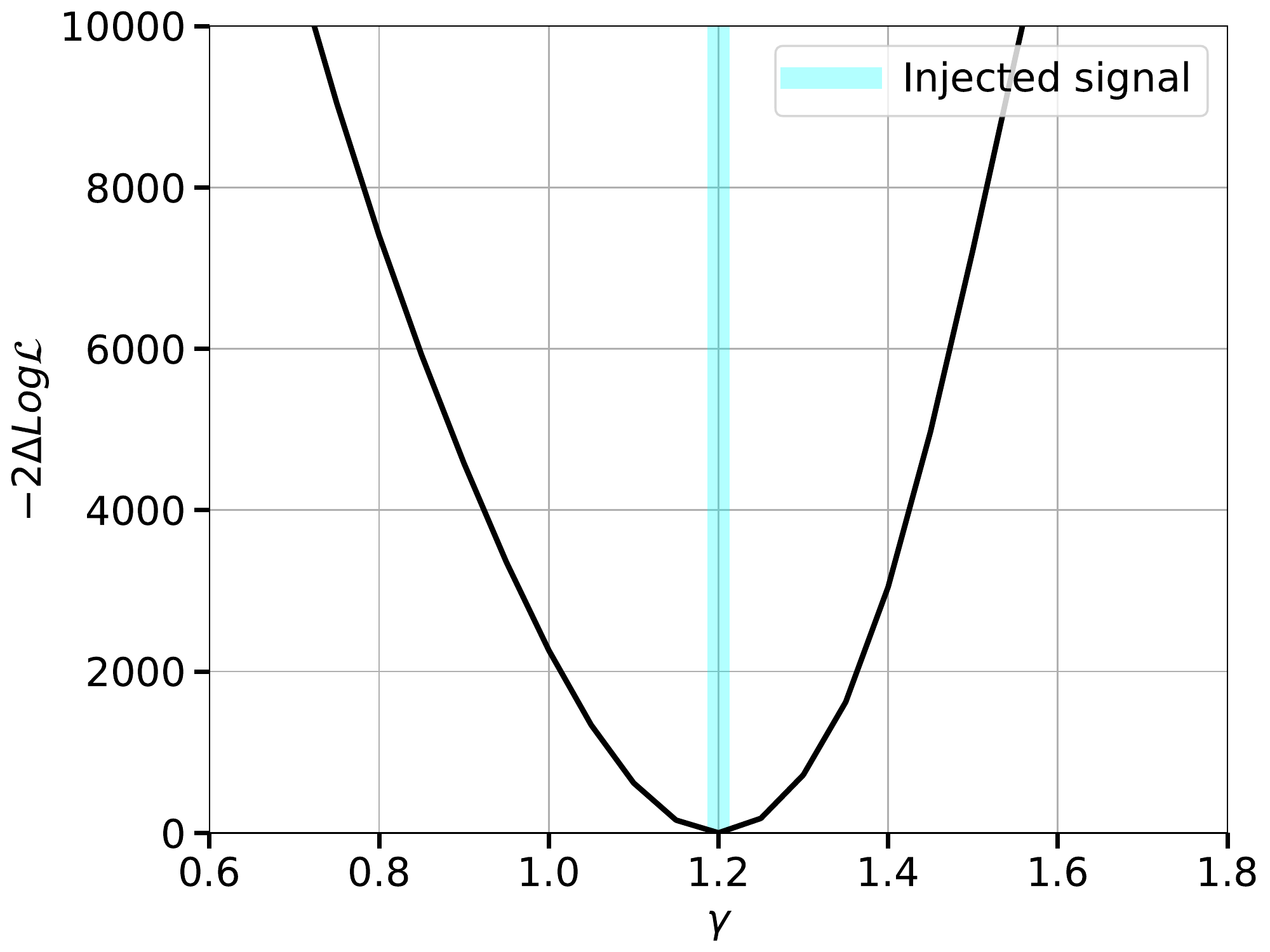}
\caption{Log-likelihood values found with a fit to the ROI using DM templates generated with different values of $\gamma$. The log-likelihood is provided as twice the difference with respect to the value obtained for $\gamma=1.2$ which provides the case that best represent the simulated data.}%, with consistent color code as the right panel.}  
\label{fig:DMgamma}
\end{figure}

%Since we have demonstrated before that we would be able to find that the GCE excess has a spatial morphology compatible with a DM signal and that is symmetric around the Galactic center 
We now study the spatial morphology of the GCE applying directly DM models modeled with a NFW density profile.
We run the analysis pipeline using DM templates generated for different values of $\gamma$ from 0.5 to 2.0.
Small (large) values of $\gamma$ implies that the spatial distribution is flatter (steeper).
For each case we extract the value of the log-likelihood ($\rm{Log}\mathcal{L}$). 
The background components we use are the same for all cases considered for $\gamma$.
Therefore, we have the likelihood profile as a function of $\gamma$ ($\rm{Log}\mathcal{L}(\gamma)$) since we marginalize the analysis over the background components.
In Fig.~\ref{fig:DMgamma} we show the function $\rm{Log}\mathcal{L}(\gamma)$ that we obtain which has a very pronounced peak around $\gamma=1.2$. Therefore, our results are perfectly compatible with the injected signal.
We apply this analysis also in different energy bins, as we did before for the annulus method, and we find very similar results to the ones presented in Fig.~\ref{fig:SBDMenergies}.

We can also check if the GCE is located in the center of the Galaxy as it is expected for a DM signal. 
In order to test this we run a fit to the simulated data using a DM template centered at different angular distances from the Galactic center. 
In particular we run the analysis with the DM emission center shifted in the range of $(-0.5^{\circ},+0.5^{\circ})$ in longitude and latitude from the Galactic center.
For each of the considered cases we run a fit to the ROI and find the $\rm{Log}\mathcal{L}$ value.
Collecting all the likelihood values for all the tested positions we thus have the likelihood as a function of longitude and latitude at which the DM template is centered ($\rm{Log}\mathcal{L}(l,b)$). The position with the highest value of the likelihood provides the best fit position of the GCE assuming that it is compatible with a DM signal.
In the case considered in this section, where we have a perfect knowledge of the ROI components, $\rm{Log}\mathcal{L}(l,b)$ has a prominent peak in the Galactic center. The change of log-likelihood moving just by $0.1^{\circ}$ from $l=b=0^{\circ}$ is roughly 350 which means a preference of about $25\sigma$ for the Galactic center position with respect to this adjacent one.

To summarize, we have demonstrated in this section that in the ideal case where the model used in the analysis is exactly the same used to create the simulation, i.e., in case we have a perfect modeling of the $\gamma$-ray sky, we are able to find the correct DM SED between $0.1-1000$ GeV, the correct GCE spatial distribution for $E>1$ GeV and choosing annulus sizes between $0.75^{\circ}$ and $1.5^{\circ}$ width and we would be able to prove that the GCE is spherically symmetric and centered in the Galactic center as it should be for a DM signal.

%(-2.5551802192785225, 0.014302380488845579, 1.1650814277089254, 0.008174015800914702, -8.103155376807758, 26.58343345929257)
%(-2.6425046926739633, 0.014735546705559877, 1.206847912367048, 0.007911952716151971, -7.9030036112593685, 21.182054391982696)
%(-2.6306389149519482, 0.014211921581194442, 1.2005999014031592, 0.007750310572483543, -8.148370284528092, 30.98484335291613)
%(-2.6200846523304477, 0.014532678800247112, 1.1957490642346196, 0.008182648870817877, -7.772948448551498, 21.843564843361907)

\section{A more realistic case: imperfect knowledge of background components}
\label{sec:realisticcase}

In a more realistic scenario, we probably do not have a perfect knowledge of the $\gamma$-ray sky when we analyze a such complicated region as the Galactic center.
This implies that an analysis of the GCE is certainly affected by mismodeling of the background components.
In order to inspect how and what properties of the GCE are affected by this circumstances, we consider two cases: a model with a missing component with respect to the one used to generate the simulations and the analysis of simulations with a different IEM model with respect to the one used to generate them.
Otherwise differently stated we consider energies in the range of $1-10$ GeV and we a DM signal modeled with $\gamma=1.2$.

\subsection{Properties of the Galactic center excess in case of a missing component}
\label{sec:missingcomp}

\begin{figure*}[t]
\includegraphics[width=0.49\textwidth]{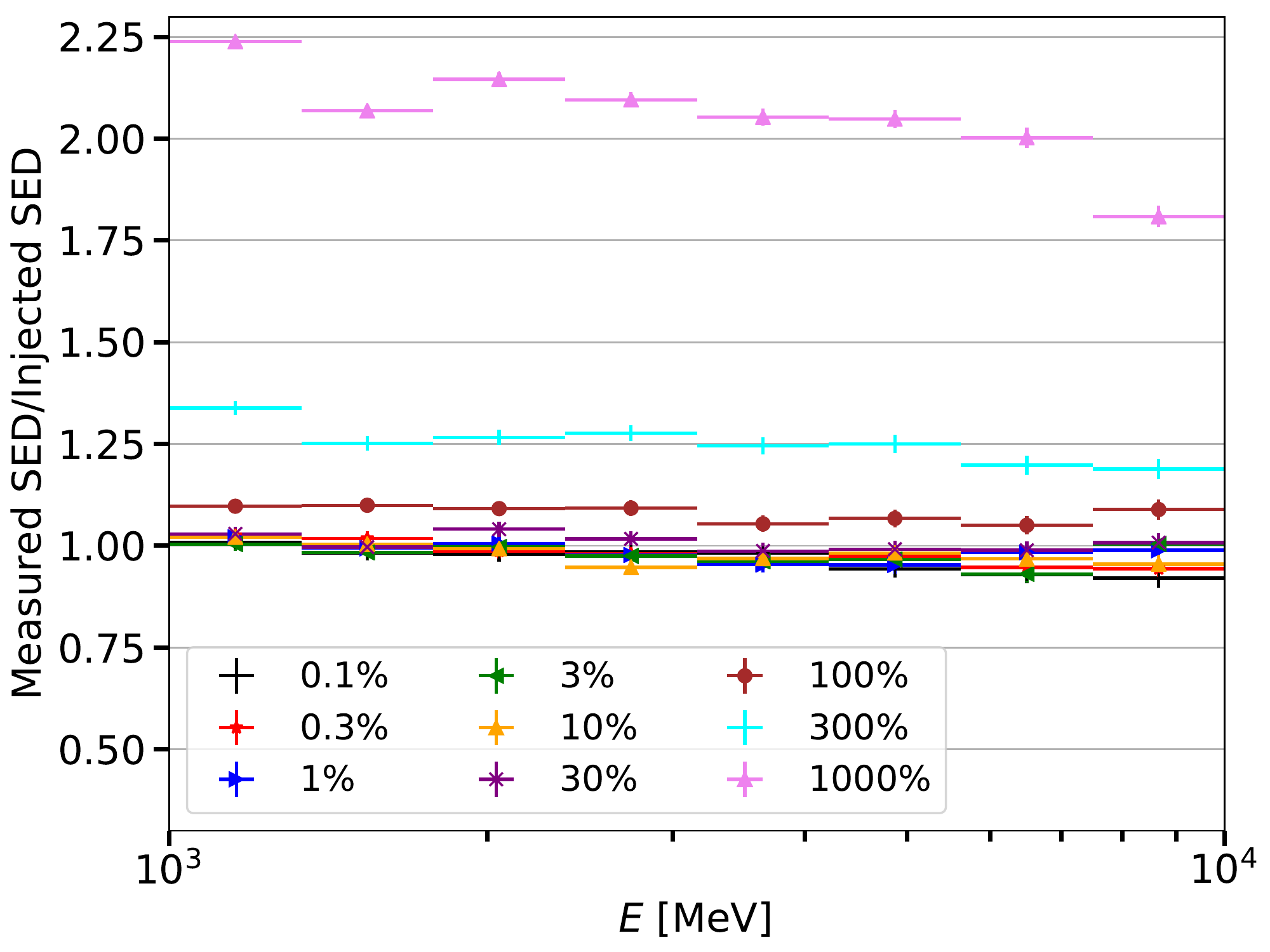}
\includegraphics[width=0.49\textwidth]{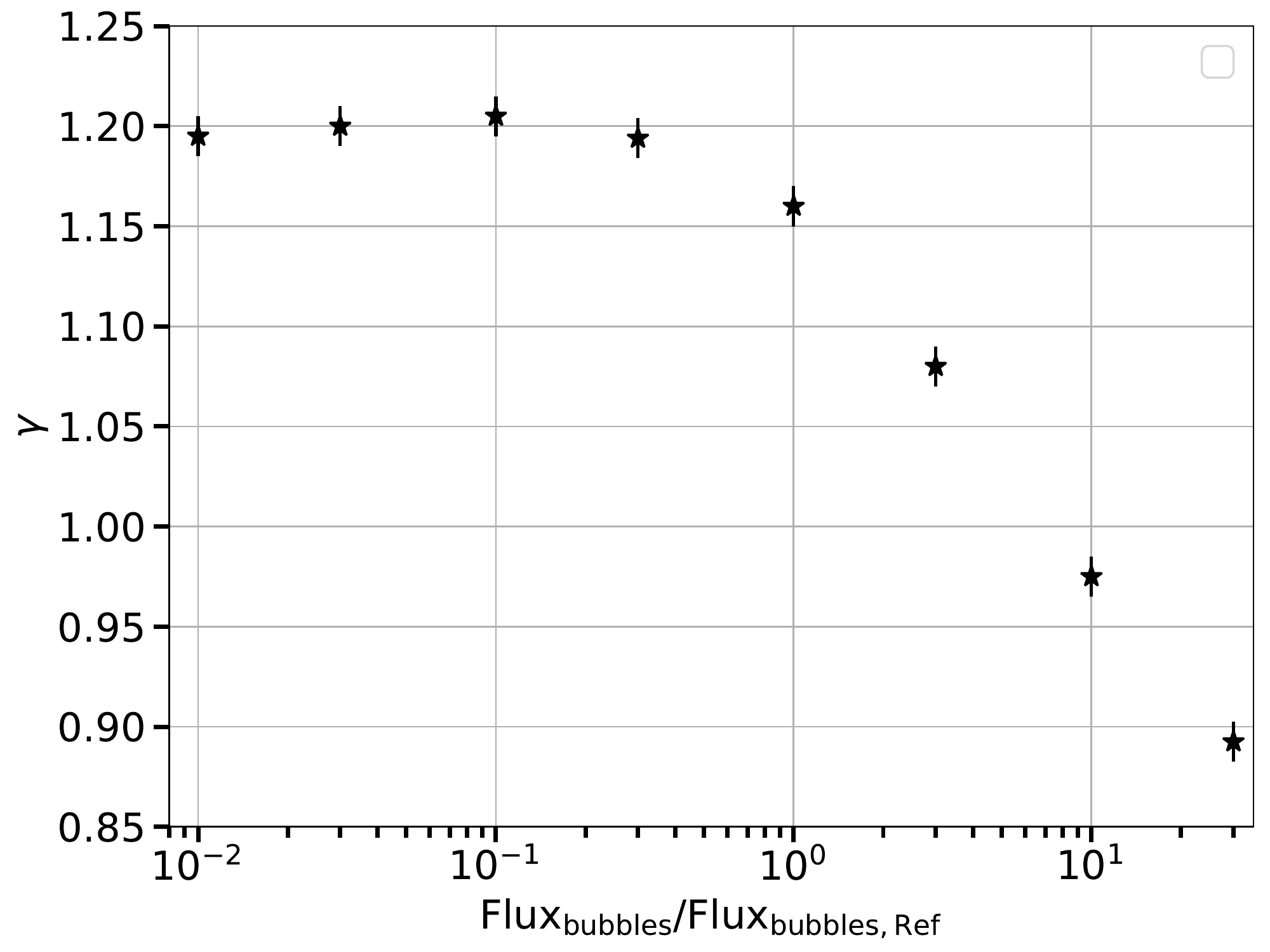}
\caption{Left Panel: Ratio of the GCE spectrum fitted with a DM template with respect to the injected DM signal for the simulations explained in Sec.~\ref{sec:missingcomp}. These ratios are reported for different simulations for which we have simulated also the low-latitude bubbles with a flux between $0.01-10$ times the reference values which is $3\times 10^{-7}$ GeV$^{-1}$ cm$^{-2}$ s$^{-1}$. Right Panel: best-fit value for $\gamma$ found for the same simulations explained before as a function of the flux simulated for the low-latitude bubbles.}%, with consistent color code as the right panel.}  
\label{fig:bubbles}
\end{figure*}

\begin{figure}[t]
\includegraphics[width=0.45\textwidth]{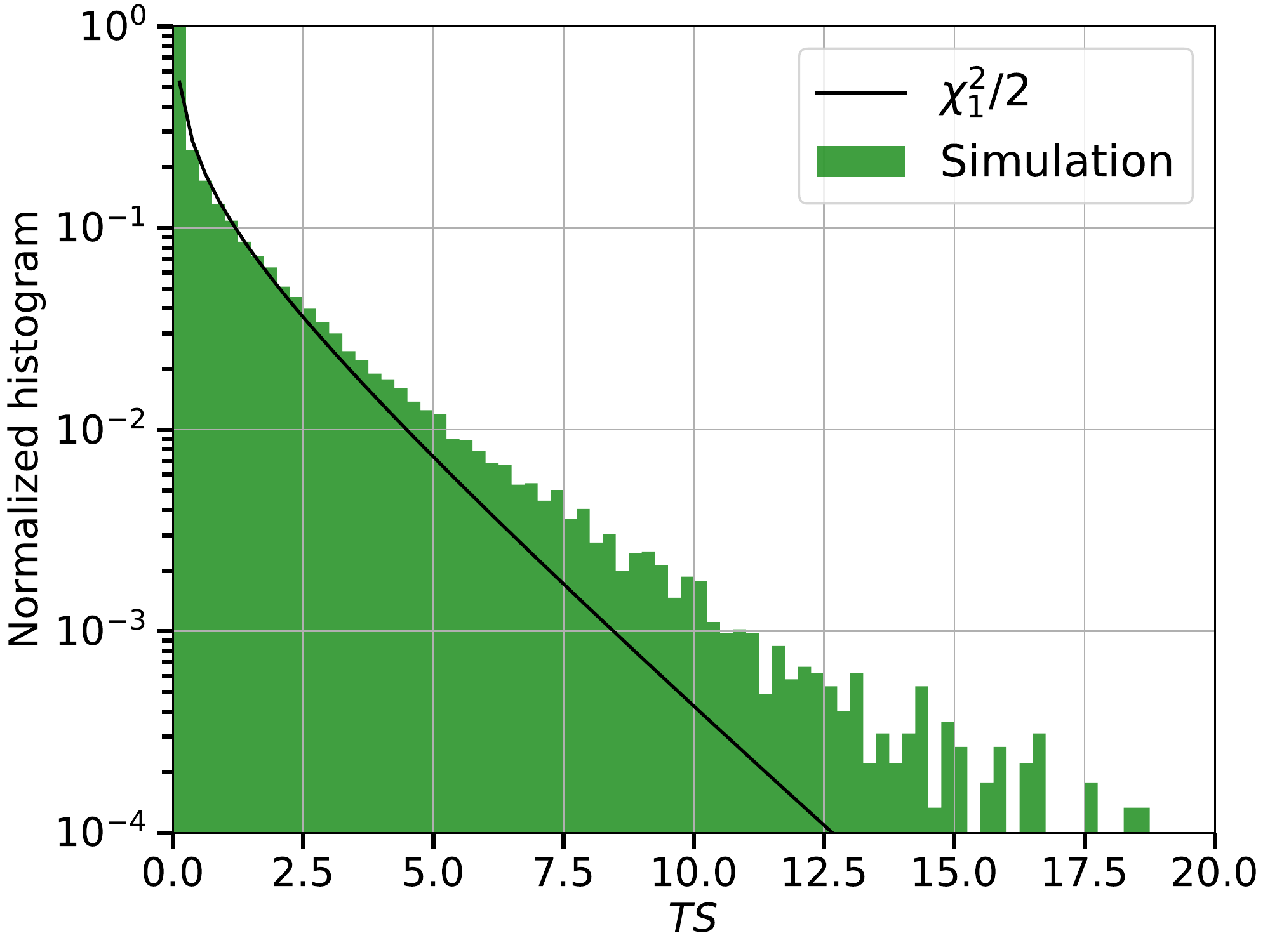}
\caption{Histogram of the $TS$ map found in the case where we simulate the low-latitude bubbles with a flux of $2\times 10^{-7}$ GeV$^{-1}$ cm$^{-2}$ s$^{-1}$ but then we do not include this component in the model during the fitting procedure to find the properties of the GCE. The $TS$ distribution is significantly different from the $\chi_1^2$ distribution for 1 degree of freedom divided by two.}%, with consistent color code as the right panel.}  
\label{fig:TSbubbles}
\end{figure}

In case the real sky contains a component that is not included in model used to fit the data, the properties of the detected GCE could be very different from the real one. 
The missing component could be a part of the IEM or a new emission in the Galaxy not properly accounted for in the model.
Here, we assume that the component missing in the model is associated to the low-latitude component of the {\it Fermi} bubbles.
This component has been carefully inspected in \cite{TheFermi-LAT:2017vmf,Herold:2019pei}.
The low-latitude bubbles are located close to the Galactic center and they extend up to about $15^{\circ}$. 
Moreover, their spectrum is roughly a few times smaller than the GCE \cite{TheFermi-LAT:2017vmf}. 
We also find that this component is the one with the largest correlation coefficient with the GCE.
It is thus ideal to use this $\gamma$-ray emission and test its influence on the GCE properties.

%We inspect how a model missing the low-latitude bubbles emission can affect the search for the GCE.
This is the procedure we follow to test this case.
We generate simulations that include the low-latitude bubbles emission.
Then, we eliminate this component from the model before running the fitting procedure and we find the properties of the GCE following the same analysis technique applied in Sec.~\ref{sec:idealcase}.
We perform simulations with low-latitude bubbles fixing the SED shape found in \cite{TheFermi-LAT:2017vmf} and varying their flux.
We will use as a reference value of the flux at 1 GeV $3\times 10^{-7}$ GeV$^{-1}$ cm$^{-2}$ s$^{-1}$ which is required to fit properly the real data from the Galactic center \cite{TheFermi-LAT:2017vmf}. We will vary it from $3\times 10^{-9}- 3\times10^{-5}$ GeV$^{-1}$ cm$^{-2}$ s$^{-1}$ which is thus between a factor of 100 times less and more intense than the reference value \cite{TheFermi-LAT:2017vmf}. 
In the following we will report the results as a function of the flux of the low-latitude bubbles injected in the simulations with respect to the reference flux of $3\times 10^{-7}$ GeV$^{-1}$ cm$^{-2}$ s$^{-1}$. 

\begin{figure}[t]
\includegraphics[width=0.49\textwidth]{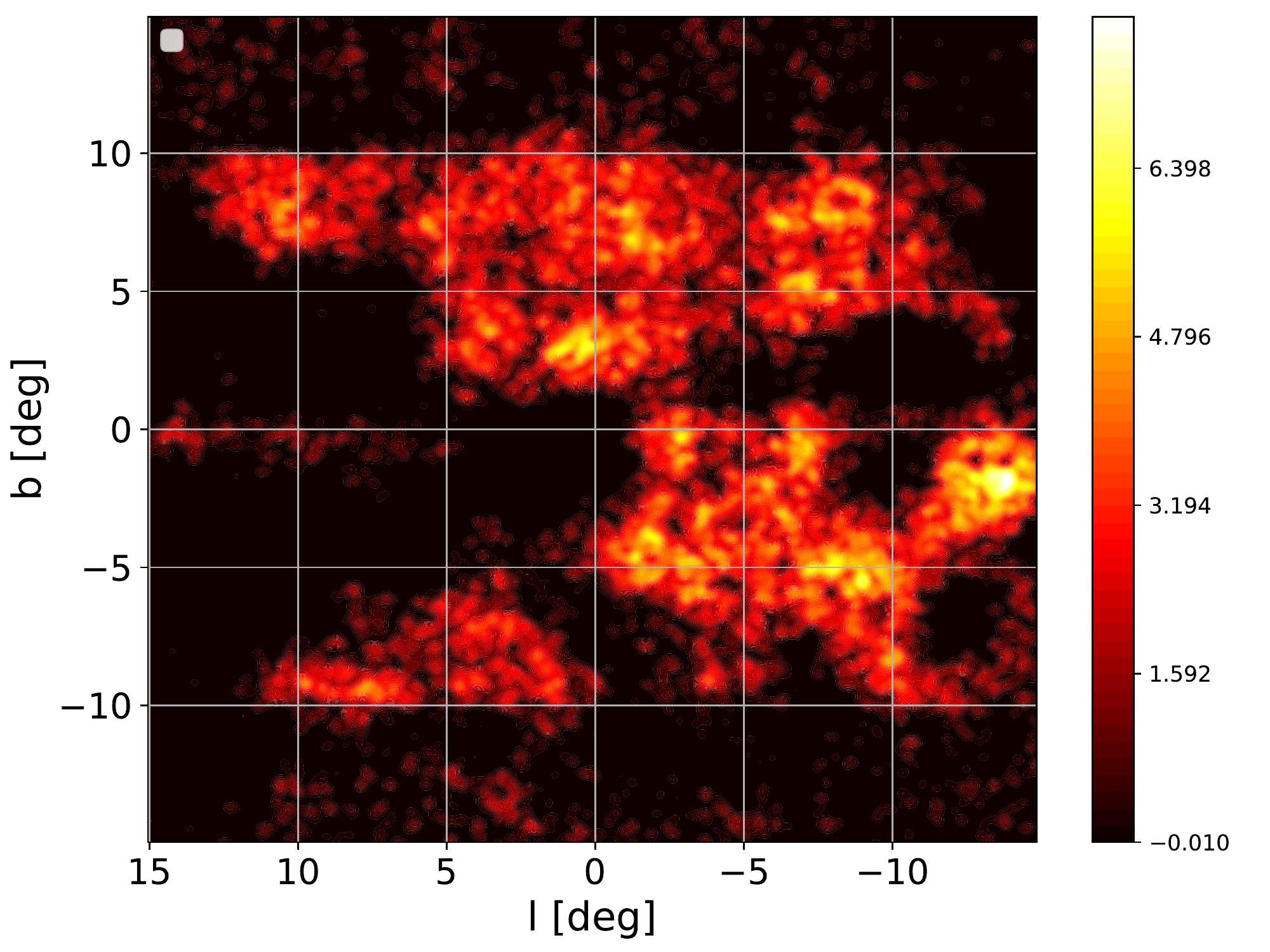}
\caption{Map of $\sqrt{TS}$ for the analysis presented in Sec.~\ref{sec:missingcomp} and obtained by simulating the data using the {\tt Baseline} IEM and the low-latitude bubbles with a flux at 1 GeV of $9\times 10^{-7}$ GeV$^{-1}$ cm$^{-2}$ s$^{-1}$ (three times the reference flux) and then analyzing the data without including this component in the background model.}  
\label{fig:TSsysbubbles}
\end{figure}

In Fig.~\ref{fig:TSsysbubbles} we show the map of $\sqrt{TS}$ obtained by simulating the data using the {\tt Baseline} IEM and the low-latitude bubbles with a flux at 1 GeV of $9\times 10^{-7}$ GeV$^{-1}$ cm$^{-2}$ s$^{-1}$ and then analyzing the data without including this component in the background model. 
The $\sqrt{TS}$ map contains values that reach maximum values of roughly $5-7 \sigma$ significance. These peaks of $TS$ affects the results for the reconstructed properties of the GCE as we will explain in more details below.

First, we investigate how the GCE spectrum changes if we neglect the presence of the low-latitude bubbles.
We report in the left panel of Fig.~\ref{fig:bubbles} this result.
The GCE spectrum starts to differ from the injected DM signal if the low-latitude bubbles have a flux larger than $30\%$ of the reference value.
%For fluxes larger than $\sim 10^{-7}$ GeV$^{-1}$ cm$^{-2}$ s$^{-1}$ the GCE spectrum starts to deviate significantly from the injected DM signal.
This is expected since the injected GCE spectrum has a flux at 1 GeV of $7\times 10^{-7}$ GeV$^{-1}$ cm$^{-2}$ s$^{-1}$ and so for low-latitude bubbles fluxes larger than $1\times 10^{-7}$ GeV$^{-1}$ cm$^{-2}$ s$^{-1}$ (i.e., $30\%$ of the reference flux) this component starts to shine with a flux that is not negligible with respect to the GCE.
Moreover, if the low-latitude bubbles flux is large enough also the shape of the GCE flux as a function of energy changes and becomes more similar to the one of the low-latitude bubbles. 
%as can be seen for the case with $10$ times the reference value. 
Indeed, for a flux larger than 3 times the reference one the spectrum is softer than $\sim E^{-2}$, which is the SED used for the DM signal.

Then, we perform the analysis in annuli to derive what is the value of the low-latitude bubbles flux above which the spatial distribution of the GCE would start to differ from the morphology of the injected DM signal. We use annulus width of $1^{\circ}$.
In the right panel of Fig.~\ref{fig:bubbles} we show the value of $\gamma$ as a function of the different fluxes of the bubble component used in the simulations.
The value we find for $\gamma$ is compatible with 1.2 for bubbles fluxes up to about $30\%$ of the reference one.
The reason for this is the same reported before for the GCE flux as a function of energy. For fluxes lower than this value, neglecting the presence of the low-latitude bubbles in the model does not introduce any significant residuals and so the value of $\gamma$ is the same of the injected signal.
On the other hand, for larger percentages the low-latitude bubbles start to have a flux that is not negligible with respect to the DM signal and so also the value found for $\gamma$ starts to differ significantly from 1.2.
In particular, for very large fluxes, $\gamma$ tends to values lower than 1 which are compatible with the spatial morphology of the low-latitude bubbles.

Finally, we test the analysis with the DM template divided into quadrants. 
%As we have seen in the ideal case presented in Sec.~\ref{sec:idealcase}, we would be able to demonstrate that the spectrum and the spatial distribution of the quadrants is the same and it is compatible with the DM signal.
Neglecting the presence of the low-latitude bubbles in the data affects significantly this study.
We verify that the GCE signal starts to be detected as asymmetric, and so the spectrum and morphology start to be different among the quadrants, if the low-latitude bubbles have a flux of about $10\%$ of the reference value.
This is explained by the fact that each quadrant has a flux that is about one fourth of the total DM signal. Therefore, the quadrant spectrum and morphology are affected by the presence of the low-latitude bubbles signal for much smaller fluxes than found before.

The presence of a missing component in the model can be discovered by looking to the histogram of the $TS$.
Indeed, in presence of residuals, the $TS$ distribution would deviate from the distribution of the $\chi_1^2/2$.
We demonstrate this by showing the $TS$ distribution we find in the analysis where we neglect the low-latitude bubbles with a flux equal to the reference value.
This is presented in Fig.~\ref{fig:TSbubbles}. At small values of $TS$ the distribution is still compatible with the null hypothesis (i.e., no additional sources) but for values $>5$ the distribution significantly deviates from the shape of $\chi_1^2/2$. This means that there are residuals left after the fitting procedure which can be due to a missing component not accounted for in the model or a slightly wrong IEM used in the analysis. In this case of course the right reason in the former.

\subsection{Properties of the Galactic center excess in case of a wrong Galactic interstellar emission model}
\label{sec:wrongmodel}

The Galactic center is probably the most complicated part of the $\gamma$-ray sky to model.
The components included in the fitting procedure could not represent perfectly the data and this can affect the properties of the
GCE that we find in the analysis.
In this section we generated simulations using the {\tt Baseline} IEM and we then use the other models to analyze the data.
We will thus verify how imperfections in the model are going to affect the detected GCE properties.
% since we have demonstrated in the previous sections that this is the most promising one to study the properties of the GCE.
Once we generate the simulations using the {\tt Baseline} IEM we substitute all its components with the ones of the other IEM models (see Sec.~\ref{sec:iemdata}).
Then, we apply the same analysis technique reported in Sec.~\ref{sec:idealcase} to find the spectrum and the morphology of the GCE using a DM template and the model independent approach that involves annuli.

\begin{figure*}[t]
\includegraphics[width=0.49\textwidth]{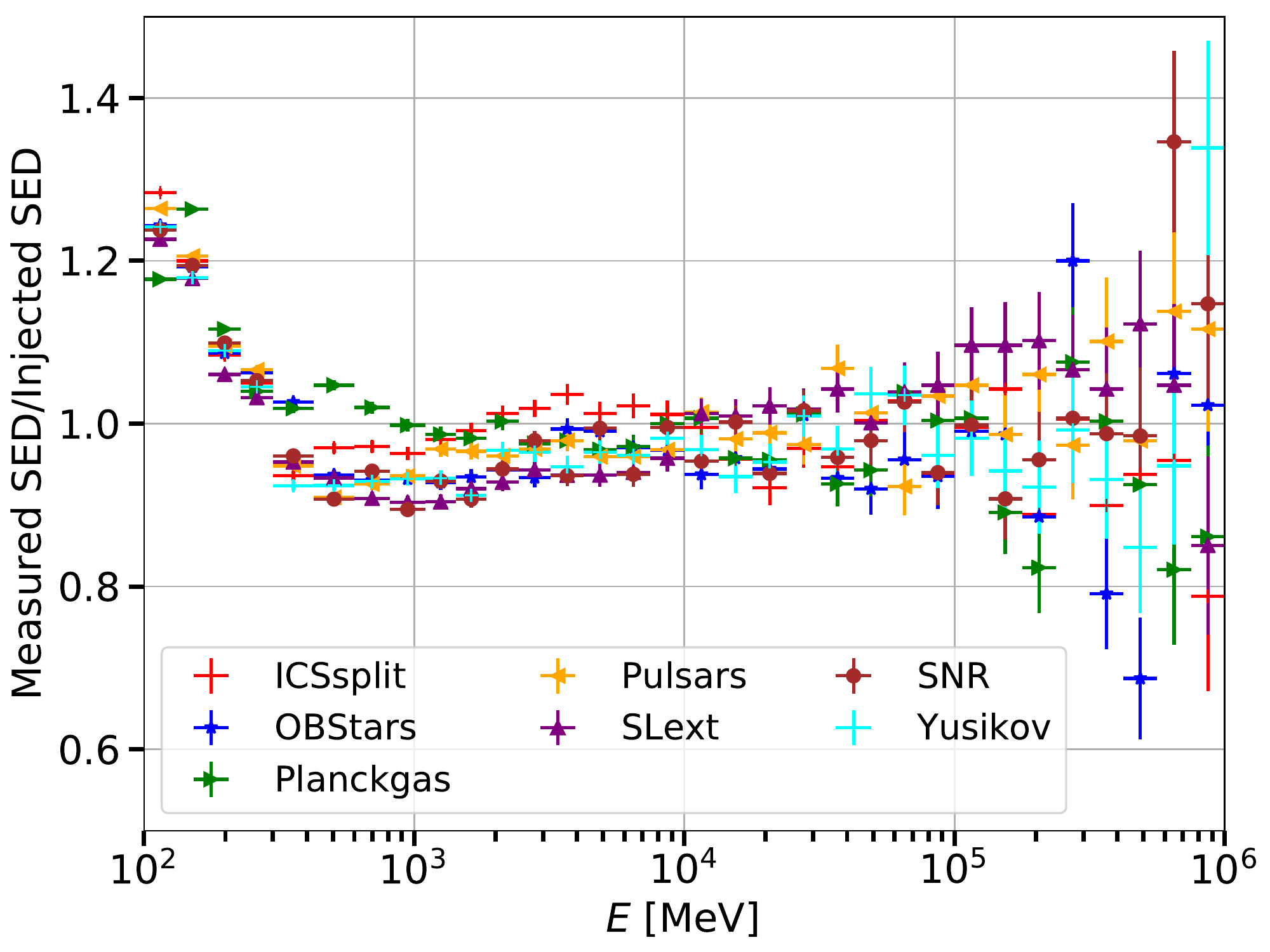}
\includegraphics[width=0.49\textwidth]{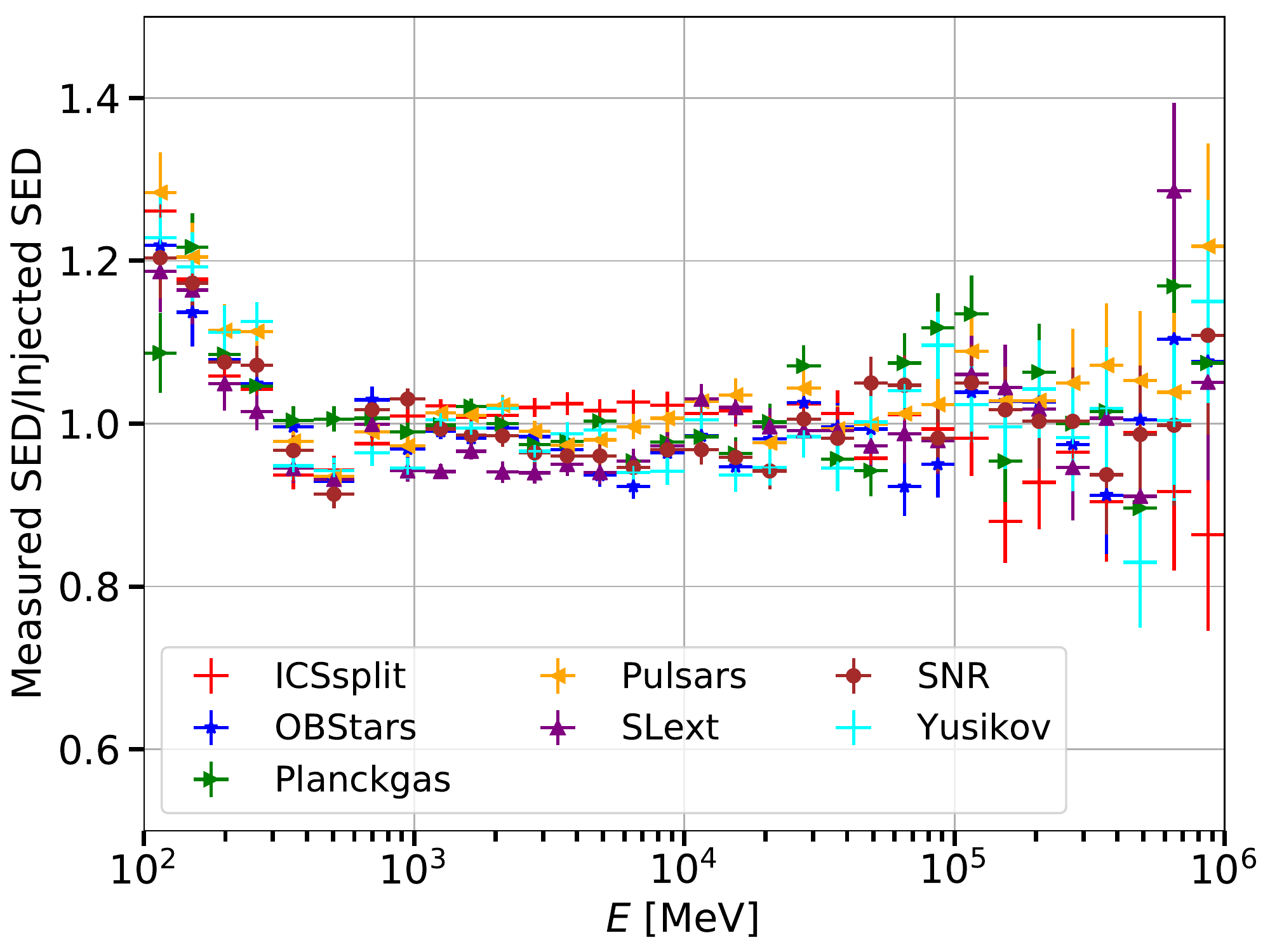}
\caption{Ratio of the measured and injected energy spectrum of the GCE found for different IEM models when we use a DM template with $\gamma=1.2$. This simulations has been performed in the energy range between 0.1-1000 GeV without (left panel) and with the weighted likelihood technique (right panel).}  
\label{fig:resultsIEMebins}
\end{figure*}

First, we run the analysis in the entire energy range from 0.1-1000 GeV using or not the weighted likelihood technique. We test thus the effect of running the analysis with the weighted likelihood technique to minimize the systematics due to the IEM.
We show in Fig.~\ref{fig:resultsIEMebins} the ratio between the GCE SED found in the analysis with respect to the injected signal and we report the results obtained using different IEMs.
Above 1 GeV the ratio is roughly compatible with 1 with differences among the IEMs of roughly $10-15\%$. On the other hand, the ratios become significantly larger than 1 at lower energies. Specifically, the lower is the energy and the larger is the ratio value reaching the largest value of $1.2-1.3$ at 100 MeV.
Performing the analysis with the weighted likelihood technique has the consequence of increasing the uncertainties of each single data point and thus reducing the discrepancy with the injected signal. In particular, the $\chi^2$ value calculated between the reconstructed and injected SED improves by a factor of about 30 by using the weighted likelihood technique with respect to using the standard analysis. This makes the results more compatible with the injected signal also below 1 GeV and the differences obtained among the different IEMs significantly smaller. Instead at energies above 1 GeV using or not this technique does not make any relevant difference. So in what follows when we will select energies between 1-1000 GeV we will use the standard maximum likelihood.

\begin{figure}[t]
\includegraphics[width=0.49\textwidth]{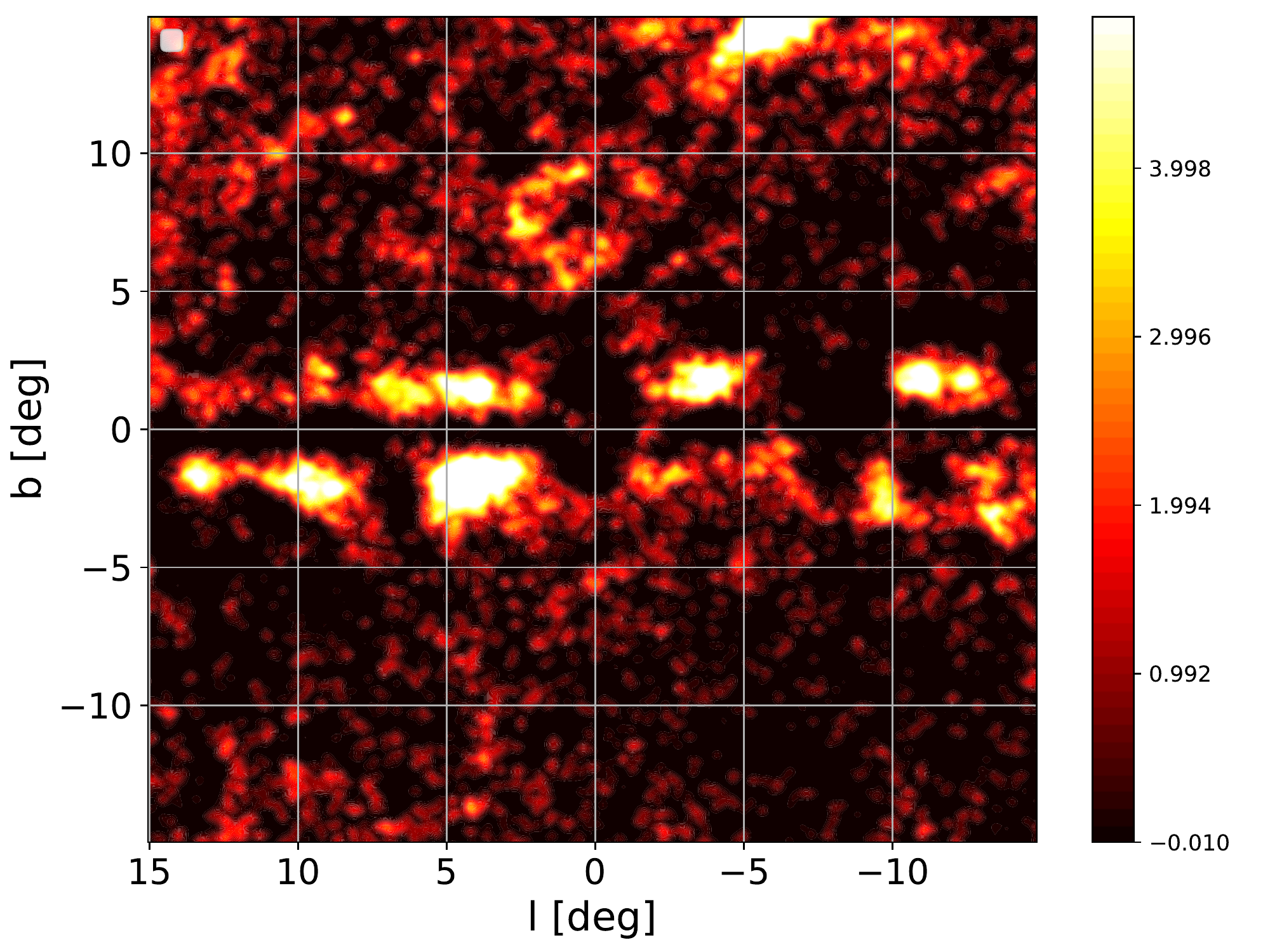}
\caption{Map of $\sqrt{TS}$ for the analysis presented in Sec.~\ref{sec:wrongmodel} and obtained by simulating the data using the {\tt Baseline} IEM and then using to analyze the data the {\tt SLext} model.}  
\label{fig:TSsysIEM}
\end{figure}

We focus now our analysis in the energy range between $1-10$ GeV for which the GCE is the most significant and the systematics due to the IEM are smaller.
In Fig.~\ref{fig:TSsysIEM} we show the map of $\sqrt{TS}$ obtained by simulating the data using the {\tt Baseline} IEM and then using to analyze the data the {\tt SLext} model. This model is among the ones that gives the largest systematics for the reconstructed GCE properties. The $\sqrt{TS}$ map contains values that are lower than 25 at latitudes $|b|>5^{\circ}$ while on the Galactic plane the mismodeling of the IEM could produce $\sqrt{TS}$ up to roughly $6-8 \sigma$ significance. These peaks of $TS$ affects the results for the reconstructed properties of the GCE as we will explain in more details below.

\begin{figure*}[t]
\includegraphics[width=0.49\textwidth]{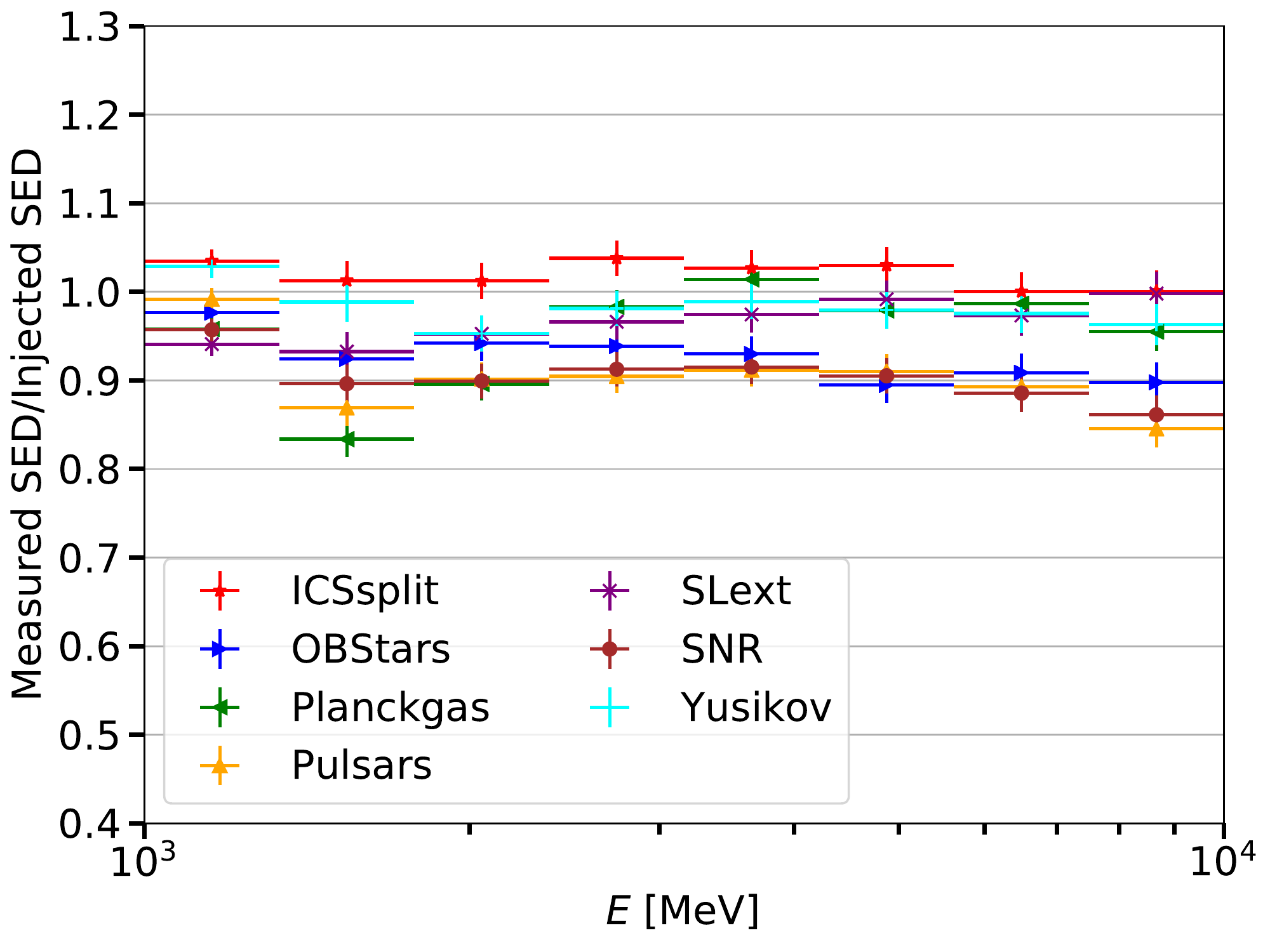}
\includegraphics[width=0.49\textwidth]{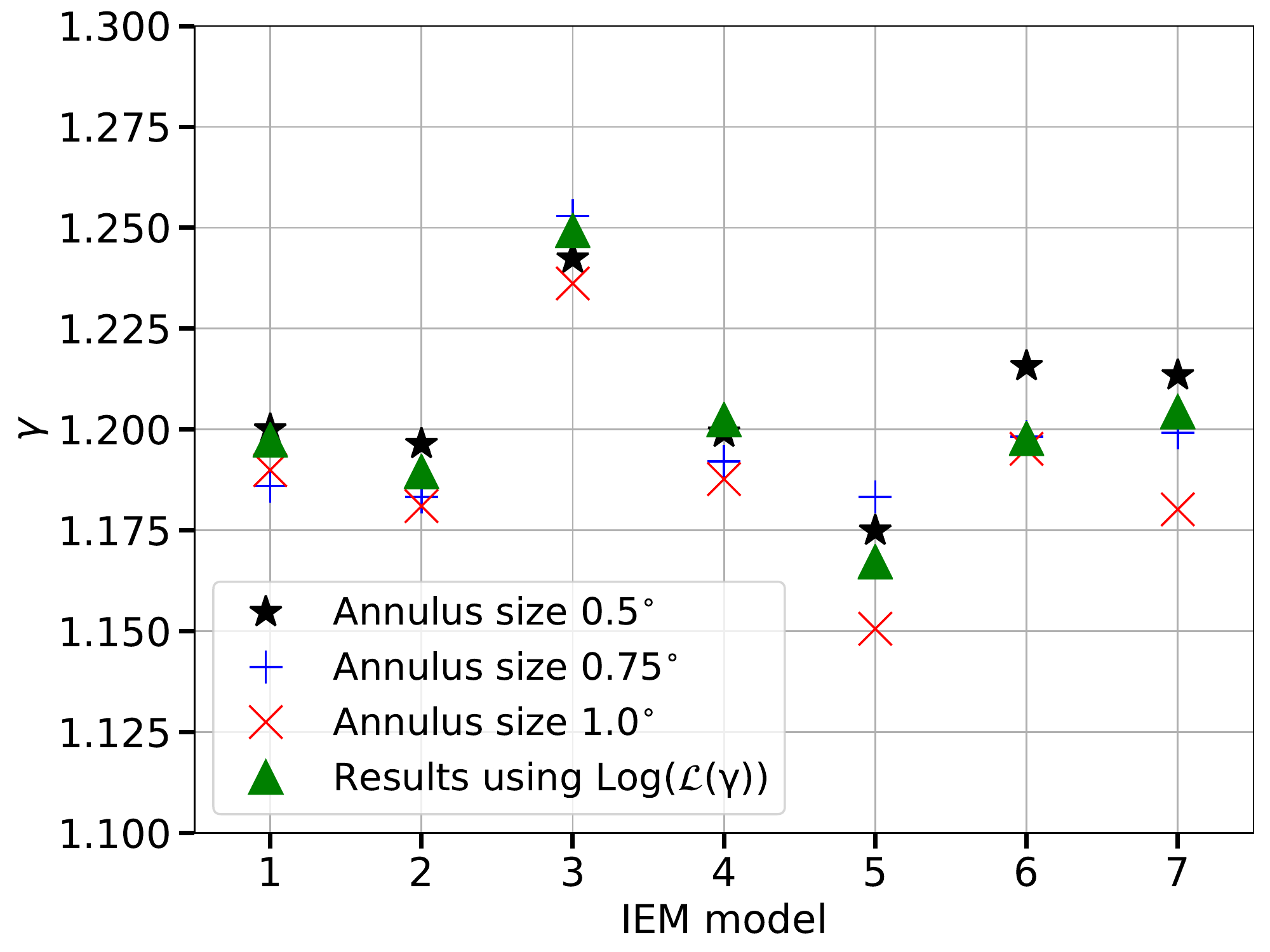}
\caption{Left Panel: Ratio of the measured and injected energy spectrum of the GCE found for different IEM models when we use a DM template with $\gamma=1.2$. Right Panel: best-fit value for $\gamma$ found by fitting the surface brightness of the GCE and considering in the analysis the same IEM models reported in the left panel. The IEM models are listed on the x axis with the same order of the left panel figure.}  
\label{fig:resultsIEM}
\end{figure*}

In Fig.~\ref{fig:resultsIEM} we show the results for the energy spectrum that we measure for the GCE using a DM template with $\gamma=1.2$. 
The spectrum we find is compared to the one of the injected signal for all the IEM we try.
There is a larger scatter with respect to the ideal case presented in Fig.~\ref{fig:DMenergy}. The differences with respect to the injected signal are at most of the order of $10-15\%$. The statistical errors are much smaller so the uncertainties in the analysis of the spectrum of the GCE in reality is probably systematic dominated. 
The differences in the results of the GCE spectrum obtained with the different IEM models are mainly a normalization factor. 
Indeed, in the energy range considered there is not significant change of slope of the GCE SED.

We also study the spatial distribution of the injected DM signal using the annulus technique.
We consider in particular annulus sizes of $0.5^{\circ}$,  $0.75^{\circ}$ and  $1^{\circ}$.
The result is presented in the right panel of Fig.~\ref{fig:resultsIEM} for the different IEM models used in the analysis.
On average we find values of $\gamma$ that are compatible with the injected signal. 
There is, however, a scatter that is due to the choice of the IEM rather than by the statistical errors of the value of $\gamma$ found for each case.
This scatter, that is at most $\sim 5\%$ of the injected value of $\gamma$, is larger than the one reported in Fig.~\ref{fig:SBDMenergies} for the same energy range.
As well as for the GCE energy spectrum thus also the study of the GCE spatial distribution is systematics dominated.

We apply also an other analysis technique to study the spatial distribution of the GCE that is based on the DM model for the $\gamma$-ray signal.
Since we have demonstrated with the previous analysis that the GCE signal is compatible with the DM template we try to find the best fit for $\gamma$ using directly a DM template.
We run the analysis with DM templates generated for different values of $\gamma$.
We take for each case the value of the log-likelihood and we profile it as a function of $\gamma$ ($\rm{Log}(\mathcal{L})(\gamma)$).
The peak of $\rm{Log}(\mathcal{L})(\gamma)$ gives the best-fit value for the DM density slope.
This analysis provides best-fit values for $\gamma$ that are comparable within the $1\sigma$ errors with the one presented in right panel of Fig.~\ref{fig:resultsIEM}.

The analysis of the GCE position gives very similar results to the ideal case presented in Sec.~\ref{sec:spatialDM}.
In particular, there are no significant systematics present in the results for the best-fit position of the GCE.

The study presented in this section demonstrates that the analysis of the spectrum and morphology of the GCE applied to real data is probably dominated by the systematic uncertainties associated to the choice of the IEM or in other words by background models that do not represent properly the data.

\section{Point sources vs diffuse emission}
\label{sec:psdiffusive}

\begin{figure*}[t]
\includegraphics[width=0.45\textwidth]{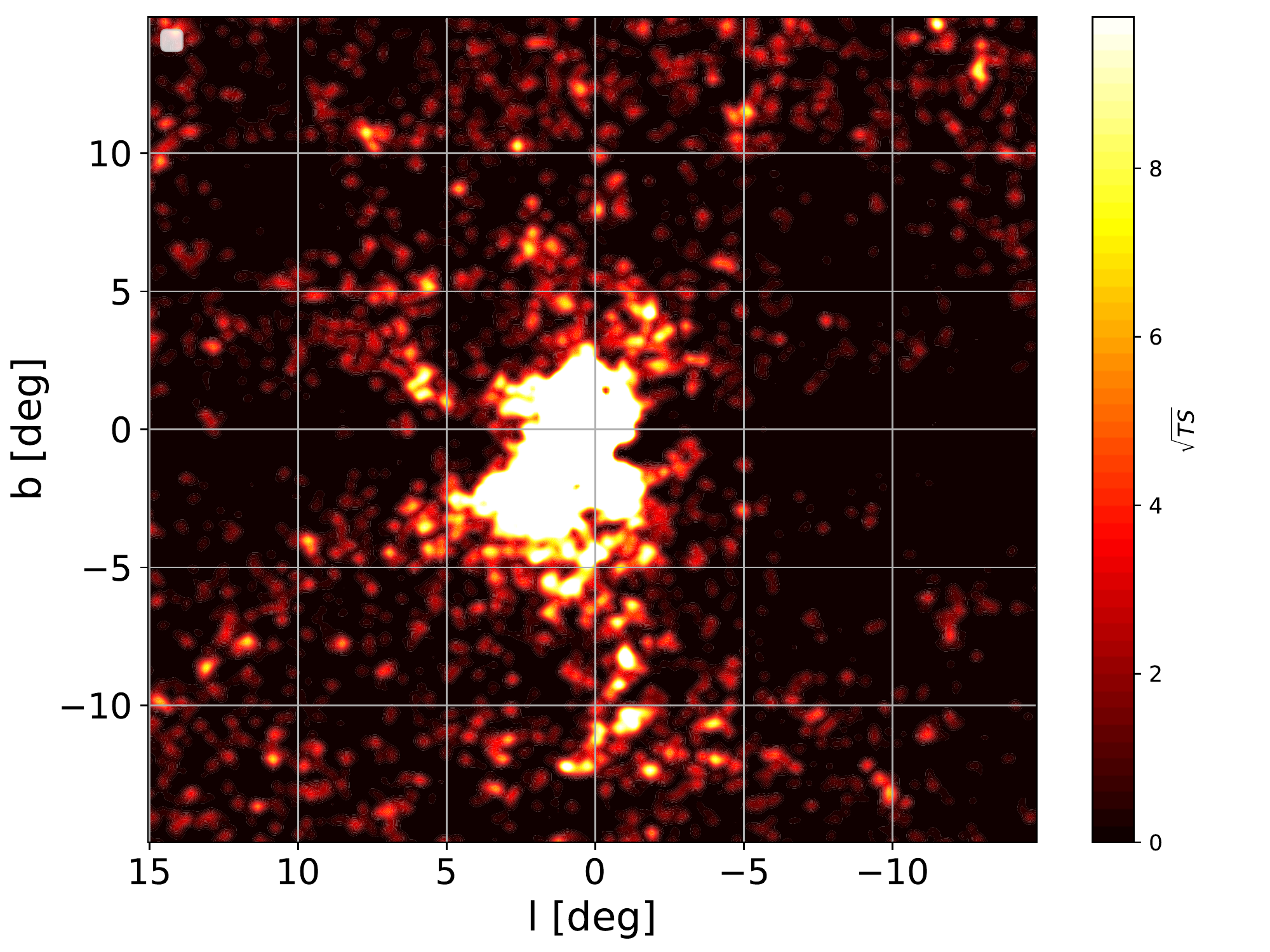}
\includegraphics[width=0.45\textwidth]{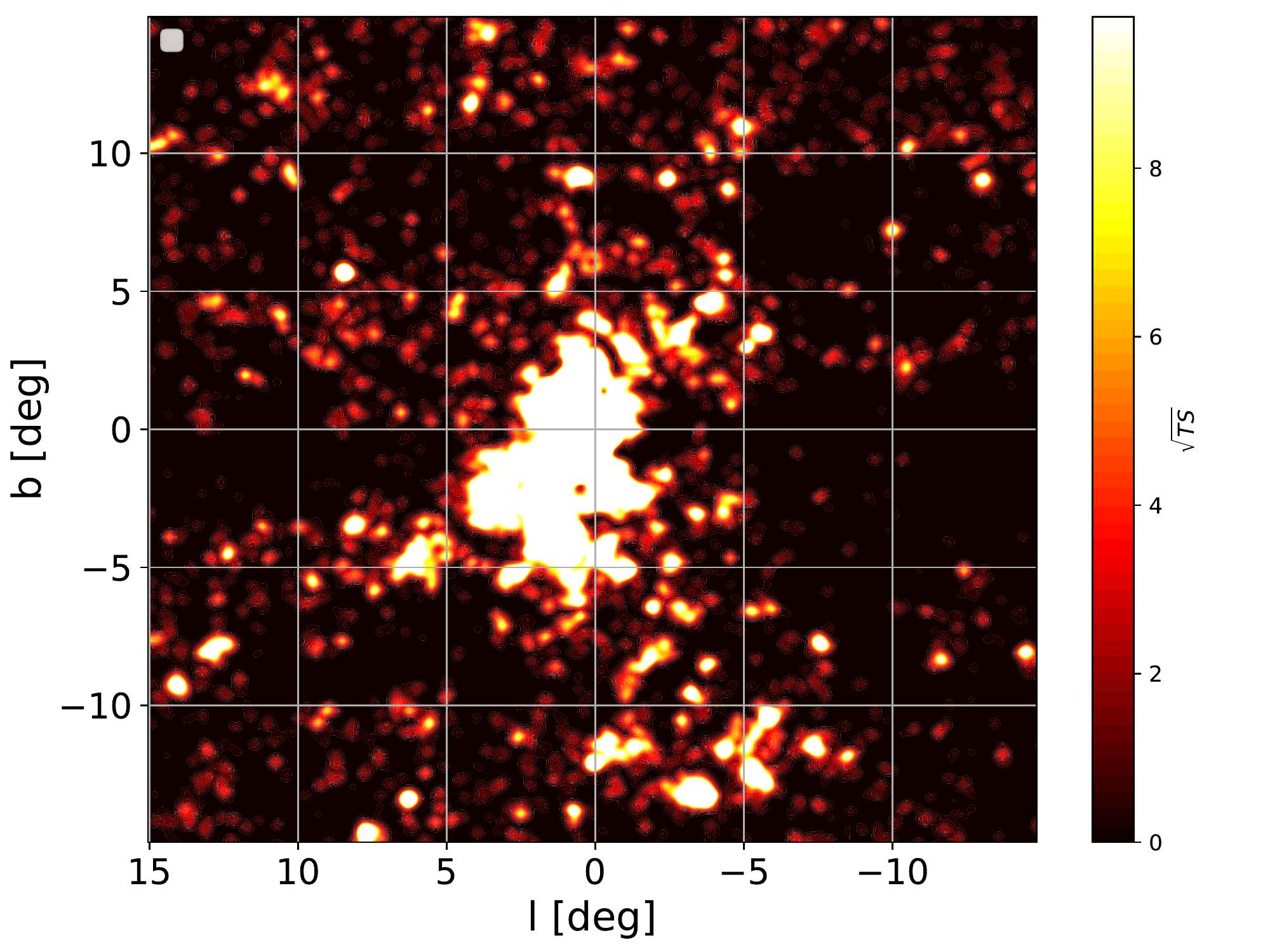}
\includegraphics[width=0.45\textwidth]{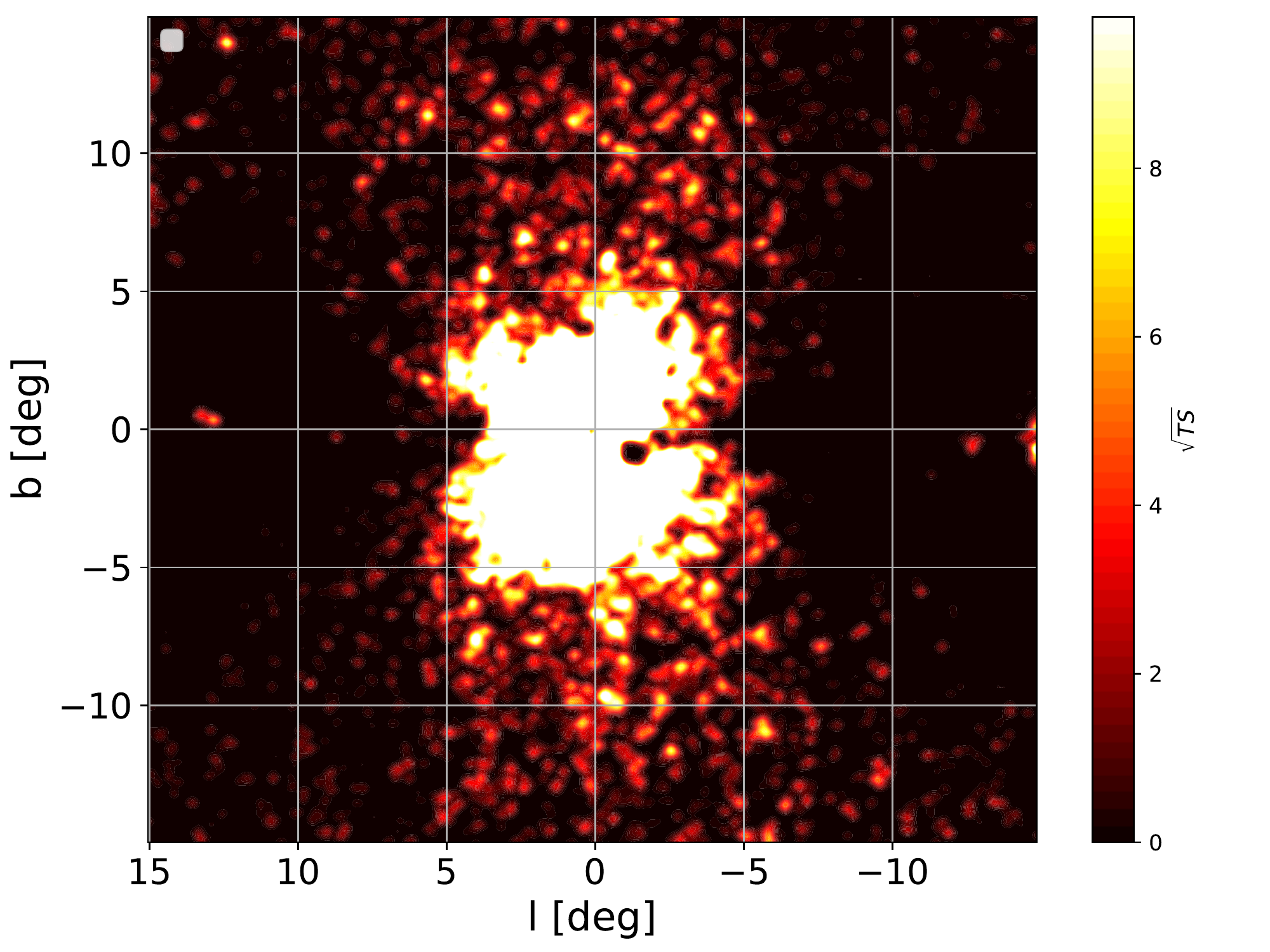}
\includegraphics[width=0.45\textwidth]{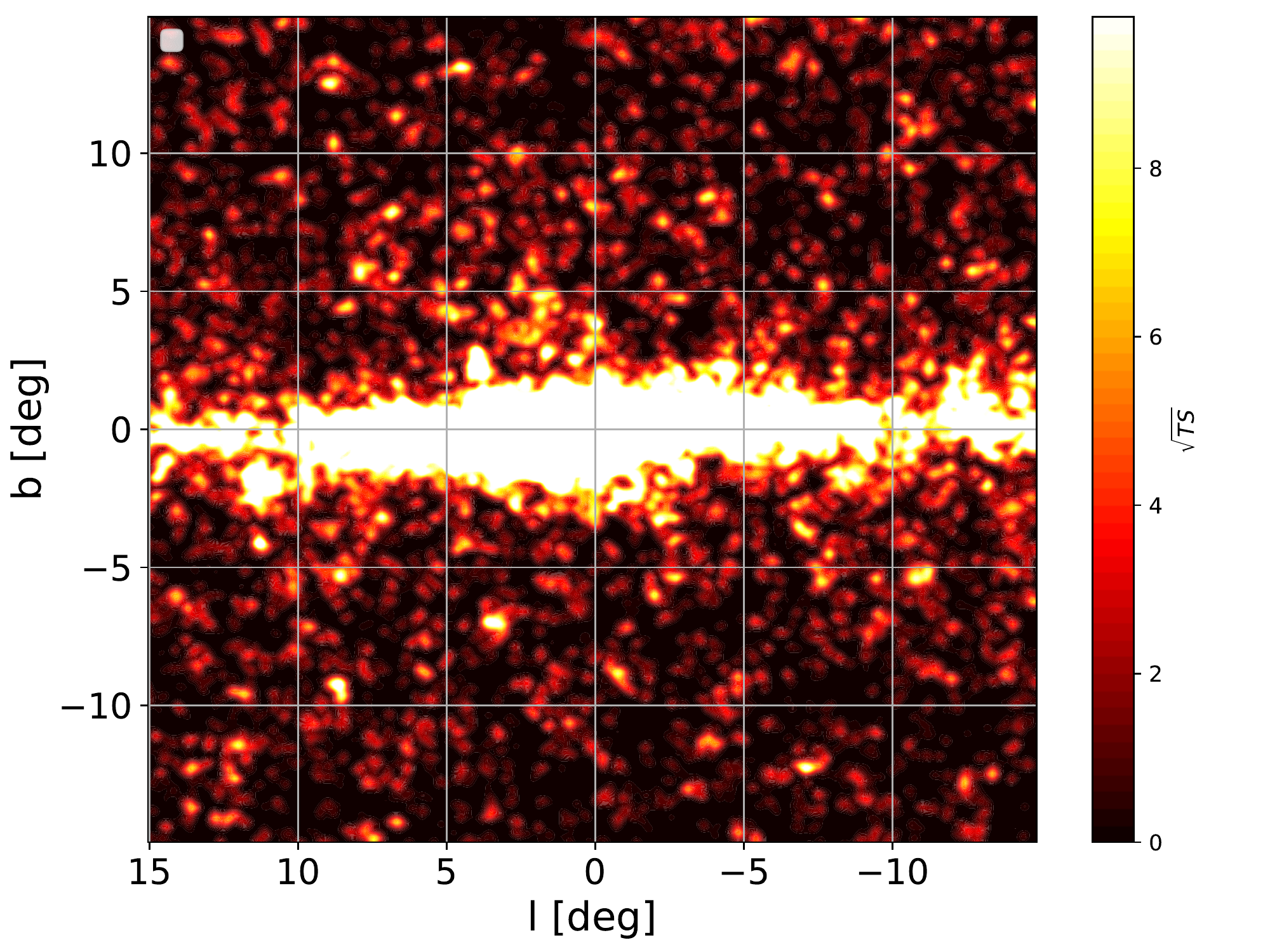}
\caption{$TS$ maps found by simulating the data with a signal from DM (top left), pulsars located in the Galactic bulge (top right), electrons and positrons injected from the Galactic bulge (bottom right) and protons produced in the CMZ (bottom right). The maps have been derived simulating the data with the signal, then eliminating it from the model, optimizing the ROI and finally running the $TS$ map. Therefore, the maps show mainly the injected signal. The color bar shows the significance (which is roughly $\sqrt{TS}$).}  
\label{fig:TSalt}
\end{figure*}

The interpretations for the GCE can be classified into two categories.
Some studies claimed the existence of a faint population of sources with a total flux and spatial morphology compatible with the GCE (see, e.g., \cite{Bartels:2015aea,Lee:2015fea}).
Others theorized an origin due to diffusive processes, such as $\gamma$ rays produced from a population of cosmic-ray electrons and positrons or protons emitted from the Galactic center or from DM particle interactions (see, e.g., \cite{Daylan:2014rsa,Petrovic:2014uda,Carlson:2014cwa,Gaggero:2015nsa}).
In this section we generate simulations where we inject a $\gamma$ ray emission produced by the following mechanisms: a point source population located in the bulge of our Galaxy, a population of electrons and positrons and cosmic-ray protons injected from the Galactic center and a DM signal.
For each of these four cases, we test how the properties of the GCE would look like putting particular emphasis on the spatial distribution of the produced signal.
The $\gamma$-ray emission generated for these processes have been normalized to have a flux at 1 GeV of $7\times 10^{-7}$ MeV$^{-1}$ cm$^{-2}$ s$^{-1}$ in order to be compatible with the GCE flux.

The DM simulation follows exactly the ones used in the previous sections ($\gamma=1.2$).
We build the simulations with point sources in order to mimic a population of pulsars located around the bulge of our Galaxy as studied in \cite{Bartels:2015aea,Lee:2015fea}.
We use a luminosity function given by a power-law $dN/dL \sim L^{-\beta}$ and we assume that $\beta= 1.8$.
We have also tested $\beta$ equal to 1.6 and 2.0 finding very similar results.
We extract from this function, $L$ values between $10^{31}$ erg/s to $10^{36}$ erg/s because this is roughly the range of luminosities found for the Galactic pulsars \cite{TheFermi-LAT:2013ssa}. We use a spatial distribution of the sources taken from a NFW with $\gamma=1.2$. 
Finally, we simulate sources until their cumulative spectrum reaches the intensity of the GCE.

For the cases with an additional population of cosmic-ray electrons and positrons or protons injected from the GCE, we follow closely the methodology reported in \cite{TheFermi-LAT:2017vmf}. Specifically, we use the models labelled as {\tt IC Bulge, bar} and {\tt CR CMZ 3D, z=8kpc}.
%one associated with the central molecular zone (CMZ) in the innermost few hundred pc ({\tt CR CMZ 3D, z=8kpc}).
For the former ({\tt IC Bulge, bar}) we assume a population of sources located in the bulge/bar in the central kpc of the Milky Way which emit electrons taken from the distribution of the old stellar population in the bulge (model B in \cite{2012A&A...538A.106R}).
These electrons and positrons injected by these sources then produce $\gamma$ rays through inverse Compton scattering on the interstellar radiation field present in the center of the Milky Way.
As an alternative, in {\tt CR CMZ 3D, z=8kpc} we consider a population of cosmic rays (mainly protons) produced from the central molecular zone (CMZ) in the innermost few hundred pc from the Galactic center. The CMZ contains very dense molecular clouds which can host intensive star formation and, as a result, a significant rate of supernovae explosions. 
As a tracer of the cosmic-ray production in the CMZ we use the distribution of molecular gas, which we model by a simplified axisymmetric version of Equation (18) in \cite{2007A&A...467..611F}. In this model the $\gamma$-ray emission is due to bremsstrahlung and $\pi^0$ decays associated to HI and H2 gas in the interstellar medium.

\begin{figure*}[t]
\includegraphics[width=0.49\textwidth]{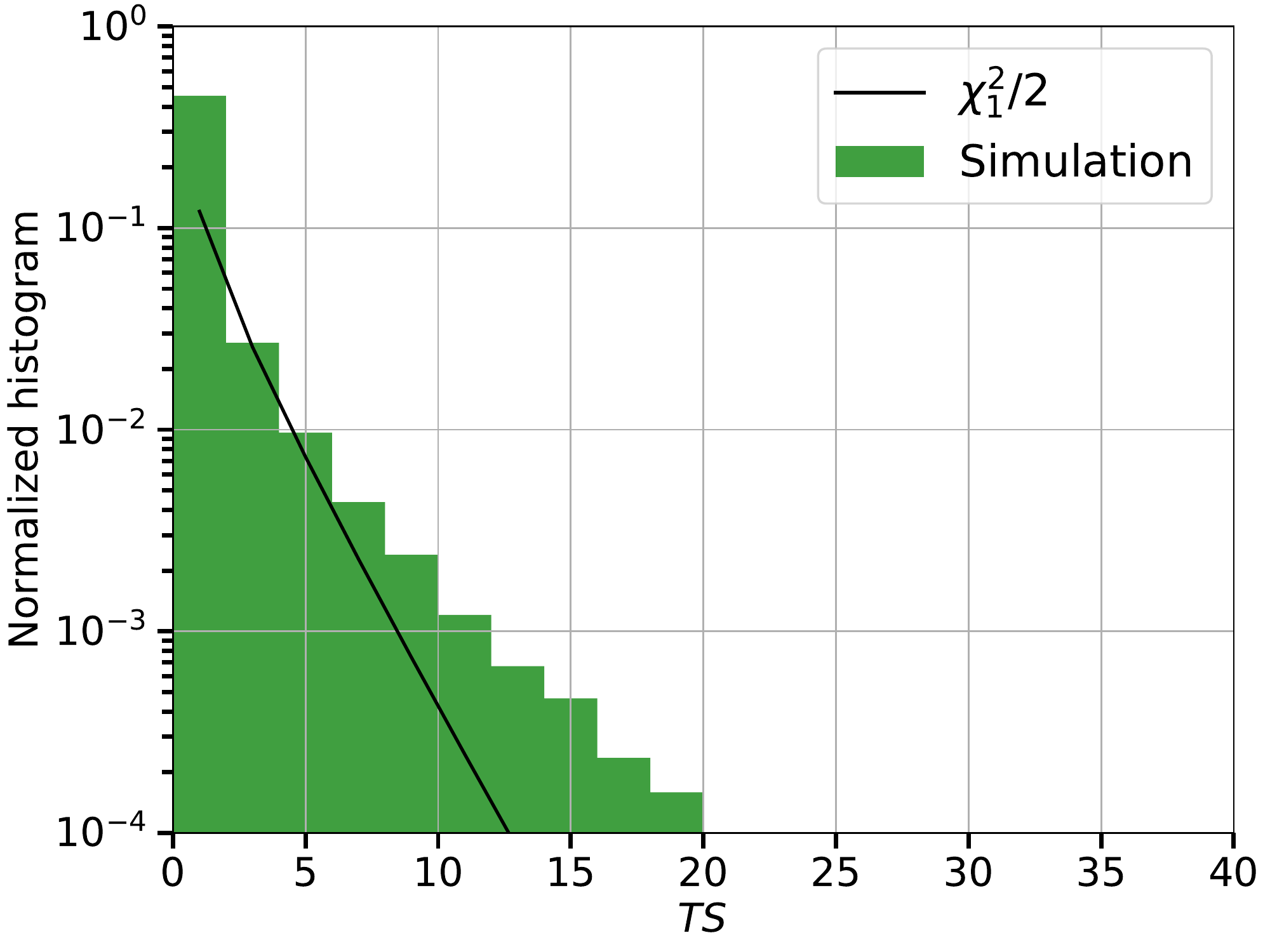}
\includegraphics[width=0.49\textwidth]{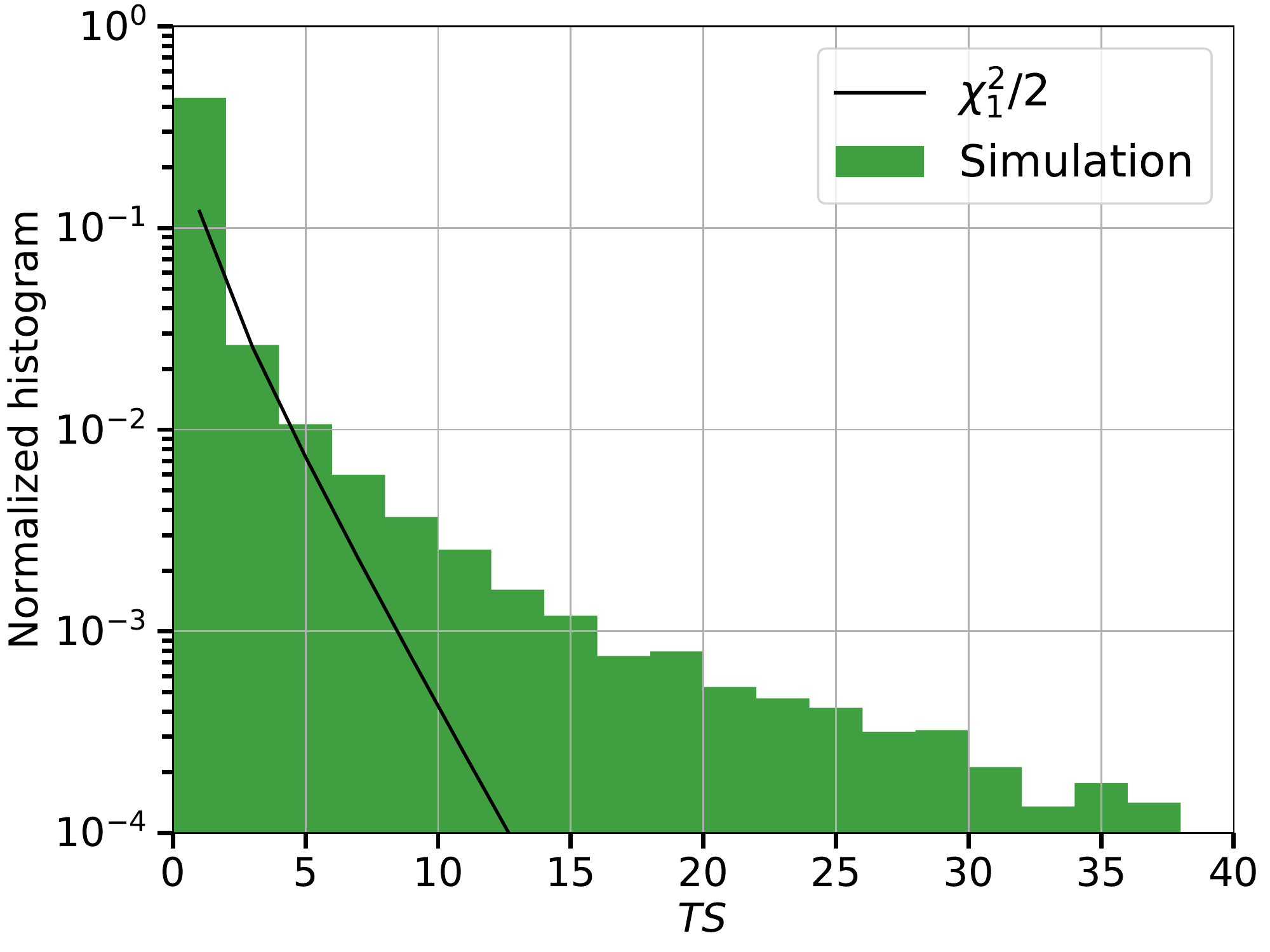}
\caption{Histogram of the $TS$ values found by masking the inner $4^{\circ}$ from the Galactic center for the maps reported in Fig.~\ref{fig:TSalt} for DM (left panel) and pulsars (right panel). Together with the $TS$ histogram we also show the distribution of the $\chi_1^2/2$. The $TS$ and $\chi_1^2/2$ distributions should be compatible in case of a perfect modeling of the data.}  
\label{fig:histo}
\end{figure*}

First, we show how the injected signal, for each of the four mechanisms mentioned above, would appear after running our analysis pipeline.
We simulate the data including the signal template, then we remove it from the model and re-optimize the ROI.
Finally, we generate a $TS$ map that we show, for each of the four cases, in Fig.~\ref{fig:TSalt}. 

As expected the signal generated by the DM and pulsar emission is roughly spherically symmetric. 
Therefore, the spatial distribution alone can not be used to distinguish these two cases.
The main difference between a pulsar and DM signal is that the former has a $TS$ map that contains more isolated peaks of high values located even a few degrees away from the Galactic center. Instead, the latter would produce a more diffusive $TS$ map and with lower values beyond a few degree from the center.
In the inner few degrees from the center the two cases are very similar because the cumulative emission of thousands of pulsars in the inner few degrees from the Galactic center generates a $\gamma$-ray signal that can not be distinguished from a diffusive emission like the one produced by DM. 
A possible method to disentangle these two mechanisms is the following. 
We mask the inner four degrees from the Galactic center and we calculate a histogram of the $TS$. This is shown in Fig.~\ref{fig:histo}.
As explained above the case with pulsars exhibits a more prominent high $TS$ tail that is due to the brightest pulsars of the Galactic bulge sample.   
In applying this method to the real data we would encounter a few challenges. 
First, if an optimization of the ROI is performed and this includes the search of new sources with a $TS>25$, all the bright spots present in the $TS$ histogram would disappear.
Indeed, all the pulsars which populate the histogram at $TS>25$ would be included in the model as new sources. 
Therefore, the difference between the DM and pulsar cases would become less evident.
The additional difficulty is due to the presence of sub-threshold Galactic and Extragalactic sources present in the data which would add an additional component of contamination for determining the presence of a population of sources in the Galactic center and associated to the Galactic bulge.

%Now, we discuss the case of cosmic-ray injected from the center of the Milky Way.
In case of electrons and positrons, the emission is roughly spherically symmetric because the process which generates the $\gamma$ rays is inverse Compton scattering on the interstellar radiation field which is expected to be roughly symmetric around the center of the Galaxy.
On the other hand the $\gamma$ ray emission from cosmic-ray protons follows the disk because the main process for the production of $\gamma$ rays is due to the interactions of the protons with the gas components and the production and subsequent decays of $\pi^0$ particles. 
Therefore, these two processes would be easily distinguished looking to the spherical symmetry or elongation on the Galactic plane of the signal.

\begin{figure*}[t]
\includegraphics[width=0.49\textwidth]{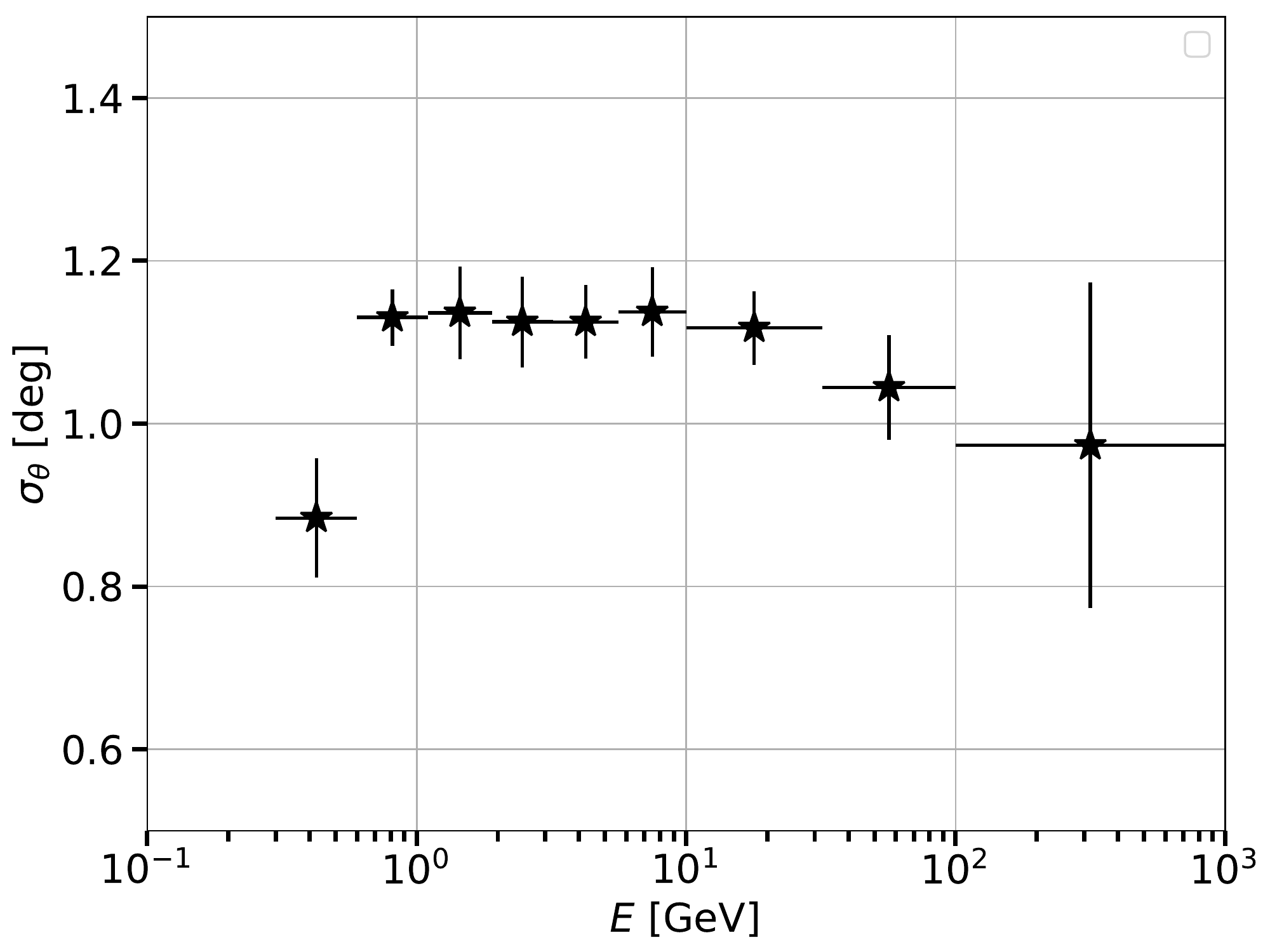}
\includegraphics[width=0.49\textwidth]{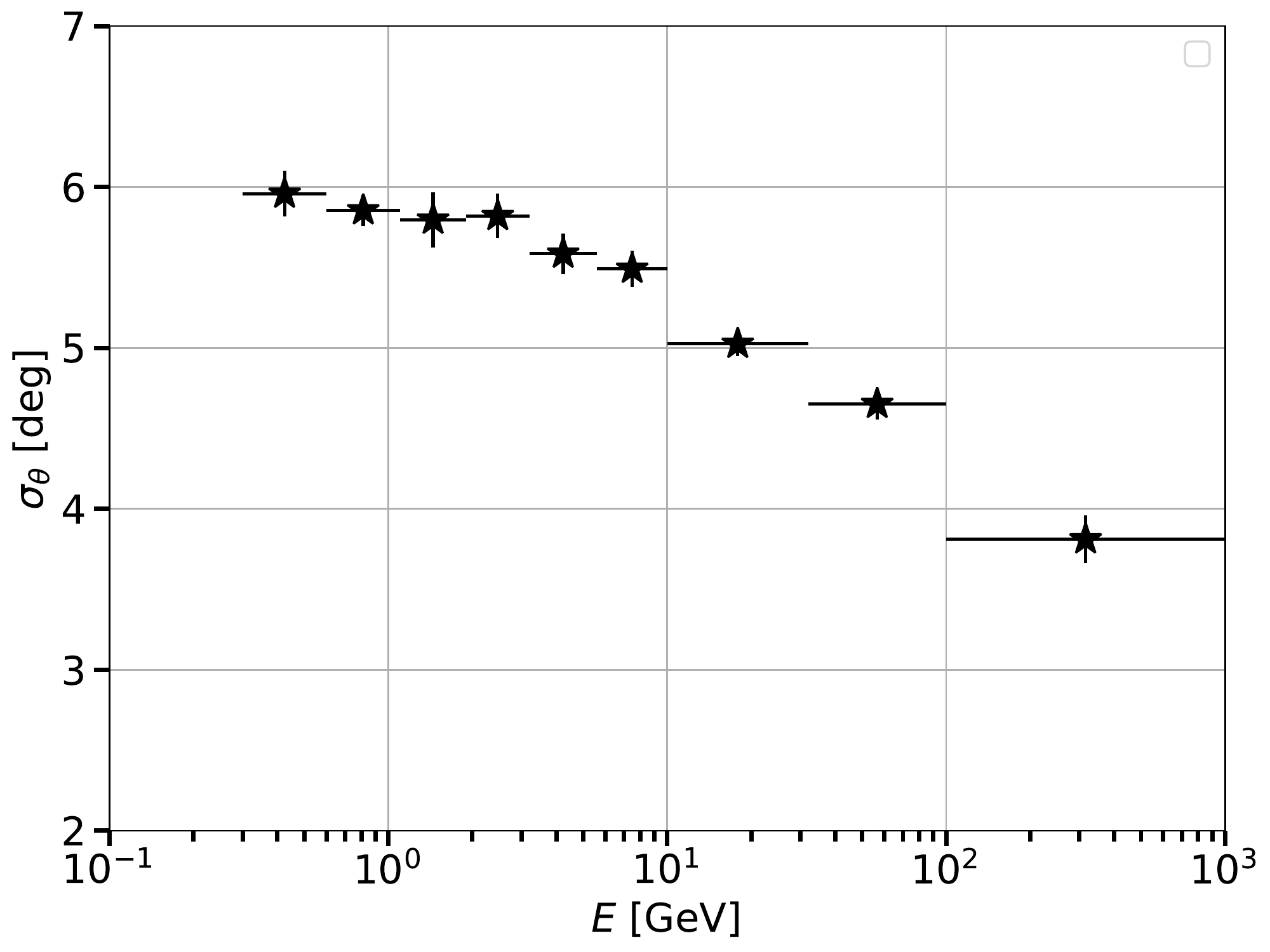}
\caption{Best-fit and errors for the parameter $\sigma_{\theta}$ (see Eq.~\ref{eq:SBfunc}) found by fitting the surface brightness data of the signal obtained with the simulations with cosmic-ray protons produced from the CMZ (left) and electrons and positrons injected in the Galactic bulge (right). See the main text for further details about these models.}  
\label{fig:extalt}
\end{figure*}

Since the nature of the Physical processes which produce $\gamma$ rays from electrons and positrons and cosmic-ray protons are different (inverse Compton for the former and $\pi^0$ decays for the latter) we expect a different evolution of the GCE spatial distribution with energy.
Electrons and positrons are affected by strong energy losses and diffusion. These two processes are highly energy dependent. 
In particular high-energy electrons and positrons have a larger propagation length but they loose energy very quickly.
Therefore, we expect that the spatial distribution of the GCE in case it is given by electrons and positrons should be smaller at high energy with respect to lower energy.
On the other hand, the $\gamma$-ray emission for $\pi^0$ decays is not expected to produce any significant change of shape as a function of energy. Indeed, in this case the spatial morphology is solely due to the distribution of gas in the Galaxy and this should not depend on energy. 

In order to verify this hypothesis we perform the analysis of the spatial distribution of the GCE using the annulus analysis.
We run the analysis in different energy bins between 0.1 and 1000 GeV and we use annuli with size of $1^{\circ}$.
Then, we calculate the surface brightness and fit the data with the following empirical function \cite{DiMauro:2019hwn}:
\begin{equation}
\frac{d\Phi_{\gamma}}{d\theta} \sim \frac{1}{\sigma_{\theta} (\theta+0.06 \cdot \sigma_{\theta})} e^{  -  \left( \frac{\theta}{\sigma_{\theta}} \right)^2},
 \label{eq:SBfunc}
\end{equation}
where $\sigma_{\theta}/2$ is the angle that contains the $80\%$ of the observed flux.
We find that this function fits better the surface brightness data with respect to the DM emission (Eq.~\ref{eq:flux}).
We have tested also a fit with the DM signal and found similar results.
We show the best-fit values for $\sigma_{\theta}$ in Fig.~\ref{fig:extalt} for the two cases.
As expected the spatial distribution in case of electrons and positrons is highly energy dependent.
In particular the size is much smaller at higher energy. We find a difference of almost a factor of 2 for the value of $\sigma_{\theta}$ between 1 GeV and 100 GeV.
On the other hand, in case of cosmic-ray protons the value of $\sigma_{\theta}$ is roughly constant with energy.
Moreover, this parameter takes much smaller values for the proton case with respect to the electron and positron scenario because the $\gamma$-ray emission produce by the former is not spherically symmetric and the annulus analysis mainly accounts for the emission in the inner few degrees from the Galactic center.
The results for $\sigma_{\theta}$ are unstable below 0.3 GeV so we decide to not show them in Fig.~\ref{fig:extalt} (see discussion in Sec.~\ref{sec:spatialDM}).

In the results presented in this section we have assumed a perfect knowledge of the IEM components. In Secs~\ref{sec:missingcomp} and \ref{sec:wrongmodel} we have demonstrated that the presence of a missing component in the IEM (we tested the absence of the low-latitude bubble flux) or the mismodeling of the IEM could produce systematic uncertainties in the reconstructed GCE spectrum and spatial morphology of the order of $10-15\%$. In fact, an imperfect modeling of the IEM could leave significant residuals of the order of $5-7\sigma$ significance (see Figs.~\ref{fig:TSsysbubbles} and \ref{fig:TSsysIEM}).
The characteristics of the GCE presented in this section could thus be slightly modified by the imperfections in the IEM model but the bulk of the signal morphologies and properties remain unchanged. Indeed, the significance of the GCE signal in case of DM, pulsars or cosmic rays in the $TS$ map is at levels much higher than $10\sigma$ significance (see Fig.~\ref{fig:TSalt}).

%%%%%%%%%%%%%%%%%%%%%%%%%%%%%%%%%%%%%%%%%%%%%%%%%%%%%%%%%%%%
\section{Conclusions}
\label{sec: conclusions}
%%%%%%%%%%%%%%%%%%%%%%%%%%%%%%%%%%%%%%%%%%%%%%%%%%%%%%%%%%%%

\begin{table*}
\begin{center}
\begin{tabular}{|c|c|c|c|}
\hline
CASE  &  Spherically symmetric & Energy dependent Morphology &  Clumpy $TS$ map  \\  
\hline
\hline
DM  &  $\mathcal{V}$ & $\mathcal{X}$ &  $\mathcal{X}$  \\  
\hline
PSR  &  $\mathcal{V}$ & $\mathcal{X}$ &  $\mathcal{V}$  \\  
\hline
$e^{\pm}$  &  $\mathcal{V}$ & $\mathcal{V}$ &  $\mathcal{X}$  \\  
\hline
CR p  &  $\mathcal{X}$ & $\mathcal{X}$ &  $\mathcal{X}$  \\  
\hline
\hline
\end{tabular}
\caption{This table summarizes the properties of the GCE if it is due to the following processes: DM, pulsars located in the Galactic bulge (PSR), cosmic-ray electrons and positrons produced from the bulge ($e^{\pm}$) or protons injected from the CMZ (CR p). We report in the table if the GCE signal produced by each of this processes would exhibit a spherically symmetric signal, an energy dependent spatial morphology or would leave a clumpy $TS$ map due to undetected point sources.}
\label{tab:prop}
\end{center}
\end{table*}

In this paper, we have investigated the detectability of the GCE using simulated {\it Fermi}-LAT data of the Galactic center region. 
In particular we have estimated the systematic uncertainties obtained in the analysis of the GCE properties due to mismodeling of the IEM model.
We have also presented model independent techniques useful to study the origin of this excess.
The results presented are particularly important considering the current debate on the origin of the GCE.

First, we have considered the ideal case where we have a perfect knowledge of the sources and IEM components that contribute to the data.
In this case we are able to reconstruct properly the energy spectrum of the GCE whether it is modeled with a unique template or it is divided into quadrants.
We also verified that we would be able to determine the GCE spatial morphology with a model independent analysis that takes into account concentric and uniform annuli fitted individually to the data.
In particular we demonstrated that by using annuli of size between $0.75^{\circ}-1.5^{\circ}$ we find the correct GCE spatial distribution at $E>0.6$ GeV.

Then, we considered the more realist scenario for which we do not have a perfect knowledge of the background components.
On this regard, we first assume we have a missing component in the model. We simulated the data with the low-latitude component of the {\it Fermi} bubbles but we did not include it in the model used to fit the data.
We find that if the flux of this missing component is roughly larger than $10^{-7}$ GeV$^{-1}$ cm$^2$ s, i.e. about $10\%-20\%$ of the GCE flux, the results for the energy spectrum and spatial morphology starts to deviate significantly from the injected signal. This implies that any $\gamma$-ray emission present in the Galactic center region, not accounted properly in the model and with a flux larger than $\sim 10^{-7}$ GeV$^{-1}$ cm$^2$ would affect significantly the results on the GCE properties. This flux is roughly a few per mille of the data measured from the Galactic center.
We also test the case where we simulated the data with the {\tt Baseline} IEM and we analyze them with other IEM models. This exercise probes the systematics on the GCE properties due to a mismodeling of the IEM model. The scatter on the energy spectrum and value of $\gamma$, i.e., the GCE spatial morphology, is $10-15\%$ and $\sim 5\%$, respectively.
These systematics are much larger than the statistical errors. Therefore, the analysis of the GCE properties in the real data are systematics dominated by the imperfections in the background model used to fit the data.

In the last part we tested four different interpretations for the GCE: a DM signal, flux from pulsars located in the Galactic bulge, $\gamma$ rays produced by electrons and positrons injected from the Galactic bulge and protons from the CMZ.
For each of these cases we simulate a signal with a flux compatible with the GCE and we inspect the properties of the produced $\gamma$ rays.
The emission from pulsars and DM is expected to be spherical symmetric. Moreover, they both would not produce any energy dependence of the spatial distribution of the GCE.
However, the former would be disentangled from the latter because the $TS$ of the signal would be clumpy. Moreover, the $TS$ histogram would exhibit a prominent tail for values $TS>16$ which is due to the brightest pulsars in the sample.
On the other hand, the signal produced from protons injected from the CMZ is the only one which would not be spherically symmetric because it would rather follow the disk of the Galaxy.
Finally, the emission produced by electrons and positrons generated from the Galactic bulge would produce a signal which is roughly spherically symmetric and it would be the only one that produces a GCE signal with a spatial extension that changes with energy, specifically smaller at higher energies.

Our analysis demonstrates that each of the processes considered above has distinctive properties that could be used to identify the origin of the GCE.
We summarize them in Tab.~\ref{tab:prop}.
We refer to a companion paper the analysis of the GCE with real data applying the analysis techniques developed in this paper.

% \begin{equation}
% Q(E, t)= L(t) \left( \frac{E}{E_0}\right)^{- \gamma_e} \exp \left(-\frac{E}{E_c} \right) ,
% \label{eq:Q_E_cont}
%\end{equation}
%\begin{figure}[t]
%\includegraphics[width=0.1\textwidth]{figures/lorimer_spiral_rings_rate1percentury.png}
%\caption{Spatial distribution of simulated pulsars in one illustrative realization. The Earth is at the center of the plot, while the Galactic center is at $d=8.5$~kpc, $l=0$. 
%The gray points indicates the position of each simulated pulsar. The concentric rings illustrate the distance rings we use to separate the contribution of simulated sources at different distances, see text for details.}%, with consistent color code as the right panel.}  
%\label{fig:ringssim}
%\end{figure}
%%%%%%%%%%%%%

\begin{acknowledgments}
The author thanks the {\it Fermi}-LAT Collaboration for insightful discussions.
MDM acknowledges support by the NASA Fermi Guest Investigator Program Cycle 12 through the Fermi Program N. 121119 (P.I.~MDM). 

\end{acknowledgments}

\bibliography{paper}

\begin{thebibliography}{43}
\expandafter\ifx\csname natexlab\endcsname\relax\def\natexlab#1{#1}\fi
\expandafter\ifx\csname bibnamefont\endcsname\relax
  \def\bibnamefont#1{#1}\fi
\expandafter\ifx\csname bibfnamefont\endcsname\relax
  \def\bibfnamefont#1{#1}\fi
\expandafter\ifx\csname citenamefont\endcsname\relax
  \def\citenamefont#1{#1}\fi
\expandafter\ifx\csname url\endcsname\relax
  \def\url#1{\texttt{#1}}\fi
\expandafter\ifx\csname urlprefix\endcsname\relax\def\urlprefix{URL }\fi
\providecommand{\bibinfo}[2]{#2}
\providecommand{\eprint}[2][]{\url{#2}}

\bibitem[{\citenamefont{Pieri et~al.}(2011)\citenamefont{Pieri, Lavalle,
  Bertone, and Branchini}}]{Pieri:2009je}
\bibinfo{author}{\bibfnamefont{L.}~\bibnamefont{Pieri}},
  \bibinfo{author}{\bibfnamefont{J.}~\bibnamefont{Lavalle}},
  \bibinfo{author}{\bibfnamefont{G.}~\bibnamefont{Bertone}}, \bibnamefont{and}
  \bibinfo{author}{\bibfnamefont{E.}~\bibnamefont{Branchini}},
  \bibinfo{journal}{Phys. Rev.} \textbf{\bibinfo{volume}{D83}},
  \bibinfo{pages}{023518} (\bibinfo{year}{2011}), \eprint{0908.0195}.

\bibitem[{\citenamefont{Goodenough and Hooper}(2009)}]{Goodenough:2009gk}
\bibinfo{author}{\bibfnamefont{L.}~\bibnamefont{Goodenough}} \bibnamefont{and}
  \bibinfo{author}{\bibfnamefont{D.}~\bibnamefont{Hooper}}
  (\bibinfo{year}{2009}), \eprint{0910.2998}.

\bibitem[{\citenamefont{Hooper and Goodenough}(2011)}]{Hooper:2010mq}
\bibinfo{author}{\bibfnamefont{D.}~\bibnamefont{Hooper}} \bibnamefont{and}
  \bibinfo{author}{\bibfnamefont{L.}~\bibnamefont{Goodenough}},
  \bibinfo{journal}{Phys. Lett.} \textbf{\bibinfo{volume}{B697}},
  \bibinfo{pages}{412} (\bibinfo{year}{2011}), \eprint{1010.2752}.

\bibitem[{\citenamefont{Boyarsky et~al.}(2011)\citenamefont{Boyarsky, Malyshev,
  and Ruchayskiy}}]{Boyarsky:2010dr}
\bibinfo{author}{\bibfnamefont{A.}~\bibnamefont{Boyarsky}},
  \bibinfo{author}{\bibfnamefont{D.}~\bibnamefont{Malyshev}}, \bibnamefont{and}
  \bibinfo{author}{\bibfnamefont{O.}~\bibnamefont{Ruchayskiy}},
  \bibinfo{journal}{Phys. Lett.} \textbf{\bibinfo{volume}{B705}},
  \bibinfo{pages}{165} (\bibinfo{year}{2011}), \eprint{1012.5839}.

\bibitem[{\citenamefont{Hooper and Linden}(2011)}]{Hooper:2011ti}
\bibinfo{author}{\bibfnamefont{D.}~\bibnamefont{Hooper}} \bibnamefont{and}
  \bibinfo{author}{\bibfnamefont{T.}~\bibnamefont{Linden}},
  \bibinfo{journal}{Phys. Rev.} \textbf{\bibinfo{volume}{D84}},
  \bibinfo{pages}{123005} (\bibinfo{year}{2011}), \eprint{1110.0006}.

\bibitem[{\citenamefont{Abazajian and Kaplinghat}(2012)}]{Abazajian:2012pn}
\bibinfo{author}{\bibfnamefont{K.~N.} \bibnamefont{Abazajian}}
  \bibnamefont{and}
  \bibinfo{author}{\bibfnamefont{M.}~\bibnamefont{Kaplinghat}},
  \bibinfo{journal}{Phys. Rev.} \textbf{\bibinfo{volume}{D86}},
  \bibinfo{pages}{083511} (\bibinfo{year}{2012}), \bibinfo{note}{[Erratum:
  Phys. Rev.D87,129902(2013)]}, \eprint{1207.6047}.

\bibitem[{\citenamefont{Gordon and Macias}(2013)}]{Gordon:2013vta}
\bibinfo{author}{\bibfnamefont{C.}~\bibnamefont{Gordon}} \bibnamefont{and}
  \bibinfo{author}{\bibfnamefont{O.}~\bibnamefont{Macias}},
  \bibinfo{journal}{Phys. Rev.} \textbf{\bibinfo{volume}{D88}},
  \bibinfo{pages}{083521} (\bibinfo{year}{2013}), \bibinfo{note}{[Erratum:
  Phys. Rev.D89,no.4,049901(2014)]}, \eprint{1306.5725}.

\bibitem[{\citenamefont{Abazajian et~al.}(2014)\citenamefont{Abazajian, Canac,
  Horiuchi, and Kaplinghat}}]{Abazajian:2014fta}
\bibinfo{author}{\bibfnamefont{K.~N.} \bibnamefont{Abazajian}},
  \bibinfo{author}{\bibfnamefont{N.}~\bibnamefont{Canac}},
  \bibinfo{author}{\bibfnamefont{S.}~\bibnamefont{Horiuchi}}, \bibnamefont{and}
  \bibinfo{author}{\bibfnamefont{M.}~\bibnamefont{Kaplinghat}},
  \bibinfo{journal}{Phys. Rev.} \textbf{\bibinfo{volume}{D90}},
  \bibinfo{pages}{023526} (\bibinfo{year}{2014}), \eprint{1402.4090}.

\bibitem[{\citenamefont{Daylan et~al.}(2016)\citenamefont{Daylan, Finkbeiner,
  Hooper, Linden, Portillo, Rodd, and Slatyer}}]{Daylan:2014rsa}
\bibinfo{author}{\bibfnamefont{T.}~\bibnamefont{Daylan}},
  \bibinfo{author}{\bibfnamefont{D.~P.} \bibnamefont{Finkbeiner}},
  \bibinfo{author}{\bibfnamefont{D.}~\bibnamefont{Hooper}},
  \bibinfo{author}{\bibfnamefont{T.}~\bibnamefont{Linden}},
  \bibinfo{author}{\bibfnamefont{S.~K.~N.} \bibnamefont{Portillo}},
  \bibinfo{author}{\bibfnamefont{N.~L.} \bibnamefont{Rodd}}, \bibnamefont{and}
  \bibinfo{author}{\bibfnamefont{T.~R.} \bibnamefont{Slatyer}},
  \bibinfo{journal}{Phys. Dark Univ.} \textbf{\bibinfo{volume}{12}},
  \bibinfo{pages}{1} (\bibinfo{year}{2016}), \eprint{1402.6703}.

\bibitem[{\citenamefont{Calore et~al.}(2015{\natexlab{a}})\citenamefont{Calore,
  Cholis, McCabe, and Weniger}}]{Calore:2014nla}
\bibinfo{author}{\bibfnamefont{F.}~\bibnamefont{Calore}},
  \bibinfo{author}{\bibfnamefont{I.}~\bibnamefont{Cholis}},
  \bibinfo{author}{\bibfnamefont{C.}~\bibnamefont{McCabe}}, \bibnamefont{and}
  \bibinfo{author}{\bibfnamefont{C.}~\bibnamefont{Weniger}},
  \bibinfo{journal}{Phys. Rev.} \textbf{\bibinfo{volume}{D91}},
  \bibinfo{pages}{063003} (\bibinfo{year}{2015}{\natexlab{a}}),
  \eprint{1411.4647}.

\bibitem[{\citenamefont{Calore et~al.}(2015{\natexlab{b}})\citenamefont{Calore,
  Cholis, and Weniger}}]{Calore:2014xka}
\bibinfo{author}{\bibfnamefont{F.}~\bibnamefont{Calore}},
  \bibinfo{author}{\bibfnamefont{I.}~\bibnamefont{Cholis}}, \bibnamefont{and}
  \bibinfo{author}{\bibfnamefont{C.}~\bibnamefont{Weniger}},
  \bibinfo{journal}{JCAP} \textbf{\bibinfo{volume}{1503}}, \bibinfo{pages}{038}
  (\bibinfo{year}{2015}{\natexlab{b}}), \eprint{1409.0042}.

\bibitem[{\citenamefont{Ajello et~al.}(2016)}]{TheFermi-LAT:2015kwa}
\bibinfo{author}{\bibfnamefont{M.}~\bibnamefont{Ajello}} \bibnamefont{et~al.}
  (\bibinfo{collaboration}{Fermi-LAT}), \bibinfo{journal}{Astrophys. J.}
  \textbf{\bibinfo{volume}{819}}, \bibinfo{pages}{44} (\bibinfo{year}{2016}),
  \eprint{1511.02938}.

\bibitem[{\citenamefont{Ackermann et~al.}(2017)}]{TheFermi-LAT:2017vmf}
\bibinfo{author}{\bibfnamefont{M.}~\bibnamefont{Ackermann}}
  \bibnamefont{et~al.} (\bibinfo{collaboration}{Fermi-LAT}),
  \bibinfo{journal}{Astrophys. J.} \textbf{\bibinfo{volume}{840}},
  \bibinfo{pages}{43} (\bibinfo{year}{2017}), \eprint{1704.03910}.

\bibitem[{\citenamefont{Aghanim et~al.}(2018)}]{Aghanim:2018eyx}
\bibinfo{author}{\bibfnamefont{N.}~\bibnamefont{Aghanim}} \bibnamefont{et~al.}
  (\bibinfo{collaboration}{Planck}) (\bibinfo{year}{2018}),
  \eprint{1807.06209}.

\bibitem[{\citenamefont{Bartels et~al.}(2016)\citenamefont{Bartels,
  Krishnamurthy, and Weniger}}]{Bartels:2015aea}
\bibinfo{author}{\bibfnamefont{R.}~\bibnamefont{Bartels}},
  \bibinfo{author}{\bibfnamefont{S.}~\bibnamefont{Krishnamurthy}},
  \bibnamefont{and} \bibinfo{author}{\bibfnamefont{C.}~\bibnamefont{Weniger}},
  \bibinfo{journal}{Phys. Rev. Lett.} \textbf{\bibinfo{volume}{116}},
  \bibinfo{pages}{051102} (\bibinfo{year}{2016}), \eprint{1506.05104}.

\bibitem[{\citenamefont{Lee et~al.}(2016)\citenamefont{Lee, Lisanti, Safdi,
  Slatyer, and Xue}}]{Lee:2015fea}
\bibinfo{author}{\bibfnamefont{S.~K.} \bibnamefont{Lee}},
  \bibinfo{author}{\bibfnamefont{M.}~\bibnamefont{Lisanti}},
  \bibinfo{author}{\bibfnamefont{B.~R.} \bibnamefont{Safdi}},
  \bibinfo{author}{\bibfnamefont{T.~R.} \bibnamefont{Slatyer}},
  \bibnamefont{and} \bibinfo{author}{\bibfnamefont{W.}~\bibnamefont{Xue}},
  \bibinfo{journal}{Phys. Rev. Lett.} \textbf{\bibinfo{volume}{116}},
  \bibinfo{pages}{051103} (\bibinfo{year}{2016}), \eprint{1506.05124}.

\bibitem[{\citenamefont{Leane and Slatyer}(2019)}]{Leane:2019uhc}
\bibinfo{author}{\bibfnamefont{R.~K.} \bibnamefont{Leane}} \bibnamefont{and}
  \bibinfo{author}{\bibfnamefont{T.~R.} \bibnamefont{Slatyer}},
  \bibinfo{journal}{Phys. Rev. Lett.} \textbf{\bibinfo{volume}{123}},
  \bibinfo{pages}{241101} (\bibinfo{year}{2019}), \eprint{1904.08430}.

\bibitem[{\citenamefont{Chang et~al.}(2019)\citenamefont{Chang, Mishra-Sharma,
  Lisanti, Buschmann, Rodd, and Safdi}}]{Chang:2019ars}
\bibinfo{author}{\bibfnamefont{L.~J.} \bibnamefont{Chang}},
  \bibinfo{author}{\bibfnamefont{S.}~\bibnamefont{Mishra-Sharma}},
  \bibinfo{author}{\bibfnamefont{M.}~\bibnamefont{Lisanti}},
  \bibinfo{author}{\bibfnamefont{M.}~\bibnamefont{Buschmann}},
  \bibinfo{author}{\bibfnamefont{N.~L.} \bibnamefont{Rodd}}, \bibnamefont{and}
  \bibinfo{author}{\bibfnamefont{B.~R.} \bibnamefont{Safdi}}
  (\bibinfo{year}{2019}), \eprint{1908.10874}.

\bibitem[{\citenamefont{Zhong et~al.}(2019)\citenamefont{Zhong, McDermott,
  Cholis, and Fox}}]{Zhong:2019ycb}
\bibinfo{author}{\bibfnamefont{Y.-M.} \bibnamefont{Zhong}},
  \bibinfo{author}{\bibfnamefont{S.~D.} \bibnamefont{McDermott}},
  \bibinfo{author}{\bibfnamefont{I.}~\bibnamefont{Cholis}}, \bibnamefont{and}
  \bibinfo{author}{\bibfnamefont{P.~J.} \bibnamefont{Fox}}
  (\bibinfo{year}{2019}), \eprint{1911.12369}.

\bibitem[{\citenamefont{Carlson and Profumo}(2014)}]{Carlson:2014cwa}
\bibinfo{author}{\bibfnamefont{E.}~\bibnamefont{Carlson}} \bibnamefont{and}
  \bibinfo{author}{\bibfnamefont{S.}~\bibnamefont{Profumo}},
  \bibinfo{journal}{\prd} \textbf{\bibinfo{volume}{90}},
  \bibinfo{pages}{023015} (\bibinfo{year}{2014}), \eprint{1405.7685}.

\bibitem[{\citenamefont{Gaggero et~al.}(2015)\citenamefont{Gaggero, Taoso,
  Urbano, Valli, and Ullio}}]{Gaggero:2015nsa}
\bibinfo{author}{\bibfnamefont{D.}~\bibnamefont{Gaggero}},
  \bibinfo{author}{\bibfnamefont{M.}~\bibnamefont{Taoso}},
  \bibinfo{author}{\bibfnamefont{A.}~\bibnamefont{Urbano}},
  \bibinfo{author}{\bibfnamefont{M.}~\bibnamefont{Valli}}, \bibnamefont{and}
  \bibinfo{author}{\bibfnamefont{P.}~\bibnamefont{Ullio}},
  \bibinfo{journal}{JCAP} \textbf{\bibinfo{volume}{1512}}, \bibinfo{pages}{056}
  (\bibinfo{year}{2015}), \eprint{1507.06129}.

\bibitem[{\citenamefont{Petrovi{\'c} et~al.}(2014)\citenamefont{Petrovi{\'c},
  Serpico, and Zaharija{\v s}}}]{Petrovic:2014uda}
\bibinfo{author}{\bibfnamefont{J.}~\bibnamefont{Petrovi{\'c}}},
  \bibinfo{author}{\bibfnamefont{P.~D.} \bibnamefont{Serpico}},
  \bibnamefont{and} \bibinfo{author}{\bibfnamefont{G.}~\bibnamefont{Zaharija{\v
  s}}}, \bibinfo{journal}{\jcap} \textbf{\bibinfo{volume}{1410}},
  \bibinfo{pages}{052} (\bibinfo{year}{2014}), \eprint{1405.7928}.

\bibitem[{\citenamefont{{Moskalenko} and {Strong}}(1998)}]{1998ApJ...493..694M}
\bibinfo{author}{\bibfnamefont{I.~V.} \bibnamefont{{Moskalenko}}}
  \bibnamefont{and} \bibinfo{author}{\bibfnamefont{A.~W.}
  \bibnamefont{{Strong}}}, \bibinfo{journal}{\apj}
  \textbf{\bibinfo{volume}{493}}, \bibinfo{pages}{694} (\bibinfo{year}{1998}),
  \eprint{astro-ph/9710124}.

\bibitem[{\citenamefont{{Strong} et~al.}(2000)\citenamefont{{Strong},
  {Moskalenko}, and {Reimer}}}]{2000ApJ...537..763S}
\bibinfo{author}{\bibfnamefont{A.~W.} \bibnamefont{{Strong}}},
  \bibinfo{author}{\bibfnamefont{I.~V.} \bibnamefont{{Moskalenko}}},
  \bibnamefont{and} \bibinfo{author}{\bibfnamefont{O.}~\bibnamefont{{Reimer}}},
  \bibinfo{journal}{\apj} \textbf{\bibinfo{volume}{537}}, \bibinfo{pages}{763}
  (\bibinfo{year}{2000}), \eprint{astro-ph/9811296}.

\bibitem[{\citenamefont{{Strong} et~al.}(2004)\citenamefont{{Strong},
  {Moskalenko}, and {Reimer}}}]{2004ApJ...613..962S}
\bibinfo{author}{\bibfnamefont{A.~W.} \bibnamefont{{Strong}}},
  \bibinfo{author}{\bibfnamefont{I.~V.} \bibnamefont{{Moskalenko}}},
  \bibnamefont{and} \bibinfo{author}{\bibfnamefont{O.}~\bibnamefont{{Reimer}}},
  \bibinfo{journal}{\apj} \textbf{\bibinfo{volume}{613}}, \bibinfo{pages}{962}
  (\bibinfo{year}{2004}), \eprint{astro-ph/0406254}.

\bibitem[{\citenamefont{{Ackermann} et~al.}(2012)}]{2012ApJ...750....3A}
\bibinfo{author}{\bibfnamefont{M.}~\bibnamefont{{Ackermann}}}
  \bibnamefont{et~al.}, \bibinfo{journal}{\apj} \textbf{\bibinfo{volume}{750}},
  \bibinfo{eid}{3} (\bibinfo{year}{2012}), \eprint{1202.4039}.

\bibitem[{\citenamefont{Lorimer et~al.}(2006)}]{Lorimer:2006qs}
\bibinfo{author}{\bibfnamefont{D.~R.} \bibnamefont{Lorimer}}
  \bibnamefont{et~al.}, \bibinfo{journal}{Mon. Not. Roy. Astron. Soc.}
  \textbf{\bibinfo{volume}{372}}, \bibinfo{pages}{777} (\bibinfo{year}{2006}),
  \eprint{astro-ph/0607640}.

\bibitem[{\citenamefont{{Schlegel} et~al.}(1998)\citenamefont{{Schlegel},
  {Finkbeiner}, and {Davis}}}]{1998ApJ...500..525S}
\bibinfo{author}{\bibfnamefont{D.~J.} \bibnamefont{{Schlegel}}},
  \bibinfo{author}{\bibfnamefont{D.~P.} \bibnamefont{{Finkbeiner}}},
  \bibnamefont{and} \bibinfo{author}{\bibfnamefont{M.}~\bibnamefont{{Davis}}},
  \bibinfo{journal}{\apj} \textbf{\bibinfo{volume}{500}}, \bibinfo{pages}{525}
  (\bibinfo{year}{1998}), \eprint{astro-ph/9710327}.

\bibitem[{\citenamefont{Porter et~al.}(2008)\citenamefont{Porter, Moskalenko,
  Strong, Orlando, and Bouchet}}]{Porter_2008}
\bibinfo{author}{\bibfnamefont{T.~A.} \bibnamefont{Porter}},
  \bibinfo{author}{\bibfnamefont{I.~V.} \bibnamefont{Moskalenko}},
  \bibinfo{author}{\bibfnamefont{A.~W.} \bibnamefont{Strong}},
  \bibinfo{author}{\bibfnamefont{E.}~\bibnamefont{Orlando}}, \bibnamefont{and}
  \bibinfo{author}{\bibfnamefont{L.}~\bibnamefont{Bouchet}},
  \bibinfo{journal}{The Astrophysical Journal} \textbf{\bibinfo{volume}{682}},
  \bibinfo{pages}{400} (\bibinfo{year}{2008}),
  \urlprefix\url{https://doi.org/10.1086%2F589615}.

\bibitem[{\citenamefont{{Yusifov} and
  {K{\"u}{\c{c}}{\"u}k}}(2004)}]{2004A&A...422..545Y}
\bibinfo{author}{\bibfnamefont{I.}~\bibnamefont{{Yusifov}}} \bibnamefont{and}
  \bibinfo{author}{\bibfnamefont{I.}~\bibnamefont{{K{\"u}{\c{c}}{\"u}k}}},
  \bibinfo{journal}{\aap} \textbf{\bibinfo{volume}{422}}, \bibinfo{pages}{545}
  (\bibinfo{year}{2004}), \eprint{astro-ph/0405559}.

\bibitem[{\citenamefont{{Case} and {Bhattacharya}}(1998)}]{1998ApJ...504..761C}
\bibinfo{author}{\bibfnamefont{G.~L.} \bibnamefont{{Case}}} \bibnamefont{and}
  \bibinfo{author}{\bibfnamefont{D.}~\bibnamefont{{Bhattacharya}}},
  \bibinfo{journal}{\apj} \textbf{\bibinfo{volume}{504}}, \bibinfo{pages}{761}
  (\bibinfo{year}{1998}), \eprint{astro-ph/9807162}.

\bibitem[{\citenamefont{{Bronfman} et~al.}(2000)\citenamefont{{Bronfman},
  {Casassus}, {May}, and {Nyman}}}]{2000A&A...358..521B}
\bibinfo{author}{\bibfnamefont{L.}~\bibnamefont{{Bronfman}}},
  \bibinfo{author}{\bibfnamefont{S.}~\bibnamefont{{Casassus}}},
  \bibinfo{author}{\bibfnamefont{J.}~\bibnamefont{{May}}}, \bibnamefont{and}
  \bibinfo{author}{\bibfnamefont{L.~{\r{A}}.} \bibnamefont{{Nyman}}},
  \bibinfo{journal}{\aap} \textbf{\bibinfo{volume}{358}}, \bibinfo{pages}{521}
  (\bibinfo{year}{2000}), \eprint{astro-ph/0006104}.

\bibitem[{\citenamefont{{Kalberla} et~al.}(2010)\citenamefont{{Kalberla},
  {McClure-Griffiths}, {Pisano}, {Calabretta}, {Ford}, {Lockman},
  {Staveley-Smith}, {Kerp}, {Winkel}, {Murphy} et~al.}}]{2010A&A...521A..17K}
\bibinfo{author}{\bibfnamefont{P.~M.~W.} \bibnamefont{{Kalberla}}},
  \bibinfo{author}{\bibfnamefont{N.~M.} \bibnamefont{{McClure-Griffiths}}},
  \bibinfo{author}{\bibfnamefont{D.~J.} \bibnamefont{{Pisano}}},
  \bibinfo{author}{\bibfnamefont{M.~R.} \bibnamefont{{Calabretta}}},
  \bibinfo{author}{\bibfnamefont{H.~A.} \bibnamefont{{Ford}}},
  \bibinfo{author}{\bibfnamefont{F.~J.} \bibnamefont{{Lockman}}},
  \bibinfo{author}{\bibfnamefont{L.}~\bibnamefont{{Staveley-Smith}}},
  \bibinfo{author}{\bibfnamefont{J.}~\bibnamefont{{Kerp}}},
  \bibinfo{author}{\bibfnamefont{B.}~\bibnamefont{{Winkel}}},
  \bibinfo{author}{\bibfnamefont{T.}~\bibnamefont{{Murphy}}},
  \bibnamefont{et~al.}, \bibinfo{journal}{\aap} \textbf{\bibinfo{volume}{521}},
  \bibinfo{eid}{A17} (\bibinfo{year}{2010}), \eprint{1007.0686}.

\bibitem[{\citenamefont{{Planck Collaboration}}(2014)}]{2014A&A...571A..11P}
\bibinfo{author}{\bibnamefont{{Planck Collaboration}}}, \bibinfo{journal}{\aap}
  \textbf{\bibinfo{volume}{571}}, \bibinfo{eid}{A11} (\bibinfo{year}{2014}),
  \eprint{1312.1300}.

\bibitem[{\citenamefont{Abdollahi et~al.}(2019)}]{Fermi-LAT:2019yla}
\bibinfo{author}{\bibfnamefont{S.}~\bibnamefont{Abdollahi}}
  \bibnamefont{et~al.} (\bibinfo{collaboration}{Fermi-LAT})
  (\bibinfo{year}{2019}), \eprint{1902.10045}.

\bibitem[{\citenamefont{{Wood} et~al.}(2017)\citenamefont{{Wood}, {Caputo},
  {Charles}, {Di Mauro}, {Magill}, {Perkins}, and {Fermi-LAT
  Collaboration}}}]{2017ICRC...35..824W}
\bibinfo{author}{\bibfnamefont{M.}~\bibnamefont{{Wood}}},
  \bibinfo{author}{\bibfnamefont{R.}~\bibnamefont{{Caputo}}},
  \bibinfo{author}{\bibfnamefont{E.}~\bibnamefont{{Charles}}},
  \bibinfo{author}{\bibfnamefont{M.}~\bibnamefont{{Di Mauro}}},
  \bibinfo{author}{\bibfnamefont{J.}~\bibnamefont{{Magill}}},
  \bibinfo{author}{\bibfnamefont{J.~S.} \bibnamefont{{Perkins}}},
  \bibnamefont{and} \bibinfo{author}{\bibnamefont{{Fermi-LAT Collaboration}}},
  in \emph{\bibinfo{booktitle}{35th International Cosmic Ray Conference
  (ICRC2017)}} (\bibinfo{year}{2017}), vol. \bibinfo{volume}{301} of
  \emph{\bibinfo{series}{International Cosmic Ray Conference}}, p.
  \bibinfo{pages}{824}, \eprint{1707.09551}.

\bibitem[{\citenamefont{{Navarro} et~al.}(1997)\citenamefont{{Navarro},
  {Frenk}, and {White}}}]{1997ApJ...490..493N}
\bibinfo{author}{\bibfnamefont{J.~F.} \bibnamefont{{Navarro}}},
  \bibinfo{author}{\bibfnamefont{C.~S.} \bibnamefont{{Frenk}}},
  \bibnamefont{and} \bibinfo{author}{\bibfnamefont{S.~D.~M.}
  \bibnamefont{{White}}}, \bibinfo{journal}{\apj}
  \textbf{\bibinfo{volume}{490}}, \bibinfo{pages}{493} (\bibinfo{year}{1997}),
  \eprint{astro-ph/9611107}.

\bibitem[{\citenamefont{{Mattox} et~al.}(1996)\citenamefont{{Mattox},
  {Bertsch}, {Chiang}, {Dingus}, {Digel}, {Esposito}, {Fierro}, {Hartman},
  {Hunter}, {Kanbach} et~al.}}]{1996ApJ...461..396M}
\bibinfo{author}{\bibfnamefont{J.~R.} \bibnamefont{{Mattox}}},
  \bibinfo{author}{\bibfnamefont{D.~L.} \bibnamefont{{Bertsch}}},
  \bibinfo{author}{\bibfnamefont{J.}~\bibnamefont{{Chiang}}},
  \bibinfo{author}{\bibfnamefont{B.~L.} \bibnamefont{{Dingus}}},
  \bibinfo{author}{\bibfnamefont{S.~W.} \bibnamefont{{Digel}}},
  \bibinfo{author}{\bibfnamefont{J.~A.} \bibnamefont{{Esposito}}},
  \bibinfo{author}{\bibfnamefont{J.~M.} \bibnamefont{{Fierro}}},
  \bibinfo{author}{\bibfnamefont{R.~C.} \bibnamefont{{Hartman}}},
  \bibinfo{author}{\bibfnamefont{S.~D.} \bibnamefont{{Hunter}}},
  \bibinfo{author}{\bibfnamefont{G.}~\bibnamefont{{Kanbach}}},
  \bibnamefont{et~al.}, \bibinfo{journal}{\apj} \textbf{\bibinfo{volume}{461}},
  \bibinfo{pages}{396} (\bibinfo{year}{1996}).

\bibitem[{\citenamefont{Herold and Malyshev}(2019)}]{Herold:2019pei}
\bibinfo{author}{\bibfnamefont{L.}~\bibnamefont{Herold}} \bibnamefont{and}
  \bibinfo{author}{\bibfnamefont{D.}~\bibnamefont{Malyshev}},
  \bibinfo{journal}{Astron. Astrophys.} \textbf{\bibinfo{volume}{625}},
  \bibinfo{pages}{A110} (\bibinfo{year}{2019}), \eprint{1904.01454}.

\bibitem[{\citenamefont{Abdo et~al.}(2013)}]{TheFermi-LAT:2013ssa}
\bibinfo{author}{\bibfnamefont{A.~A.} \bibnamefont{Abdo}} \bibnamefont{et~al.}
  (\bibinfo{collaboration}{Fermi-LAT}), \bibinfo{journal}{Astrophys. J. Suppl.}
  \textbf{\bibinfo{volume}{208}}, \bibinfo{pages}{17} (\bibinfo{year}{2013}),
  \eprint{1305.4385}.

\bibitem[{\citenamefont{{Robin} et~al.}(2012)\citenamefont{{Robin}, {Marshall},
  {Schultheis}, and {Reyl{\'e}}}}]{2012A&A...538A.106R}
\bibinfo{author}{\bibfnamefont{A.~C.} \bibnamefont{{Robin}}},
  \bibinfo{author}{\bibfnamefont{D.~J.} \bibnamefont{{Marshall}}},
  \bibinfo{author}{\bibfnamefont{M.}~\bibnamefont{{Schultheis}}},
  \bibnamefont{and}
  \bibinfo{author}{\bibfnamefont{C.}~\bibnamefont{{Reyl{\'e}}}},
  \bibinfo{journal}{\aap} \textbf{\bibinfo{volume}{538}}, \bibinfo{eid}{A106}
  (\bibinfo{year}{2012}), \eprint{1111.5744}.

\bibitem[{\citenamefont{{Ferri{\`e}re}
  et~al.}(2007)\citenamefont{{Ferri{\`e}re}, {Gillard}, and
  {Jean}}}]{2007A&A...467..611F}
\bibinfo{author}{\bibfnamefont{K.}~\bibnamefont{{Ferri{\`e}re}}},
  \bibinfo{author}{\bibfnamefont{W.}~\bibnamefont{{Gillard}}},
  \bibnamefont{and} \bibinfo{author}{\bibfnamefont{P.}~\bibnamefont{{Jean}}},
  \bibinfo{journal}{\aap} \textbf{\bibinfo{volume}{467}}, \bibinfo{pages}{611}
  (\bibinfo{year}{2007}), \eprint{astro-ph/0702532}.

\bibitem[{\citenamefont{Di~Mauro et~al.}(2019)\citenamefont{Di~Mauro, Manconi,
  and Donato}}]{DiMauro:2019hwn}
\bibinfo{author}{\bibfnamefont{M.}~\bibnamefont{Di~Mauro}},
  \bibinfo{author}{\bibfnamefont{S.}~\bibnamefont{Manconi}}, \bibnamefont{and}
  \bibinfo{author}{\bibfnamefont{F.}~\bibnamefont{Donato}}
  (\bibinfo{year}{2019}), \eprint{1908.03216}.

\end{thebibliography}

\end{document}